\documentclass[preprint,3p,review]{elsarticle}

\usepackage[numbers]{natbib}

\usepackage[english]{babel}
\usepackage{amssymb}
\usepackage{amsmath}
\usepackage{subcaption}
\newsavebox{\measurebox}
\usepackage[export]{adjustbox}[2011/08/13]
\usepackage{graphicx}

\usepackage[colorlinks=true, allcolors=blue]{hyperref}

\usepackage{mathrsfs}
\usepackage{booktabs}
\usepackage{multirow}
\usepackage{svg}
\usepackage[np]{numprint}
	\npthousandsep{,}
	\npdecimalsign{.}
\usepackage[normalem]{ulem}
\usepackage{MnSymbol,bbding,pifont}
\usepackage{lineno}
\usepackage{setspace}
\usepackage{todonotes}
\usepackage{ulem}

\newcommand{\red}[1]{\textcolor{black}{#1}}

\newcommand{\blue}[1]{\textcolor{black}{#1}}

\journal{Computers $\&$ Fluids}

\begin{document}

\begin{frontmatter}
%\todo[author=Jiayi, color=red!30, inline]{Write here.}
%\todo[author=Didier, color=green!30, inline]{Write here.}
%\todo[author=Guillaume, color=yellow!30, inline]{Write here.}
%\todo[author=Pierre-Emmanuel, color=cyan!30, inline]{Write here.}

\title{Revisiting Tensor Basis Neural Networks for Reynolds stress modeling: application to plane channel and square duct flows}

\author[1,2]{Jiayi Cai}
\author[1]{Pierre-Emmanuel Angeli}
\author[3]{Jean-Marc Martinez}
\author[3]{Guillaume Damblin}
\author[2]{Didier Lucor}

\affiliation[1]{organization={Université Paris-Saclay, CEA},%Department and Organization
            addressline={Service de Thermo-hydraulique et de Mécanique des Fluides}, 
            city={Gif-sur-Yvette},
            postcode={91191}, 
            country={France}}
            
\affiliation[2]{organization={Université Paris-Saclay, CNRS, Laboratoire Interdisciplinaire des Sciences du Numérique},%Department and Organization
            city={Orsay},
            postcode={91405}, 
            country={France}}            

\affiliation[3]{organization={Université Paris-Saclay, CEA},%Department and Organization
            addressline={Service de Génie Logiciel pour la Simulation}, 
            city={Gif-sur-Yvette},
            postcode={91191}, 
            country={France}}
            
\begin{abstract}
Recent advances in computing power devoted to Computational Fluid Dynamics (CFD) have opened up a new avenue for addressing the turbulence closure problem through Machine Learning (ML). Several Tensor Basis Neural Network (TBNN) frameworks aimed at enhancing turbulence Reynolds-Averaged Navier-Stokes (RANS) modeling have recently been proposed in the literature as data-driven constitutive models for systems with known invariance properties. However, persistent ambiguities remain regarding the physical adequacy of applying the General Eddy Viscosity Model (GEVM) within these frameworks. This work aims at investigating this aspect in an \textit{a priori} stage for better predictions of the Reynolds stress anisotropy tensor, while preserving the Galilean and rotational invariances. In particular, we propose a general framework providing optimal tensor basis models for two types of canonical flows with increasing complexity: Plane Channel Flow (PCF) and Square Duct Flow (SDF). Subsequently, deep neural networks based on these optimal models are trained using state-of-the-art strategies to achieve a balanced and physically sound prediction of the full anisotropy tensor. \red{\textit{A priori} results obtained by the proposed framework are in very good agreement with the reference \red{Direct Numerical Simulations (DNS)} data.} Notably, our shallow network with three layers excels in providing accurate predictions of the anisotropy tensor for PCF at unobserved friction Reynolds numbers, both in interpolation and extrapolation scenarios. The learning of the SDF case is more challenging because of its physical nature but also due to a lack of training data at various regimes. We propose a numerical strategy to alleviate this problem based on Transfer Learning (TL). In order to more efficiently generalize to an unseen intermediate $\mathrm{Re}_\tau$ regime, we take advantage of our prior knowledge acquired from a training with a larger and wider dataset. Our results indicate the potential of the developed network model, and demonstrate the feasibility and efficiency of the TL process in terms of training data size and training time. Based on these results, we believe there is a promising future by integrating these neural networks into an adapted in-house RANS solver.
\end{abstract}


\end{frontmatter}

%\linenumbers

\section{Introduction}
Reynolds-Averaged Navier-Stokes (RANS) is the most extensively utilized method for engineering analysis and for design in a wide range of industries, primarily due to its high cost-effectiveness and decent accuracy, compared with Direct Numerical Simulation (DNS) and Large Eddy Simulation (LES). However, there exist several limitations in RANS models, especially for configurations with separation effects and secondary flows, as documented in the literature \cite{durbin2018}. Hence, it remains an area of active research and development to pursue more accurate RANS closure models.

The modeling problem encountered in the RANS approach is known as the closure problem, which is caused by the appearance of the unknown Reynolds Stress Tensor (RST) in the RANS equations due to the time-averaging operator. As a result, additional models need to be included in order to close the system of equations. Traditionally, these models were developed by the combination of physical knowledge and experimental coefficient calibration on simple flow configurations. For example, the commonly used $k - \epsilon$ model includes five coefficients calibrated for plane jets and simple shear flows, yet notably not suitable for axisymmetric jets \cite{launder1974}. 

In recent years, the development of computing resources has shed new light on RANS modeling to improve classical RANS closure models, via Machine Learning (ML) techniques (see \cite{duraisamy2019, brunton2020} for a comprehensive review). There are mainly two approaches: via the closure coefficients, and via the RST. 

The former approach centers on the re-calibration of the closure coefficients via Bayesian approach \cite{beck1998} and aims at establishing reliable models with quantified error estimates. Various investigations have been conducted in this direction encompassing a variety of turbulence models: the Spalart-Allmaras, the Launder-Sharma $k-\epsilon$, and the $k-\omega$ models, and more \cite{cheung2011, edeling2014}. 

On the other hand, the latter approach is intended to provide more accurate predictions directly on the RST itself. The published literature in this field is dominated by supervised algorithms, among which Neural Networks (NNs) and Random Forests (RFs) are the most widely utilized. Ling \textit{et al.} \cite{ling2016} designed a special network structure called Tensor Basis Neural Network (TBNN) to incorporate the General Eddy Viscosity Model (GEVM) \cite{pope1975} for Reynolds anisotropy tensor predictions. Sáez de Ocáriz Borde \textit{et al.} \cite{saezdeocarizborde2021, saezdeocarizborde2022} developed a Convolutional Neural Network (CNN) in order to better capture non-local effects. Quattromini \textit{et al.} \cite{quattromini2023} employed a Graph Neural Network (GNN) to predict the RST, aiming to surpass the constraints of traditional NNs (over-fitting, necessity of a large amount of data, lack of generalizability) and CNNs (necessity of structured mesh). Xiao et co-workers \cite{wang2017, wu2018} adopted the RF algorithm to learn the discrepancies between the RANS-predicted RST and the DNS data. Analogously to the TBNN framework, Kaandorp and Dwight \cite{kaandorp2020} proposed the Tensor Basis Random Forest (TBRF), which performs comparably with the TBNN and is simpler to train. Other less popular machine learning algorithms can also be found in the community. The team of Weatheritt and Sandberg \cite{weatheritt2016, weatheritt2017, zhao2020} used the Gene Expression Programming (GEP) to provide an Explicit Algebraic Reynolds Stress Model (EARSM) and performed \textit{a priori} and/or \textit{a posteriori} tests on different turbomachinery flows. McConkey \textit{et al.} \cite{mcconkey2023} employed the eXtreme Gradient Boosting (XGBoost) algorithm due to its superiority on tabular data. In their work, the low-tuning cost and good performance of XGBoost were demonstrated, compared to NNs and RFs upon a huge dataset \cite{mcconkey2021}. 

The present study is primarily driven by the promising outcomes achieved in previous machine-learning-assisted RANS modeling frameworks. Of particular relevance to this work is the research conducted by Ling \textit{et al.} \cite{ling2016}, and related publications \cite{kaandorp2020, zhang2018, fang2020}. The major advantage of their TBNN framework is that it guarantees Galilean and rotational invariances. In Ling \textit{et al.}'s paper, they conducted TBNN training on an extensive RANS and LES/DNS database, encompassing various flow types (e.g., channel flow, duct flow, square cylinder flow, etc.). Subsequently, they tested the model on two flows that are particularly sensitive to accurate Reynolds stress anisotropy modeling: a similar duct flow (but within a distinct turbulent regime) and a new case involving flow over a wavy wall (not included in the database). Although \textit{a priori} TBNN results demonstrated improved predictions compared to standard RANS, including the new extrapolated flow case, TBNN predictions did not achieve the level of accuracy seen in DNS references. Moreover, a series of studies have been carried out on the basis of Ling et al.'s work, yet lacking adequate caution concerning the application of Pope's GEVM \cite{pope1975}. Despite its generalized form, this model is only valid for a high Reynolds number nearly homogeneous flow where local effects dominate. Therefore, in its standard form, it does not take the Reynolds number into account. \red{Furthermore, two model forms were introduced in the original paper with distinct integrity bases: one two-dimensional form and one general three-dimensional form \cite{pope1975}. 
However, it has been observed that earlier research often universally applied the general model version without making necessary distinctions 
 \cite{ling2016, fang2020}. 
 It is essential to note that the proper application condition for these two forms is subtle. Applying the former to two-dimensional flows and the latter to three-dimensional flows may seem straightforward, but is not the proper usage and requires clarification.}

In order to tackle the aforementioned limitations, we revisit the TBNN framework with more focus on crucial physics model aspects, and propose a general augmented TBNN (referred to as aTBNN in the following) framework coupled with modern training strategies: 
\begin{itemize}
    \item Additional quantities are included in the GEVM, accounting for non-local effects and the Reynolds number dependence of turbulent flows.
    \item \red{The integrity basis is discussed separately for two types of canonical flows with increasing complexity, aiming to clarify the proper usage of the two forms of Pope's GEVM and identify an optimized tensor basis for the current flow cases.}
    \item The training process is corporated with state-of-the-art strategies to optimize a multi-part loss function with the contribution of each component of the anisotropy tensor.
\end{itemize}

Various aTBNN models are established in the present study, individually trained, and subsequently evaluated on two distinct flow configurations: the Plane Channel Flow (PCF) and the Square Duct Flow (SDF), for which DNS data are available \cite{moser1999, kaneda2021, hoyas2022, pirozzoli2018}. In particular, unlike the approach taken in Ling \textit{et al.}'s study, where the input features of their networks consist of RANS quantities \cite{ling2016}, this work employs DNS data as both inputs and outputs. Indeed, a major concern within the ML-RANS framework revolves around the capability to accurately reconstruct the mean velocity following the integration of the ML model with the RANS solver. Several studies have demonstrated that, even when injecting DNS quantities directly into the RANS solver, the overall predictions on the velocity field remain unsatisfactory \cite{poroseva2016, thompson2016}. This issue may be attributed to data-model inconsistency or ill-conditioning problems, as highlighted in previous research \cite{beck2021, wu2019b, amarloo2022b}. Therefore, the coupling framework between the ML model and the RANS solver holds significant importance. While it falls outside the scope of the current study, it is a subject of consideration for our ongoing \textit{a posteriori} validation efforts. Three major coupling frameworks are prevalent in the existing literature: namely the frozen and the iterative substitutions \cite{yin2022b}, and the fully online training framework \cite{zhao2020}. Ling \textit{et al.} employed the first type of coupling, utilizing RANS quantities as their inputs. This corrective approach resulted in qualitatively more accurate velocity predictions in their \textit{a-posteriori} tests compared to standard RANS. However, there still remained a quantitative discrepancy with DNS results. For our future studies, we have chosen the second iterative substitution framework, utilizing DNS data as network inputs and outputs. In this way, the developed NN model can be seen as a replacement of classical closure models \cite{kaandorp2020}. While less explored in the existing literature, this approach has the potential to yield results that are more physically relevant and accurate, provided that the learning process for mapping high-fidelity mean flow features to RST true value is effectively managed \cite{yin2022b}.

Moreover, neural network models in previous studies were mainly assessed one unobserved friction Reynolds number $\mathrm{Re}_\tau$ at a time. This evaluation was conducted either in an interpolation case, where the tested $\mathrm{Re}_\tau$ is within the range of the learning database, or conversely, in an extrapolation case. In our work, with the benefit of newly available DNS databases on PCF \cite{kaneda2021, hoyas2022}, we will, for the first time, be able to concurrently evaluate the predictive performance of our aTBNN models in both interpolation and extrapolation scenarios. This is challenging, yet it is of utmost importance, as the prediction model must demonstrate accuracy in both scenarios to be practically useful. For the second learning flow, the SDF, only interpolation capacity is evaluated because of the lack of DNS databases at different Reynolds numbers. Transfer Learning (TL) \cite{pan2010} is proposed in this work to mitigate the data lack issue, aiming to transfer knowledge among networks trained by datasets at different Reynolds numbers. This technique has been successfully applied in previous works involving CNN models, resulting in promising outcomes \cite{guastoni2020, guastoni2021, guan2022}. For a comprehensive review of TL's application in data-driven turbulence modeling, readers can refer to Subel \textit{et al.}'s work \cite{subel2023}. 

The remainder of this article is organized as follows. Section~\ref{sec:metho} presents the governing equations of RANS modeling, the GEVM, and the TBNN. The deep learning framework is thoroughly described in Section~\ref{sec:framework}, from data collection to training strategies. Particular attention is paid to the application of the GEVM to our two learning flows. Results are then presented and discussed in Section~\ref{sec:results}. Finally, the main conclusions and perspectives of the work are given in Section~\ref{sec:conclu}. 

\section{Methodology}\label{sec:metho}
In this section, we first provide a theoretical background on RANS modeling, notably the GEVM proposed by Pope \cite{pope1975}. Afterwards, we give a brief review of Ling \textit{et al.}'s TBNN \cite{ling2016}.

\subsection{RANS modeling}\label{sec:RANS}
The incompressible Navier-Stokes equations for a Newtonian fluid with unvarying viscosity can be stated as:
\begin{singlespace} 
\begin{align} 
    \begin{cases}
\dfrac{\partial u_i}{\partial x_i}=0 \\[10pt]
\dfrac{\partial u_i}{\partial t}+{u}_j\dfrac{\partial {u}_i}{\partial x_j}=-\dfrac{1}{\rho}\dfrac{\partial {p}}{\partial x_i}+\nu\dfrac{\partial ^2 {u}_i}{\partial x_j \partial x_j} \label{eq:NS}
    \end{cases}   
\end{align}
\end{singlespace} 
\noindent where $u_i$, $x_i$, $t$, $\rho$, $p$ and $\nu$ are the components of the velocity vector, the spatial coordinates, the time, the density, the pressure and the kinematic viscosity, respectively.

By decomposing the velocity and pressure into their mean and fluctuating components ($u_i=\overline{u}_i+u^\prime_i$, $p=\overline{p}+p^\prime$) and averaging the equations, we obtain the RANS equations:
\begin{singlespace} 
\begin{align} 
    \begin{cases}
\dfrac{\partial {\overline{u}_i}}{\partial x_i}=0 \\[10pt]
\dfrac{\partial \overline{u}_i}{\partial t}+\overline{u}_j\dfrac{\partial \overline{u}_i}{\partial x_j}=-\dfrac{1}{\rho}\dfrac{\partial \overline{p}}{\partial x_i}+\nu\dfrac{\partial ^2 \overline{u}_i}{\partial x_j \partial x_j}-\dfrac{\partial \overline {u^\prime_iu^\prime_j}}{\partial x_j}. \label{eq:RANS}
    \end{cases}   
\end{align}
\end{singlespace} 

The major difficulty in RANS modeling is to relate the unknown RST $\mathscr{R}_{ij}=\overline {u^\prime_iu^\prime_j}$ appearing in Eq.~\eqref{eq:RANS} to the mean flow field. The Linear Eddy Viscosity Model (LEVM) is the most widely used model to tackle this closure problem, and can be approximated as:
\begin{equation}
\overline{u^\prime_iu^\prime_j}=-\nu_t\left(\dfrac{\partial \overline{u}_i}{\partial x_j}+\dfrac{\partial \overline{u}_j}{\partial x_i}\right)+\dfrac{2}{3}k\delta_{ij}, \label{eq:BousR}
\end{equation}

\noindent where $\nu_t$ and $k$ are the eddy viscosity and turbulent kinetic energy, respectively; $\delta_{ij}$ denotes the Kronecker delta. This expression is known as the Boussinesq assumption \cite{boussinesq1897a}.

However, this simple linear relationship between the RST and the mean velocity gradients turns out to be inaccurate for some complex flows, especially involving secondary flows and curvature effects. An amount of Non-Linear Eddy Viscosity Models (NLEVM) at higher orders have been developed to capture these effects. For the sake of example, a Quadratic Eddy Viscosity Model (QEVM) simplified from Craft \textit{et al.}'s cubic model \cite{craft1996} can be written as:
\begin{equation}
\label{eq:craft}
\begin{split}
b_{ij}=&-\dfrac{\nu_t}{k}S_{ij}+C_1\dfrac{\nu_t}{\epsilon}\left(S_{ik}S_{kj}-\dfrac{1}{3}S_{kl}S_{kl}\delta_{ij}\right)+C_2\dfrac{\nu_t}{\epsilon}\left(R_{ik}S_{kj}+\dfrac{1}{3}R_{jk}S_{ki}\right) \\
&+C_3\dfrac{\nu_t}{\epsilon}\left(R_{ik}R_{jk}-\dfrac{1}{3}R_{kl}R_{kl}\delta_{ij}\right),
\end{split}
\end{equation}

\noindent where $\epsilon$ is the turbulent dissipation rate; $C_1$, $C_2$ and $C_3$ are parameters usually taken as constants, which are calibrated on experiments or high-fidelity simulations upon simple flow configurations; $b_{ij}$, $S_{ij}$ and $R_{ij}$ are the Reynolds stress anisotropy, mean strain-rate and rotation-rate tensors, respectively:
\begin{equation}
b_{ij}=\dfrac{\overline{u^\prime_iu^\prime_j}}{2k}-\dfrac{1}{3}\delta_{ij}\label{eq:defb}
\end{equation} 
\begin{equation}
S_{ij}=\dfrac{1}{2}\left(\dfrac{\partial \overline{u}_i}{\partial x_j}+\dfrac{\partial \overline{u}_j}{\partial x_i}\right)\label{eq:defS}
\end{equation}
\begin{equation}
R_{ij}=\dfrac{1}{2}\left(\dfrac{\partial \overline{u}_i}{\partial x_j}-\dfrac{\partial \overline{u}_j}{\partial x_i}\right)\label{eq:defR}
\end{equation}

By combining Eq.~\eqref{eq:BousR}, Eq.~\eqref{eq:defb} and Eq.~\eqref{eq:defS}, one can rewrite the LEVM shown in Eq.~\eqref{eq:BousR} as follows:
\begin{equation}
\label{eq:BousB}
b_{ij}=-\dfrac{\nu_t}{k}S_{ij}
\end{equation}
\noindent which is equivalent to Craft \textit{et al.}'s cubic model shown in Eq.~\eqref{eq:craft} in first-order approximation. 

Specially in the RANS standard $k-\epsilon$ model \cite{launder1974}, the eddy viscosity can be expressed as:
\begin{equation}
\label{eq:nu}
\nu_t = C_\mu \dfrac{k^2}{\epsilon}
\end{equation}
\noindent where $C_\mu$ is a calibrated parameter, generally taken positively as 0.09.

\subsection{General Eddy Viscosity Model (GEVM)} \label{sec:EVM}
One of the most generalized NLEVM was proposed by Pope \cite{pope1975} in order to extend the universality of RANS closure models. Pope's approach focused on the Reynolds stress anisotropy tensor $\textbf{b}$ and postulated that it can be expressed as a function of normalized tensors $\textbf{S}^*$ and $\textbf{R}^*$ for a homogeneous flow:
\begin{equation}
\textbf{b}=\textbf{b}(\textbf{S}^*, \textbf{R}^*), \label{eq:hypoPope}
\end{equation}

\noindent where $\textbf{S}^*$ and $\textbf{R}^*$ are respectively the mean strain-rate and the rotation-rate tensors normalized by a turbulent time scale formed with the turbulent kinetic energy and dissipation rate:
\begin{equation}
S_{ij}^*=\dfrac{1}{2}\dfrac{k}{\epsilon}\left(\dfrac{\partial \overline{u}_i}{\partial x_j}+\dfrac{\partial \overline{u}_j}{\partial x_i}\right) \label{eq:defS*}
\end{equation}
\begin{equation}
R_{ij}^*=\dfrac{1}{2}\dfrac{k}{\epsilon}\left(\dfrac{\partial \overline{u}_i}{\partial x_j}-\dfrac{\partial \overline{u}_j}{\partial x_i}\right) \label{eq:defR*}
\end{equation}

By furtherly supposing the function \eqref{eq:hypoPope} as a polynomial function and applying the Cayley-Hamilton theorem, Pope obtained the following GEVM model of $\textbf{b}$, which expresses as a series of $n$ finite tensor polynomials:
\begin{equation}
\textbf{b}(\textbf{S}^*, \textbf{R}^*)=\sum_{n}g^{(n)}\left(\lambda_1^*, \lambda_2^*...\right)\textbf{T}^{*(n)} \label{generalPope}
\end{equation}
\noindent where $g^{(n)}$ are coefficient functions depending on physical independent invariants $\lambda_i^*$ and $\textbf{T}^{*(n)}$ are basis tensors depending on $\textbf{S}^*$ and $\textbf{R}^*$.

For general flows, there are five invariants and ten tensors $(1\leq n \leq10)$:
\begin{equation}
\textbf{b}=\sum_{n=1}^{10} g^{(n)}\left(\{\lambda_1^*\}_{i=1,2...,5}\right)\textbf{T}^{*(n)} \label{eq:Pope3D}
\end{equation}

\noindent with
\begin{equation}
\lambda_1^*=\text{tr}(\textbf{S}^{*2})\, ,\quad \lambda_2^*=\text{tr}(\textbf{R}^{*2})\, ,\quad \lambda_3^*=\text{tr}(\textbf{S}^{*3})\, ,\quad \lambda_4^*=\text{tr}(\textbf{R}^{*2}\textbf{S})\, ,\quad \lambda_5^*=\text{tr}(\textbf{R}^{*2}\textbf{S}^{*2})\label{eq:PopeInvariant3D}
\end{equation}
\begin{align} 
    \begin{cases}
     	\textbf{T}^{*(1)} =  \textbf{S}^{*} &  \textbf{T}^{*(2)} =  \textbf{S}^{*}\textbf{R}^{*}-\textbf{R}^{*}\textbf{S}^{*}  \\
      	\textbf{T}^{*(3)} =  \textbf{S}^{*2}-\dfrac{\lambda_1^*}{3}\textbf{I}_3 & \textbf{T}^{*(4)} =  \textbf{R}^{*2}-\dfrac{\lambda_2^*}{3}\textbf{I}_3 \\
      	\textbf{T}^{*(5)} =  \textbf{R}^{*}\textbf{S}^{*2}-\textbf{S}^{*2}\textbf{R}^{*} & \textbf{T}^{*(6)} =  \textbf{R}^{*2}\textbf{S}^{*}+\textbf{S}^{*}\textbf{R}^{*2}-\dfrac{2\lambda_4^*}{3}\textbf{I}_3\\
      	\textbf{T}^{*(7)}=\textbf{R}^{*}\textbf{S}^{*}\textbf{R}^{*2}-\textbf{R}^{*2}\textbf{S}^{*}\textbf{R}^{*} & \textbf{T}^{*(8)}=\textbf{S}^{*}\textbf{R}^{*}\textbf{S}^{*2}-\textbf{S}^{*2}\textbf{R}^{*}\textbf{S}^{*} \\
      	\textbf{T}^{*(9)}=\textbf{R}^{*2}\textbf{S}^{*2}+\textbf{S}^{*2}\textbf{R}^{*2}-\dfrac{2\lambda_5^*}{3}\textbf{I}_3 & \textbf{T}^{*(10)}=\textbf{R}\textbf{S}^{*2}\textbf{R}^{*2}-\textbf{R}^{*2}\textbf{S}^{*2}\textbf{R}\\
     \end{cases}\label{eq:PopeTensor3D}
\end{align}
\noindent where $\textbf{I}_3$ denotes the identity tensor. 

It can be noticed that Pope's model is a generalized form of LEVM shown in Eq. \eqref{eq:BousB} in first-order approximation and the QEVM shown in Eq. \eqref{eq:craft} in second-order approximation. In particular, given Eqs. \eqref{eq:BousB} and \eqref{eq:nu}, the coefficient function $g^{(1)}$ is identified with $-C_\mu$ and should therefore be negative.

Specifically for flows where the mean velocity and the variation of mean quantities in one direction are zero, Pope proposed a simplified model version with only two invariants and a basis of three tensors $(0\leq n \leq2)$:
\begin{equation}
\textbf{b}=\sum_{n=0}^{2} g^{(n)}\left(\lambda_1^*, \lambda_2^*\right)\textbf{T}^{*(n)} \label{eq:2D_Pope}
\end{equation}

\noindent with
\begin{equation}
\lambda_1^*=\text{tr}(\textbf{S}^{*2})\, ,\quad \lambda_2^*=\text{tr}(\textbf{R}^{*2})\label{eq:PopeInvariant2D}
\end{equation}
\begin{equation}
    \begin{cases}

     	\textbf{T}^{*(0)} =  \dfrac{1}{2}\textbf{I}_2 - \dfrac{1}{3}\textbf{I}_3\\
     	\textbf{T}^{*(1)} =  \textbf{S}^{*} \\
      	\textbf{T}^{*(2)} =  \textbf{S}^{*}\textbf{R}^{*}-\textbf{R}^{*}\textbf{S}^{*}\\
    \end{cases}\label{eq:PopeTensor2D}
\end{equation}
\noindent where $\textbf{I}_2=\mathrm{diag}(1,1,0)$ or its permutations depending on the characterizing direction of the flow. If, for example, there is zero mean velocity and invariance along the $x_3$ direction, then $\textbf{I}_2=\mathrm{diag}(1,1,0)$. One might notice that the invariants and basis tensors here are the same as those of the general model shown in Eqs.~\eqref{eq:PopeInvariant3D} and \eqref{eq:PopeTensor3D},  except for the choice of $\textbf{T}^{*(0)}$ instead of $\textbf{T}^{*(3)}$. In fact, it can be easily demonstrated that $\textbf{T}^{*(3)}=-\lambda_1^*\textbf{T}^{*(0)}$ under the restricted condition of this simplified case.

Similarly to Navier-Stokes equations, RANS equations and any other traditional closure models, Pope's GEVM, given in Eq.~\eqref{generalPope}, satisfies the Galilean and rotational invariances, which means that this model remains identical while undergoing a rectilinear and uniform motion or a rotation at a constant angle. The consideration of these invariances is fundamental since they are properties to which the fluid flow physically obeys. \red{However, it is essential to note that Pope's model relies on the fundamental assumption that the Reynolds stresses are locally determined by rates of strain and local scalar quantities. This assumption becomes questionable in inhomogeneous flow regions, such as the near-wall region, as discussed in \cite{pope2000d}, where non-local effects, should be considered.} This issue sometimes referred to as a multi-value problem \cite{liu2021a, jiang2021} will be addressed hereinafter.

\subsection{Tensor Basis Neural Network (TBNN)}\label{sec:TBNN}
\begin{figure}
    \centering
    \includegraphics[width=0.5\linewidth]{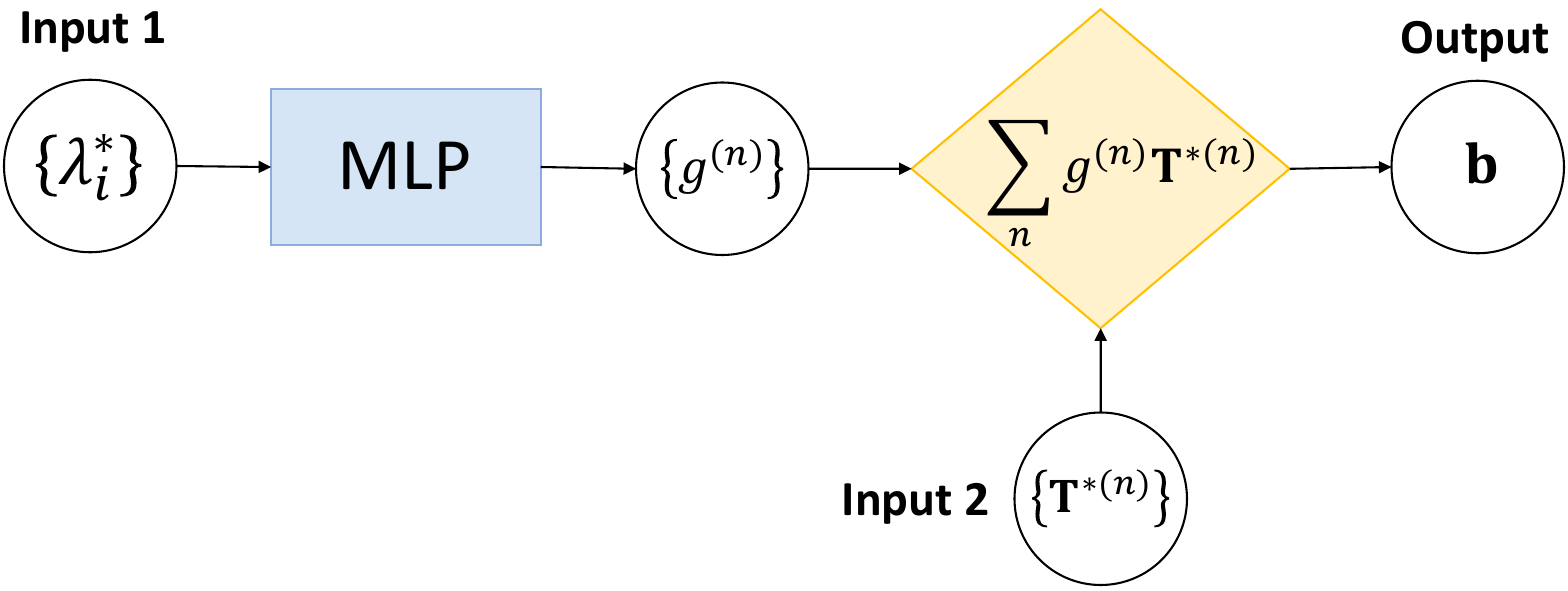}
    \caption{Architecture of the TBNN.}
    \label{fig:tbnn_ling}
\end{figure}

Ling \textit{et al.} \cite{ling2016} designed the TBNN, the architecture of which is illustrated in Figure \ref{fig:tbnn_ling}. The TBNN architecture can be regarded as a deep learning interpretation of Pope's GEVM with 5 invariants and 10 tensor shown in Eq.~\eqref{eq:Pope3D}. Two input layers are provided in the TBNN, one containing the invariants $\lambda_1, ..., \lambda_5$ and the other composed of the tensors $\textbf{T}^{(n)}$ for $n=1, ..., 10$. The first input layer is followed by 8 hidden layers, with 30 nodes per layer, in order to learn the ten coefficients functions $g^{(n)}$ for $n=1, ..., 10$ of Eq.~\eqref{eq:Pope3D} in the final hidden layer. The latter is then merged with a basis tensor input layer by element-wise multiplications so as to give final predictions on Reynolds stress anisotropy tensor. This particular architecture ensures the preservation of both Galilean and rotational invariances, as Pope's model does. 

In their original paper, the TBNN was trained on an extensive RANS and LES/DNS database, encompassing various canonical flows (e.g., channel flow, duct flow, square cylinder flow, etc.). Subsequently, they tested the model on two flows that are
particularly sensitive to accurate Reynolds stress anisotropy modeling: a similar duct flow (but within a distinct turbulent regime) and a new case involving flow over a wavy wall (not included in the database). Despite the wide variety of flows under consideration, both \textit{a priori} predictions on Reynolds stress anisotropy tensor and \textit{a posteriori} results on mean velocity yielded by TBNN were more accurate than traditional RANS models and a generic neural network of the Multi-Layer Perceptron (MLP) type that did not embed invariance properties.

A series of studies have been conducted, drawing inspiration from the aforementioned TBNN framework \cite{kaandorp2020, zhang2018, fang2020}. However, we have identified ongoing ambiguities regarding the application of Pope's GEVM. These issues will be addressed in the subsequent sections, focusing on two canonical learning flows.

\section{Deep learning framework for Reynolds anisotropy tensor predictions}\label{sec:framework}
\subsection{Flow cases}\label{sec:flow}
\begin{figure}[h]
\centering
\includegraphics[scale=0.25]{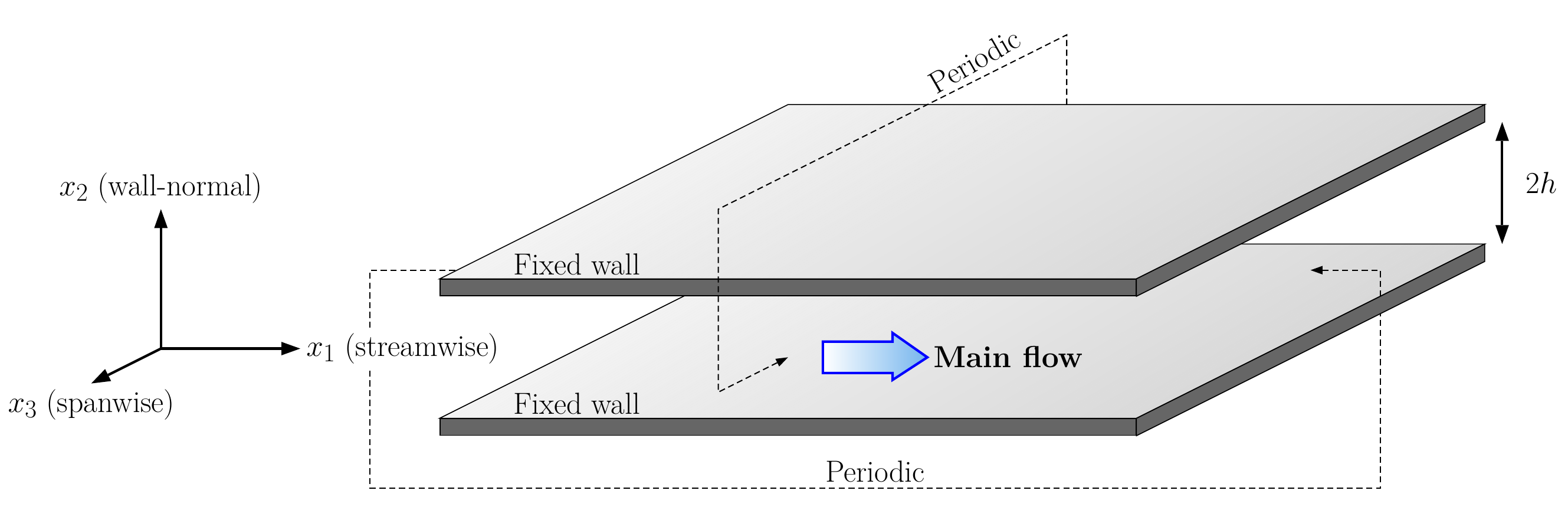}
\caption{Sketch of a PCF configuration.}
\label{fig:sketch_channel_flow}
\end{figure}

The first training flow configuration refers to the Plane Channel Flow (PCF) which is a flow between two parallel plates separated at a distance $2h$. The streamwise direction is aligned with the $x_1\;(x)$ axis, while the wall-normal and spanwise directions are along $x_2\;(y)$ and $x_3\;(z)$, respectively. A sketch of the flow can be seen in Figure \ref{fig:sketch_channel_flow}. This configuration has been largely investigated and high-fidelity simulation data are readily available in the literature \cite{moser1999, kaneda2021, hoyas2022}. Due to its simplicity, this flow is often chosen as the starting point for researchers to consolidate newly developed models, including some ML-RANS closure models proposed in recent years \cite{saezdeocarizborde2021, zhang2018, fang2020}. 

\begin{figure}[h]
\centering
\includegraphics[scale=0.15]{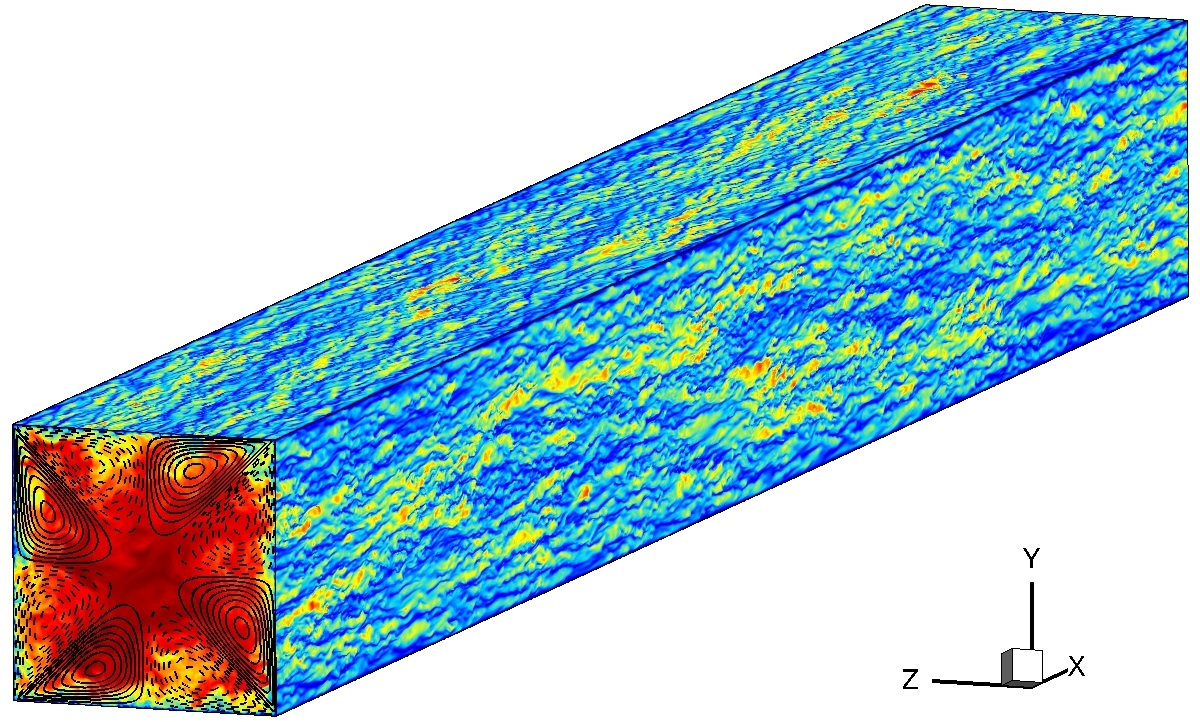}
\caption{Sketch of a SDF configuration, reprinted from \cite{pirozzoli}.}
\label{fig:square_duct}
\end{figure}

The second flow is the Square Duct Flow (SDF), which represents a flow passing through a square section, as illustrated in Figure~\ref{fig:square_duct}. The streamwise direction is still by the direction $x_1\;(x)$, and the square section is situated in the plane $(x_1, x_2)$, i.e. $(y, z)$. On one hand, this is an intriguing case from a scientific perspective, since the anisotropy of the Reynolds stress induces secondary flows over the cross-section, which can not be accurately represented by Boussinesq's hypothesis \cite{speziale1987a}. Nonetheless, various studies indicate that, despite their relative weakness, secondary motions have a notable impact on the overall structure of the mean flow \cite{gessner1965, gessner1973}. Consequently, it is in our interest to improve the prediction of the anisotropy of the RST to better reproduce these effects. On the other hand, this learning flow presents a much more challenging task, as the assumptions of Pope's simplified model version are no longer valid. The general model version should therefore be employed, which will be discussed in the subsequent sections.

\subsection{DNS dataset}
The dataset used in the present work consists of DNS datasets of (a) PCF at seven different friction Reynolds numbers $\mathrm{Re}_\tau=[550; \np{1000}; \np{2000}; \np{4000}; \np{5200}; \np{8000}; \np{10000}]$ \cite{moser1999, kaneda2021, hoyas2022}, and (b) SDF at four different friction Reynolds numbers $\mathrm{Re}_\tau=[150; 250; 500; \np{1000}]$ \cite{pirozzoli2018}. A summary of the dataset sizes is provided in Table~\ref{tab:dataset}. Note that data points for PCF are distributed in one dimension along the $x_2\;(y)$ axis, while data points for SDF are two-dimensional data within a square section, hence the large difference in their dataset sizes.

\begin{figure}[h!]
\begin{subfigure}{0.32\textwidth}
\centering
\includegraphics[width=\linewidth]{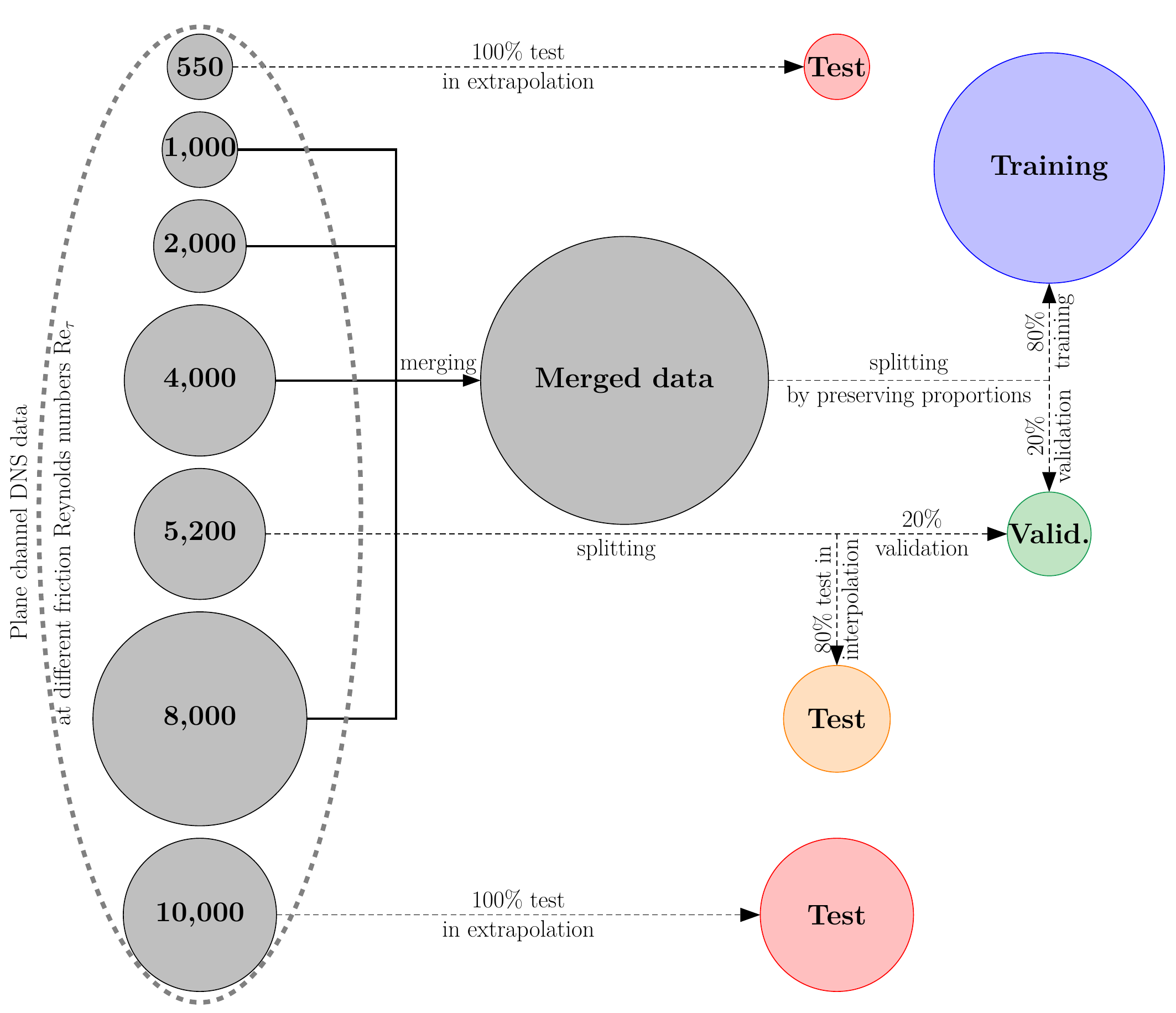}
\caption{PCF}
\label{fig:data_PCF}
\end{subfigure}%
\hfill
\begin{subfigure}{.32\textwidth}
\centering
\includegraphics[width=\linewidth]{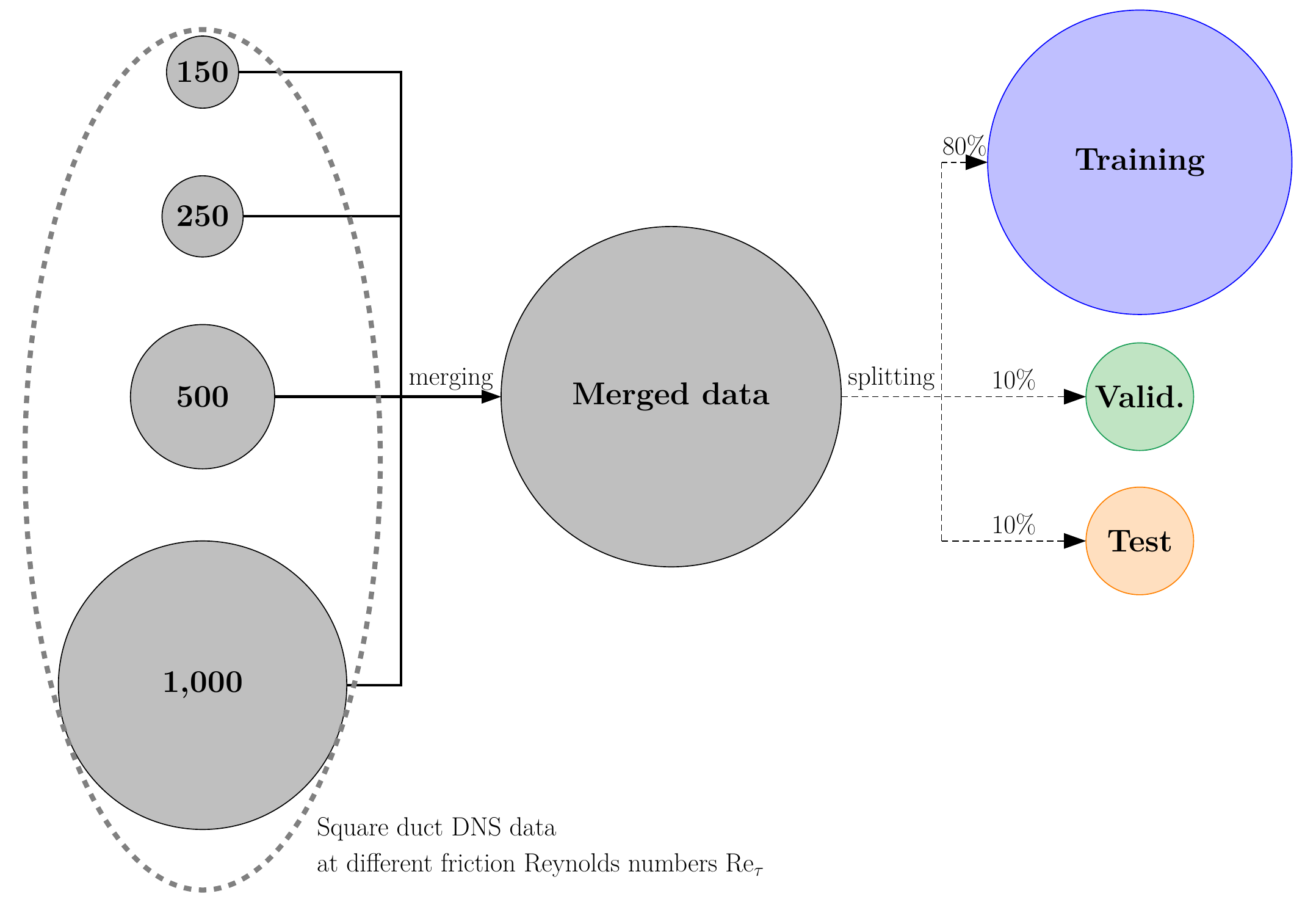}
\caption{SDF: Random-mix}
\label{fig:data_random_mix}
\end{subfigure}%
\hfill
\begin{subfigure}{.32\textwidth}
\centering
\includegraphics[width=\linewidth]{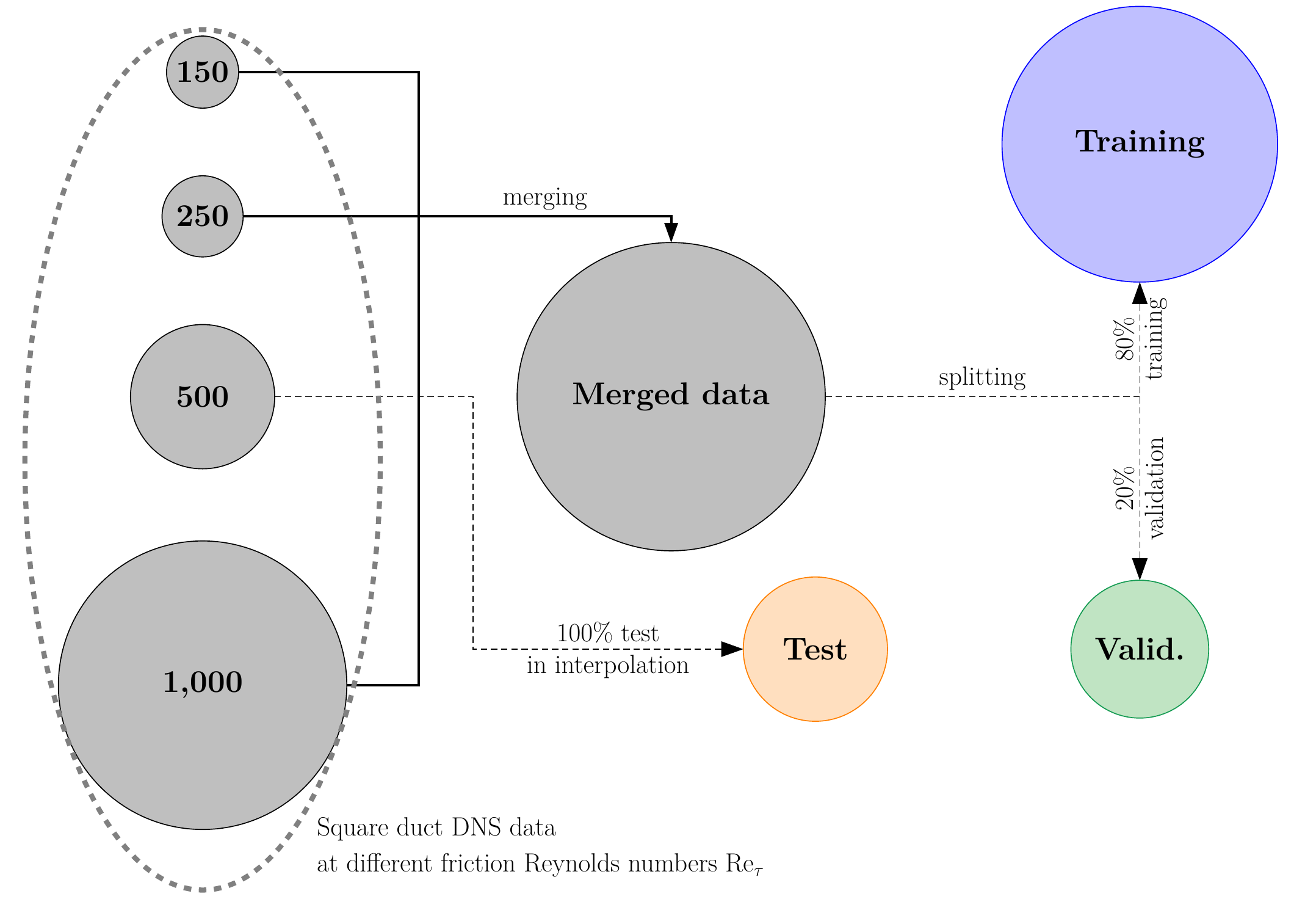}
\caption{SDF: Interpolation}
\label{fig:data_interpolation}
\end{subfigure}

\caption{Different data-splittings for case studies on PCF and SDF. The size of each bubble is proportional to the data size. }
\label{fig:data_split}
\end{figure}

Various data-splittings are employed in order to conduct case studies on each flow configuration. For case studies on PCF, data at $\mathrm{Re}_\tau=[550; \np{10000}]$ are only used in test set, data at $\mathrm{Re}_\tau=\np{5200}$ are split randomly into 80\% test data and 20\% validation data, to avoid over-fitting. The remaining data are divided randomly into 80\% training data and 20\% validation data. The test set here is used to evaluate the predictive capacity of our aTBNN models at unobserved friction Reynolds numbers during training, both for interpolation and extrapolation. An illustration of the data split process for PCF study is shown in Figure \ref{fig:data_PCF}. For case studies on SDF, two different data-splitting methods are applied:
\begin{enumerate}
    \item Random-mix: all data at $\mathrm{Re}_\tau = [150; 250; 500; \np{1000}]$ are split randomly into 80\% of training, 10\% of validation and 10\% of test, as illustrated in Figure~\ref{fig:data_random_mix}.
    \item Interpolation: data at $\mathrm{Re}_\tau = 500$ are used in test set, the rest of the data are split randomly into 80\% of training and 20\% of validation, as illustrated in Figure~\ref{fig:data_interpolation}.
\end{enumerate}

We would like to draw the reader's attention to the data imbalance problem that can be seen in Table~\ref{tab:dataset}. This is a common issue when dealing with data originating from different sources, as is the case in our study where different DNS datasets on PCF are merged together. It comes from the fact that it is common practice to employ a greater number of grid points in turbulence simulations at higher Reynolds numbers in order to capture small flow scales. Moreover, data collected for one specific Reynolds number is not spatially balanced: indeed we find much more data in the near-wall region and less data far from the wall. Addressing such data imbalance problems typically involves techniques such as re-sampling and re-weighting \cite{yang2021}. For example, Jiang \textit{et al.} \cite{jiang2021} employed beforehand a (k-means) clustering algorithm to better balance their data. % to categorize data into these two distinct regions, then oversampled data in the subregion where the data are less abundant.
In the present study, we opt to maintain the original data size, as well as data distribution for simplicity. This decision is, to some extent, grounded in the belief that this data imbalance, to some extent, mirrors the complexity of turbulent flows and might guide the NN to capture this complexity. However, we ensure the same preprocessing on all data from different sources. 

\begin{table}[h!]
\centering
\caption{\label{tab:dataset}DNS data size for PCF and SDF at each friction Reynolds number.}
\begin{tabular}{lccccccccccc}
\toprule
& \multicolumn{7}{c}{\textbf{PCF}} & \multicolumn{4}{c}{\textbf{SDF}}\\
$\mathrm{Re}_\tau$ & 550 & \np{1000} & \np{2000} & \np{4000} & \np{5200} & \np{8000} & \np{10000} & 150 & 250 & 500 & \np{1000}\\\hline
Data size & 191 & 255 & 383 & \np{1023}  & 767 & \np{2047} & \np{1050} & 4096 & 5184 & \np{16384} & \np{65536} \\
Reference & \cite{moser1999}  & \cite{moser1999} & \cite{moser1999} & \cite{kaneda2021} & \cite{moser1999} & \cite{kaneda2021} & \cite{hoyas2022} & \multicolumn{4}{c}{\cite{pirozzoli2018}}\\
\bottomrule
\end{tabular}
\end{table}

\subsection{Neural network architectures}
Two types of NN architectures are independently investigated in this work, namely: MLP and TBNN. The former one is a classic NN architecture, for which details can be found in \cite{Goodfellow-et-al-2016}. The objective is to generate direct predictions of the anisotropy tensor in the output layer. This architecture is relatively straightforward to implement and offers a high degree of flexibility.

The latter TBNN architecture is previously reviewed in \ref{sec:TBNN}, and will be augmented in the present study in order to make it more suitable for PCF and SDF configurations. This augmentation is achieved by a thorough application of Pope's GEVM on the two considered flows coupled with state-of-the-art training strategies, which will be discussed in the following sections. 

\subsubsection{Application of Pope's GEVM on PCF} \label{sec:pope_pcf}
We note that the characteristics in the $x_3$ direction of our considered PCF: $\overline{u}_3=0$ and $\dfrac{\partial }{\partial x_3}=0$, satisfy the conditions of Pope's simplified model version given in Eq.~\eqref{eq:2D_Pope}. In order to write the Reynolds stress anisotropy tensor as shown in Eq.~\eqref{eq:2D_Pope}, we first give the expressions of the normalized mean strain-rate, rotation-rate and Reynolds stress anisotropy tensors as follows:
\begin{singlespace} 
\begin{gather}
\textbf{S}^* = \dfrac{1}{2}
\begin{bmatrix}
0 & \alpha & 0 \\
\alpha & 0 & 0 \\
0 & 0 & 0
\end{bmatrix}
\;\text{,}\;
\textbf{R}^* = \dfrac{1}{2}
\begin{bmatrix}
0 & \alpha & 0 \\
-\alpha & 0 & 0 \\
0 & 0 & 0
\end{bmatrix}
\;\text{and}\;
\textbf{b} = 
\begin{bmatrix}
\;b_{11} & b_{12} & 0 \\
\;b_{12} & b_{22} & 0 \\
\;0 & 0 & b_{33}
\end{bmatrix}   \label{eq:SRB_channel}
\end{gather} 
\end{singlespace} 

\noindent where $\alpha = \dfrac{k}{\epsilon}\dfrac{d\overline{u}_1}{dx_2}$ is the normalized mean velocity gradient, as well as the only nonzero mean velocity statistics.

Substituting Eq.~\eqref{eq:SRB_channel} into Eq.~\eqref{eq:PopeInvariant2D} and Eq.~\eqref{eq:PopeTensor2D} leads to:
\begin{equation}
\lambda_1^*=\text{tr}(\textbf{S}^{*2})=\dfrac{\alpha^2}{2} \, ,\quad \lambda_2^*=\text{tr}(\textbf{R}^{*2})=-\dfrac{\alpha^2}{2} \\\label{eq:lambdaChannel}
\end{equation}

\noindent and

\begin{singlespace} 
\begin{align} 
    \begin{cases}
     	\textbf{T}^{*(0)} =  \dfrac{1}{2}\textbf{I}_2 -  \dfrac{1}{3}\textbf{I}_3&=
    	    \begin{bmatrix}
   			1/6 & 0 & 0 \\
   			0 & 1/6 & 0 \\
   			0 & 0 & -1/3 \\
   			\end{bmatrix}
            \\[10pt]   			    			
     	\textbf{T}^{*(1)} =  \textbf{S}^{*} & = 
     	  \begin{bmatrix}
   			0 & \alpha/2 & 0 \\
   			\alpha/2 & 0 & 0 \\
   			0 & 0 & 0 \\
   			\end{bmatrix}
   			\\[10pt]
      	\textbf{T}^{*(2)} =  \textbf{S}^{*}\textbf{R}^{*}-\textbf{R}^{*}\textbf{S}^{*} & = 
      	    \begin{bmatrix}
   			\alpha^2/2 & 0 & 0 \\
   			0 & -\alpha^2/2 & 0 \\
   			0 & 0 & 0 
   			\end{bmatrix}\\[10pt]
    \end{cases}\label{eq:TChannel}
\end{align}
\end{singlespace}

\noindent As it is obvious that $\lambda_1^*=-\lambda_2^*$, only one invariant is relevant in our case and we will keep $\lambda_1^*$ in the following. For reminder, $\textbf{I}_2$ is taken here as $\mathrm{diag}(1,1,0)$, as required by Pope's model. From now on, we denote by $\textbf{T}^{*(03)}$ this $\textbf{T}^{*(0)}$ for referring to the location of zero in the diagonals of the chosen $\textbf{I}_2$. It is worth remarking that other higher order $\textbf{T}^{*(n)}$ are found to be zero-tensors by developing them in the PCF case, which also proves the appropriate choice of Pope's simplified model.

The expression of the Reynolds stress anisotropy tensor shown in Eq.~\eqref{eq:2D_Pope} can therefore be rewritten as:
\begin{equation}
\begin{split}
\textbf{b}& =g^{(0)}(\lambda_1^*)\textbf{T}^{*(0)} + g^{(1)}(\lambda_1^*)\textbf{T}^{*(1)} + g^{(2)}(\lambda_1^*)\textbf{T}^{*(2)}\\
& =g^{(0)}(\alpha)\textbf{T}^{*(0)} + g^{(1)}(\alpha)\textbf{T}^{*(1)} + g^{(2)}(\alpha)\textbf{T}^{*(2)} \label{eq:Pope_Channel}    
\end{split}
\end{equation}

Here, we have clarified the first ambiguity remaining in the literature around the application of Pope's GEVM to the PCF: only one invariant and three tensors are indeed necessary, and they depend merely on $\alpha$. Particularly in our domain of interest, it was found some works attempted to apply the general model version, shown in Eq.~\eqref{eq:Pope3D}, with five invariants and ten tensors, to PCF \cite{fang2020}. 

Another concern related to the choice of the constant tensor $\textbf{T}^{*(0)}$ has been identified. We notice that $\textbf{T}^{*(0)}$ obtained by two other possible permutations of $\textbf{I}_2$ can also form an integrity basis with $\textbf{T}^{*(1)}$ and $\textbf{T}^{*(2)}$ in the case of PCF, since we have:
\begin{equation}
\textbf{T}^{*(01)} = \mathrm{diag}(-1/3,1/6,1/6) = -\dfrac{1}{2}\textbf{T}^{*(03)} - \dfrac{1}{4\lambda_1^*}\textbf{T}^{*(2)} 
\end{equation}
and
\begin{equation}
\textbf{T}^{*(02)} = \mathrm{diag}(1/6,-1/3,1/6) = -\dfrac{1}{2}\textbf{T}^{*(03)} + \dfrac{1}{4\lambda_1^*}\textbf{T}^{*(2)} 
\end{equation}

\noindent provided that $\lambda_1^* \not= 0$, which is correct everywhere except at the two singular points, either on the wall or at the channel center.

Substituting Eq.~\eqref{eq:TChannel} into Eq.~\eqref{eq:Pope_Channel}, we obtain the expression of the Reynolds stress anisotropy tensor components as follows, the three systems of equations using $\textbf{T}^{*(01)}$, $\textbf{T}^{*(02)}$ and $\textbf{T}^{*(03)}$, respectively:
\begin{singlespace} 
\begin{align}
\overset{\text{\normalsize\textbf{{with}} $\vphantom{\dfrac{a}{b}}{\mathbf{T}^{*(01)}}$:}}{
    \begin{cases}
    b_{11} = -\dfrac{1}{3}g^{(0)} - \dfrac{\alpha^2}{2}g^{(2)} \\[10pt]
    b_{12} = \dfrac{\alpha}{2}g^{(1)} \\[10pt]
    b_{22} = \dfrac{1}{6}g^{(0)} + \dfrac{\alpha^2}{2}g^{(2)} \\[10pt]
    b_{33} = \dfrac{1}{6}g^{(0)}
    \end{cases}}
\overset{\text{\normalsize\textbf{{with}} $\vphantom{\dfrac{a}{b}}{\mathbf{T}^{*(02)}}$:}}{
\text{or}
    \begin{cases}
    b_{11} = \dfrac{1}{6}g^{(0)} - \dfrac{\alpha^2}{2}g^{(2)}\\[10pt]
    b_{12} = \dfrac{\alpha}{2}g^{(1)} \\[10pt]
    b_{22} = -\dfrac{1}{3}g^{(0)} + \dfrac{\alpha^2}{2}g^{(2)} \\[10pt]
    b_{33} = \dfrac{1}{6}g^{(0)}
    \end{cases}}
\overset{\text{\normalsize\textbf{{with}} $\vphantom{\dfrac{a}{b}}{\mathbf{T}^{*(03)}}$:}}{
\text{or}
    \begin{cases}
    b_{11} = \dfrac{1}{6}g^{(0)} - \dfrac{\alpha^2}{2}g^{(2)} \\[10pt]
    b_{12} = \dfrac{\alpha}{2}g^{(1)} \\[10pt]
    b_{22} = \dfrac{1}{6}g^{(0)} + \dfrac{\alpha^2}{2}g^{(2)} \\[10pt]
    b_{33} = -\dfrac{1}{3}g^{(0)}
    \end{cases}} \label{eq:bijPope}       
\end{align}
\end{singlespace} 

\begin{figure}[h!]
\centering
\includegraphics[width=0.4\textwidth,center]{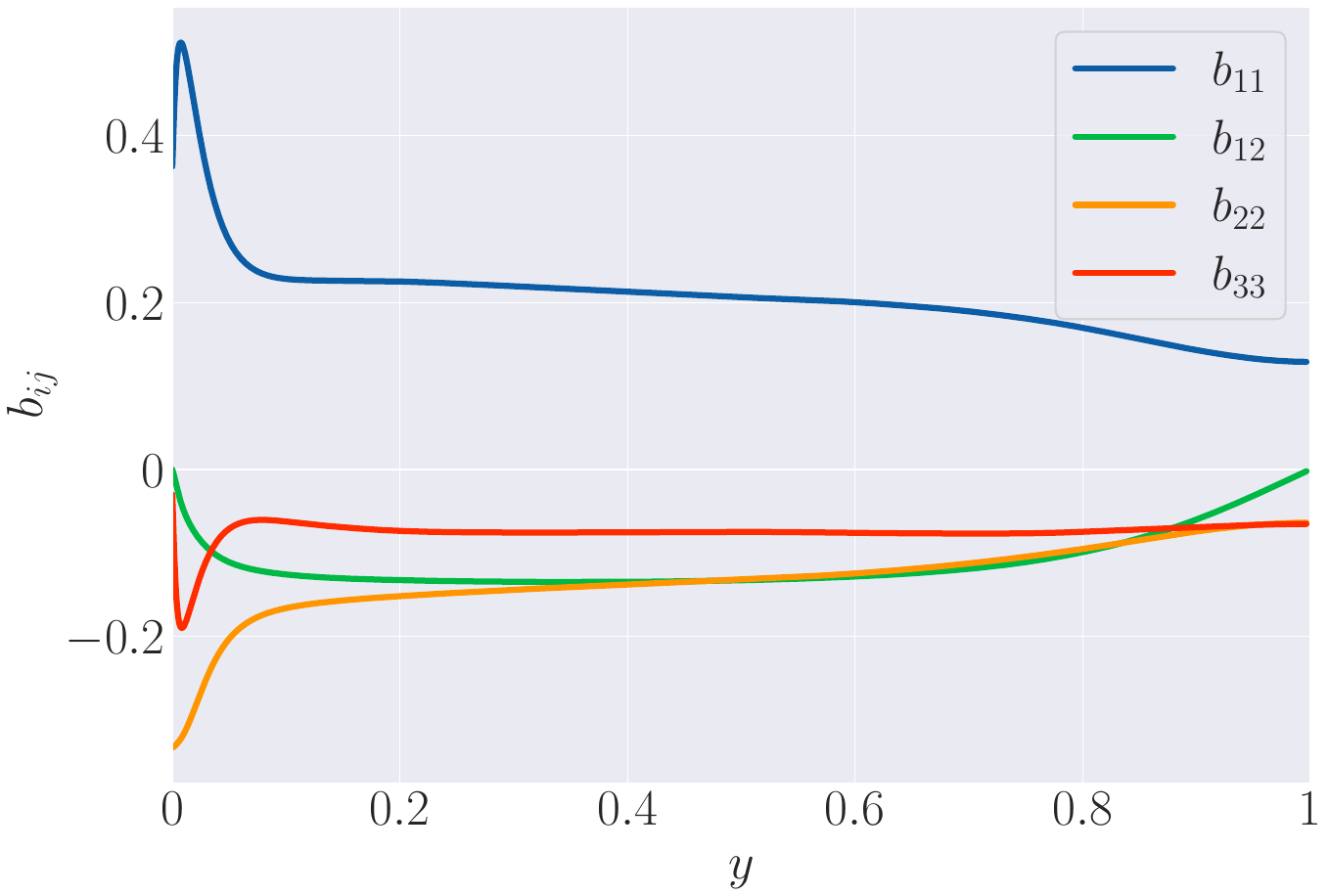}
\caption{Reynolds stress anisotropy tensor $b_{ij}$ as a function of the wall distance $y$ in DNS dataset for PCF at $\mathrm{Re}_\tau=\np{1000}$ \cite{moser1999}.}
\label{fig:b_diag}
\end{figure}

An arising question pertains to the optimal selection of $\textbf{T}^{*(0)}$, among these three alternatives. We especially note that the value of $\textbf{T}^{*(0)}$ should represent the value of the anisotropy tensor at the channel center, because the other two tensors are both zero at this location where $\dfrac{du_1}{dx_2} \Bigr|_{\substack{x_2=h}} = 0$. In Figure \ref{fig:b_diag}, we plot each nonzero component of the Reynolds stress anisotropy tensor as a function of the wall distance $y$, at the friction Reynolds number $\mathrm{Re}_\tau=\dfrac{u_{\tau}h}{\nu}=\np{1000}$, using the DNS data of \cite{moser1999}. Here, $u_{\tau}=\sqrt{\nu \dfrac{d\overline{u}_1}{dx_2}}$ is the friction velocity. Interestingly, it can be observed that $b_{12}\approx0$ and $b_{22}\approx b_{33}\approx -b_{11}/2$ at the channel center. This observation also holds true for other DNS data at different $\mathrm{Re}_\tau$. Accordingly, it turns out that only $\textbf{T}^{*(01)}$ is proportional to the Reynolds stress anisotropy tensor at the channel center. Hence, it physically makes sense to include $\textbf{T}^{*(01)}$ in the basis tensors, which is contradictory with the choice of $\textbf{T}^{*(0)}=\textbf{T}^{*(03)}$ in Pope's statement. Despite the number of literature studies applying Pope's model to PCF, this issue has never been discussed to our knowledge. To this end, we would like to examine this issue by testing the present aTBNN models using different $\textbf{T}^{*(0)}$, including a new generalized $\textbf{T}^{*(0)}$ discussed hereinafter. \\ 

\noindent \textit{Generalized} $\textbf{T}^{*(0)}$ \\
 In order to resolve the uncertainty surrounding the $\textbf{T}^{*(0)}$ tensor, a generalized form of $\textbf{T}^{*(0)}$ is proposed in our work, which is written as a linear combination of each alternative of $\textbf{T}^{*(0)}$: 
\begin{equation}
\textbf{T}^{*(0)}_{\text{gen}}=g_{01}\textbf{T}^{*(01)}+g_{02}\textbf{T}^{*(02)}+g_{03}\textbf{T}^{*(03)} \label{eq:newT0_g0}
\end{equation}
\noindent where $g_{01}$, $g_{02}$ and $g_{03}$ are coefficient functions depending on $\alpha$, instead of some fixed constants, in order to make the generalization as broad as possible under Pope's framework.

A more compact formulation of Eq.~\eqref{eq:newT0_g0} can be obtained as:
\begin{singlespace}
\begin{equation} 
    \begin{split}
\textbf{T}^{*(0)}_{\text{gen}}& =
     	    \begin{bmatrix}
   			-\dfrac{1}{3} g_{01} + \dfrac{1}{6} g_{02} + \dfrac{1}{6} g_{03} & 0 & 0 \\
   			0 & \dfrac{1}{6} g_{01} - \dfrac{1}{3} g_{02} + \dfrac{1}{6} g_{03} & 0 \\
   			0 & 0 & \dfrac{1}{6} g_{01} + \dfrac{1}{6} g_{02} - \dfrac{1}{3} g_{03} \\
   			\end{bmatrix}
   			 = 
     	    \begin{bmatrix}
   			f_{01} & 0 & 0 \\
   			0 & f_{02} & 0 \\
   			0 & 0 & f_{03} \\
   			\end{bmatrix}\label{eq:newT0}
   	\end{split}
\end{equation}   	
\end{singlespace}

where $f_{01}$, $f_{02}$ and $f_{03}$ are functions of $\alpha$ as $g_{01}$, $g_{02}$ and $g_{03}$ are, with $f_{01}+f_{02}+f_{03}=0$ to preserve the zero-trace of the Reynolds stress anisotropy tensor.

Meanwhile, we notice that the information of $\alpha$ in $\textbf{T}^{*(2)}$ is also included in $\textbf{T}^{*(0)}_{\text{gen}}$, since they are both diagonal tensors. The tensor basis can therefore be reduced to only $\textbf{T}^{*(0)}_{\text{gen}}$ and $\textbf{T}^{*(1)}$:

\begin{singlespace} 
\begin{equation}
\textbf{b}=\textbf{T}^{*(0)}_{\text{gen}}(\alpha) + g^{(1)}(\alpha)\textbf{T}^{*(1)} \label{eq:Pope_Channel_gen}    
\end{equation}
which can be developed into the following system of equations, giving the expression of $\textbf{T}^{*(0)}_{\text{gen}}$ shown in Eq.~\eqref{eq:newT0}:
\begin{align} 
    \begin{cases}
    b_{11} = f_{01} \\[10pt]
    b_{12} = \dfrac{\alpha}{2}g^{(1)} \\[10pt]
    b_{22} = f_{02}\\[10pt]
    b_{33} = - (f_{01}+f_{02})
    \end{cases} \label{eq:bijF}  
\end{align}
\end{singlespace}
In summary, we obtain four representations of the Reynolds stress anisotropy tensor for PCF, shown in Eq.~\eqref{eq:bijPope} and Eq.~\eqref{eq:bijF}, respectively, using either one of the three constant $\textbf{T}^{*(0)}$, or the newly proposed $\textbf{T}^{*(0)}_{\text{gen}}$.

\subsubsection{Application of Pope's GEVM on SDF}\label{sec:popeSDF}
Unlike the case of PCF, the normalized mean strain-rate $\textbf{S}^*$ and rotation-rate $\textbf{R}^*$ tensors are more complicated in the case of SDF, because no mean velocity component is zero due to the secondary flow and there is only one periodicity direction along the streamwise direction. \red{To derive them, we first denote the dimensionless velocity gradients as follows:}
\red{
\begin{equation}
\alpha_{ij} = \dfrac{k}{\epsilon} \dfrac{\partial \overline{u}_i}{\partial x_j}
\end{equation}
}
\red{We have straightforwardly $\alpha_{11}=\alpha_{21}=\alpha_{31}=0$ because of periodicity, and $\alpha_{22}=-\alpha_{33}$ because of mass conservation. Consequently,}
\red{
\begin{equation}
  \nabla{\overline{\textit{\textbf{u}}}}^* =
\begin{bmatrix}
\alpha_{11} & \alpha_{12} & \alpha_{13} \\
\alpha_{21} & \alpha_{22} & \alpha_{23} \\
\alpha_{31} & \alpha_{32} & \alpha_{33}
\end{bmatrix}    
=
\begin{bmatrix}
0 & \alpha_{12} & \alpha_{13} \\
0 & \alpha_{22} & \alpha_{23} \\
0 & \alpha_{32} & -\alpha_{22}
\end{bmatrix}   \label{eq:gradU}  
\end{equation}
}
\red{Combining Eqs.~\eqref{eq:defS*}, ~\eqref{eq:defR*} and ~\eqref{eq:gradU}, leads to:}
\red{
\begin{singlespace}
\begin{equation}
\textbf{S}^* = \dfrac{1}{2}
\begin{bmatrix}
0 & \alpha_{12} & \alpha_{13} \\
\alpha_{12} & 2\alpha_{22} & \alpha_{23}+\alpha_{32} \\
\alpha_{13} & \alpha_{23}+\alpha_{32} & -2\alpha_{22}
\end{bmatrix}
:=
\begin{bmatrix}
0 & S_{12}^* & S_{13}^* \\
S_{12}^* & S_{22}^* & S_{23}^* \\
S_{13}^* & S_{23}^* & -S_{22}^*
\end{bmatrix}\label{eq:Ssquare}
\end{equation}
\end{singlespace} 
}
\noindent \red{and}
\red{
\begin{singlespace}
\begin{equation}
\textbf{R}^* = \dfrac{1}{2}
\begin{bmatrix}
0 & \alpha_{12} & \alpha_{13} \\
-\alpha_{12} & 0 & \alpha_{23}-\alpha_{32} \\
-\alpha_{13} & \alpha_{32}-\alpha_{23} & 0
\end{bmatrix}
:=
\begin{bmatrix}
0 & S_{12}^* & S_{13}^* \\
-S_{12}^* & 0 & R_{23}^* \\
-S_{13}^* & -R_{23}^* & 0
\end{bmatrix}\label{eq:Rsquare}
\end{equation}
\end{singlespace}
}
\red{It appears from the above equations that unlike the PCF case with only one independent parameter $\alpha_{12}$ (previously denoted $\alpha$), there remain five independent parameters $\alpha_{12}^*, \alpha_{13}^*, \alpha_{22}^*, \alpha_{23}^*, \alpha_{23}^*$ (or alternatively $S_{12}^*, S_{13}^*, S_{22}^*, S_{23}^*, R_{23}^*$) in the SDF case, resulting in a complexity jump between these two learning flows. In any case, the expressions of the basis tensors in Eq.~\eqref{eq:PopeTensor3D} for the SDF case cannot be further simplified and are not written here because of their cumbersomeness. Developing the invariants in Eq.~\eqref{eq:PopeInvariant3D} leads to:}
\red{
\begin{singlespace} 
\begin{align} 
\begin{cases}
\lambda_1^*&=2\left( S_{12}^{*2}+S_{13}^{*2}+S_{22}^{*2}+S_{23}^{*2}\right) \\
\\
\lambda_2^*&=-2\left( S_{12}^{*2}+S_{13}^{*2}+R_{23}^{*2}\right) \\
\\
\lambda_3^*&=3S_{22}^*\left(S_{12}^{*2}-S_{13}^{*2}\right)+6S_{12}^*S_{13}^*S_{23}^* \\
\\
\lambda_4^*&=-S_{22}^*\left(S_{12}^{*2}-S_{13}^{*2}\right)-2S_{12}^*S_{13}^*S_{23}^* \\
\\
\lambda_5^*&=-\left(S_{12}^{*2}+S_{13}^{*2}\right)\left(2S_{12}^{*2}+2S_{13}^{*2}+S_{22}^{*2}+S_{23}^{*2}\right)-R_{23}^{*2}\left(S_{12}^{*2}+S_{13}^{*2}-2S_{22}^{*2}-2S_{23}^{*2}\right)\\
&+2R_{23}^*\left(S_{12}^{*2}S_{23}^*-2S_{12}^*S_{13}^*S_{22}^*-S_{13}^{*2}S_{23}^*\right)
\end{cases}\label{eq:invariant_square}
\end{align}
\end{singlespace}
}
\red{Invariants 1 and 2 are independent because they do not contain the same variables. Invariants 1 and 3 are also independent due to the $S_{12}^*S_{13}^*S_{23}^*$ cross product, as are invariants 1 and 5 due to the $S_{12}^*S_{13}^*S_{22}^*R_{23}^*$ cross product. Thus we observe that there are four independent invariants, because $\lambda_3^*+3\lambda_4^*=0$ implying that one of them ($\lambda_3^*$ or $\lambda_4^*$) could be indifferently removed from the SDF model. However, this is a minor simplification, and in the following of the study we prefer to retain the five invariants for greater generality.}

\red{The symmetries impose that all the $\alpha_{ij}$ at the center of the duct's cross-section are theoretically zero, even though numerical errors (among which discretization errors and imperfect convergence of statistics) exist in DNS, as shown in Table~\ref{tab:center_sdf_aij}. According to the base model, it would lead all the diagonal components of the anisotropy tensor to be zero at the center, which is in contradiction with the DNS data gathered in Table~\ref{tab:center_sdf}. As for the PCF, we remark that $b_{22}\approx b_{33}\approx -b_{11}/2$ at the section's center, and we are therefore led to introduce the constant tensor $\textbf{T}^{*(01)}$ in the basis in order to properly account for the physics at this location. In the following of the article, the performance of aTBNN models without constant tensor or with different permutations of its diagonal will be compared for the sake of completeness.}

\begin{table}[h]
\centering
\caption{\label{tab:center_sdf_aij} $\alpha_{ij}$ values at the center of the duct's cross-section form square duct flow DNS data \cite{pirozzoli2018}.}
\begin{tabular}{lcccccc}
\toprule
$\mathrm{Re}_\tau$ & $\alpha_{12}$ & $\alpha_{13}$ & $\alpha_{22}$ & $\alpha_{23}$ & $\alpha_{32}$ & $\alpha_{33}$ \\\hline
150 &  9.568e-02 &  6.528e-02 & -4.597e-03 &  1.850e-03 &  1.553e-03 &  1.553e-03 \\
250 &1.024e-01 &  5.955e-02 &  8.528e-03 &  3.253e-03 & -2.153e-03 & -2.153e-03 \\
500 &  2.553e-03 &  2.385e-03 & -1.617e-04 & -2.219e-04 &  2.059e-04 &  2.059e-04 \\ 
\np{1000} & 2.147e-03 & -6.704e-04 & -6.194e-04 &  3.744e-04 & -7.543e-05 & -7.543e-05 \\
\bottomrule
\end{tabular}
\end{table}

\begin{table}[h]
\centering
\caption{\label{tab:center_sdf} $b_{ij}$ values at the center of the duct's cross-section form square duct flow DNS data \cite{pirozzoli2018}.}
\begin{tabular}{lcccccc}
\toprule
$\mathrm{Re}_\tau$ & $b_{11}$ & $b_{12}$ & $b_{13}$ & $b_{22}$ & $b_{23}$ & $b_{33}$ \\\hline
150 & -3.453e-01 & -1.515e-02 & -6.793e-03 & 1.736e-01 & -5.351e-04 & 1.717e-01 \\
250 & -2.506e-01 & -1.082e-02 & -1.092e-04 & 1.235e-01 &  5.879e-04 & 1.271e-01 \\
500 & -1.611e-01 &  1.695e-03 & -3.301e-03 & 8.030e-02 &  1.819e-04 & 8.079e-02 \\ 
\np{1000} & -1.023e-01 & -8.497e-03 &  6.241e-03 & 5.430e-02 & -1.403e-04 & 4.799e-02 \\
\bottomrule
\end{tabular}
\end{table}

Regarding the anisotropy tensor $\textbf{b}$, clearly it can no longer be represented by one single variable. What's more, every component $b_{ij}$ is non-zero. Considering its symmetric and zero-trace characteristics, at least five components should be predicted: $b_{11}$, $b_{12}$, $b_{13}$, $b_{22}$ and $b_{23}$.   

\subsubsection{Input feature selection}
A diverse set of input features has been employed in prior TBNN frameworks. In Ling \textit{et al.}'s original study \cite{ling2016}, five invariants and ten tensors are used. However, as in the previous analysis on PCF in Section~\ref{sec:pope_pcf}, the model can be simplified to two invariants and three tensors. Zhang \textit{et al.} \cite{zhang2018} tested it and demonstrated its considerable advantages compared to the full model. Pursuing this simplified framework, the input feature set can be further reduced to the variable $\alpha = \dfrac{k}{\epsilon}\dfrac{d\overline{u}_1}{dx_2}$ as shown in Eq.~\eqref{eq:Pope_Channel}, which has also been used in Fang \textit{et al.}'s study on PCF, upon an MLP model though \cite{fang2020}. 

\begin{figure}[h!]
\centering
\begin{subfigure}{.8\textwidth}
\centering
\includegraphics[width=\textwidth,center]{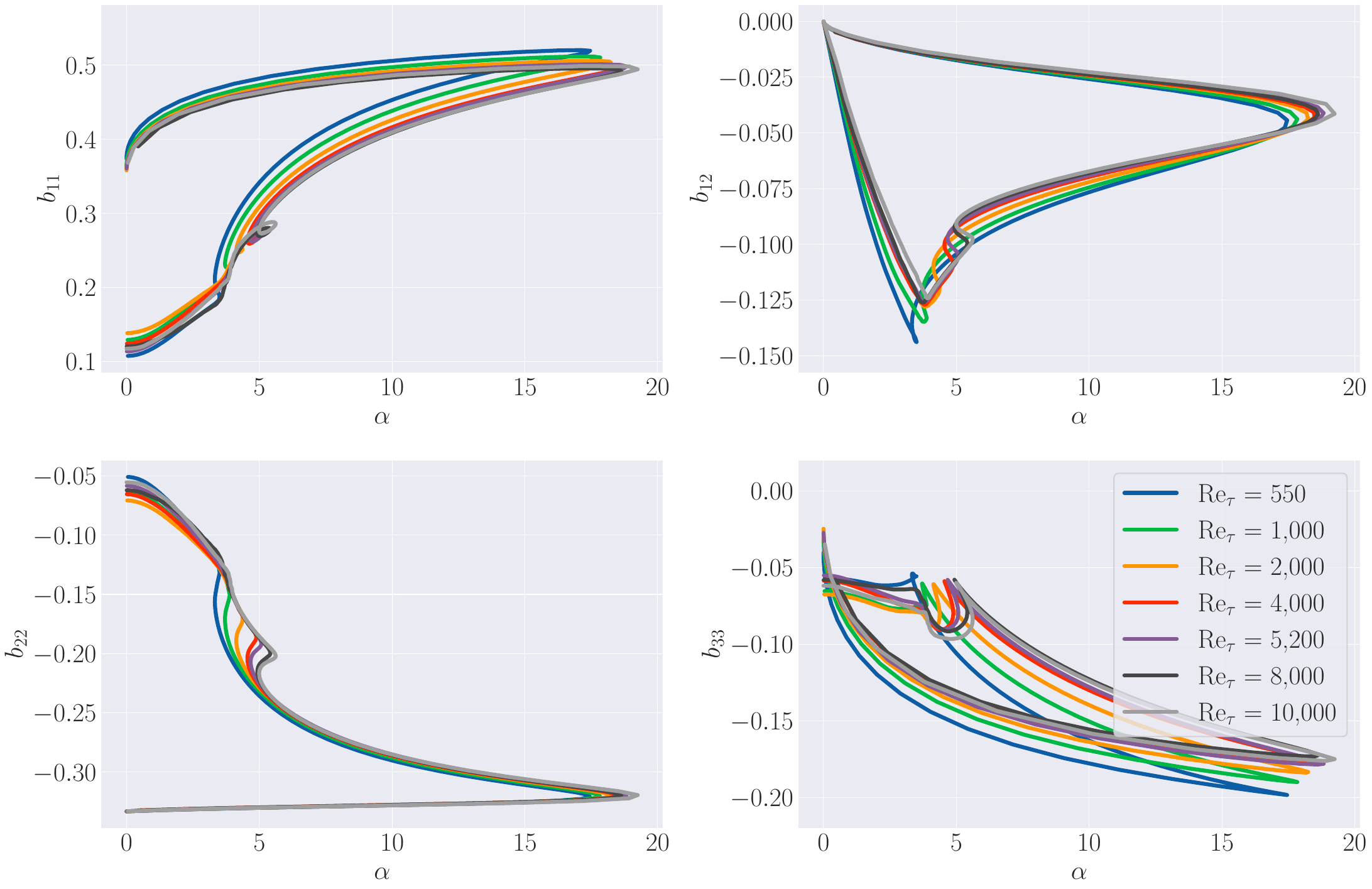}
  \caption{}
\label{fig:b_alpha}
    \end{subfigure}%
\hfill
\begin{subfigure}{.8\textwidth}
\includegraphics[width=\textwidth,center]{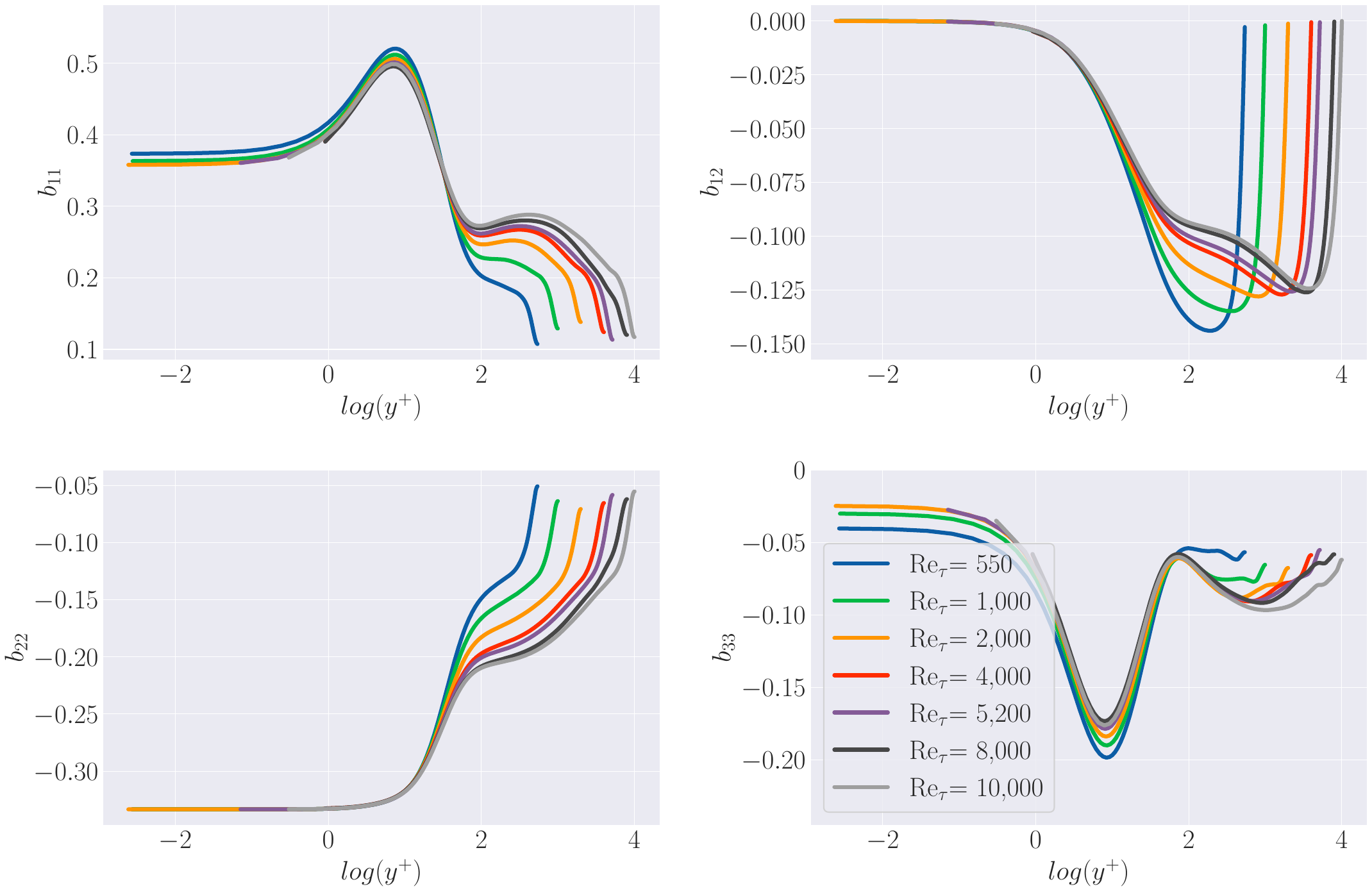}
  \caption{}
\label{fig:b_y+}
\end{subfigure}

\caption{Anisotropy tensor $b_{ij}$ against (a) characteristic variable $\alpha$ (b) dimensionless wall distance $y^+$ in DNS datasets for PCF at different frictions Reynolds numbers \cite{moser1999, kaneda2021, hoyas2022}.}
\end{figure}

In addition to this single variable given by Pope's model, we believe that other input features should be included, that truthfully reflect the complexity of the turbulent flow. Figure~\ref{fig:b_alpha} shows the true relationship between each component $b_{ij}$ and $\alpha$ given by DNS datasets on PCF. Obviously, Eq.~\eqref{eq:Pope_Channel} derived from Pope's model is inconsistent with Figure~\ref{fig:b_alpha}: none of the $b_{ij}$ components can establish a unique mapping from $\alpha$. This is usually called as a multi-valued issue and has been reported in various works \cite{liu2021a, jiang2021}. Jiang \textit{et al.} \cite{jiang2021} trained two TBNN frameworks upon the plane channel flow at $\mathrm{Re}_\tau=5200$, one over the whole dataset, and the other included only the subset of data far from the wall. Predictions from the former framework over the whole dataset are much worse and exhibit noticeable oscillations, showing that Pope's model cannot correctly represent the whole flow regime due to multi-valued issue. 

Besides, we remark from Figure~\ref{fig:b_alpha} that $b_{ij}$ exhibits some level of dependence on the Reynolds number, which can not be revealed by Pope's model. Consequently, in order to address these challenges and to forecast $b_{ij}$ accurately, it becomes imperative to include additional representative input features. Different attempts have been made in previous works: Fang \textit{et al.} \cite{fang2020} took the dimensionless wall distance $y^+=\dfrac{yu_\tau}{\nu}$ as extra input for near wall considerations; Jiang \textit{et al.} \cite{jiang2021} proposed to use the turbulent Reynolds number $\mathrm{Re}_t = k^2 / (\nu \epsilon)$ which is commonly used in the damping functions of low-Reynolds-number models. Indeed, these two different parameters construct a unique mapping between them. We propose in this work to rely on $y^+$, and add the friction Reynolds number $\mathrm{Re}_\tau$ into the input feature set, as a classifier of data originating from flows with different turbulence levels. As Figure~\ref{fig:b_y+} shows, there is a unique functional mapping from $y^+$ to any targeted $b_{ij}$ at one given $\mathrm{Re}_\tau$. Thus, by introducing $y^+$ and $\mathrm{Re}_\tau$ into the input feature set, we should successfully overcome the multi-valued problem in conventional Pope's model in the case of PCF. Similarly, for the case of SDF, considering the dimensional increase, $z^+$ is further included, and $\alpha$ is replaced by the five invariants presented in Eq.~\eqref{eq:invariant_square}. \blue{It should be noted that $y^+$ and $z^+$ used here in the SDF case, are two scalar values, representing the dimensionless wall distances to the two walls, $y=0$ and $z=0$, respectively. Hence, we denote them as $y^+$ and $z^+$ for convenience, because the two distances are in our case aligned with the $y$ and $z$ axes. As distances, they do not depend on the definition of the two axes, and remain unchanged even when the coordinate system is rotated. One should only be careful to represent them correctly in a new coordinate system, in order to make sure that they remain distances relative to the $y=0$ and $z=0$ axes in the original coordinate system.}

\subsubsection{Physics model-based neural network architectures}
Considering both the prior analysis on the $\textbf{T}^{*(0)}$ and the input feature selection, an augmented version of Pope's model is obtained for PCF: 
\begin{equation}\label{eq:aTBNN1}
\textbf{b} =\sum_{n=0}^2 g^{(n)}\left(\alpha, y^+, \mathrm{Re}_\tau \right)\textbf{T}^{*(n)} 
\end{equation}
for the model with a constant $\textbf{T}^{*(0)}$, and
\begin{equation}\label{eq:aTBNN2}
\textbf{b}=\textbf{T}^{*(0)}_{\text{gen}}(\alpha, y^+, \mathrm{Re}_\tau) + g^{(1)}(\alpha, y^+, \mathrm{Re}_\tau)\textbf{T}^{*(1)}   
\end{equation}
for the model using the newly proposed $\textbf{T}^{*(0)}_{\text{gen}}$. 

For SDF, we have: 
\begin{equation}\label{eq:aTBNN3}
\textbf{b} =g^{(0)}\textbf{T}^{*(01)} + \sum_{n=1}^{5} g^{(n)}\left(\{\lambda_i^*\}_{i=1,2,...,5}, y^+, z^+, \mathrm{Re}_\tau \right)\textbf{T}^{*(n)} 
\end{equation}
which is a novel model of 11 tensors, including a constant $\textbf{T}^{*(0)}$.
\blue{These augmented models for PCF and SDF preserve the Galilean and rotational invariances of the original model. Galilean invariance refers to the invariance to rectilinear uniform motion, namely a Galilean invariant remains the same if expressed in one or another Galilean frame of reference (see definition in \cite{pope2000c}). Obviously, the proposed input features are Galilean invariant since they are all dimensionless scalar values, remaining therefore the same in different inertial frames. Rotational invariance refers to invariance under an arbitrary rotation of the coordinate system by a constant angle, namely a change in orientation of the frame of reference, meaning that here is no need to specify a reference frame of validity. Sometimes these two properties are grouped together under the generic term of Galilean invariance \cite{ling2016c}. Mathematically, the model $\textbf{b}(\textbf{T}^{*(0)}_{\text{gen}}, \alpha, y^+, \mathrm{Re}_\tau, \cdots )$ has also this property if we have for any rotation matrix $\textbf{Q}$ (see definition in \cite{ling2016c}):}

\begin{equation}
\textbf{Q} \textbf{b}(\textbf{T}^{*(0)}_{\text{gen}}, \alpha, y^+, \mathrm{Re}_\tau, \cdots ) \textbf{Q}^T = \textbf{b}(\textbf{Q}\textbf{T}^{*(0)}_{\text{gen}}\textbf{Q}^T, \textbf{Q} \alpha \textbf{Q}^T, \textbf{Q} y^+ \textbf{Q}^T, \textbf{Q} \mathrm{Re}_\tau \textbf{Q}^T, \cdots)
\label{Eq=Rotational_invariance}
\end{equation}

\blue{Notably, all the scalar entries of the model are rotational invariant(i.e. independent of the coordinate frame), now comprising the normalized mean velocity gradient $\alpha$ and $\lambda_1^{*}$ to $\lambda_5^{*}$ for the PCF and SDF respectively, the dimensionless wall distances $y^+$, $z^+$ and the friction Reynolds number $\mathrm{Re}_\tau$. The introduction of $\textbf{T}^{*(0)}_{\text{gen}}$ also preserves the rotational invariance since it can be expressed as a linear combination of the tensors diag(1,1,0), diag(1,0,1), diag(0,0,1) and the identity tensor diag(1,1,1).}

\begin{figure}[h!]
\begin{subfigure}{.33\textwidth}
\centering
\includegraphics[width=\linewidth]{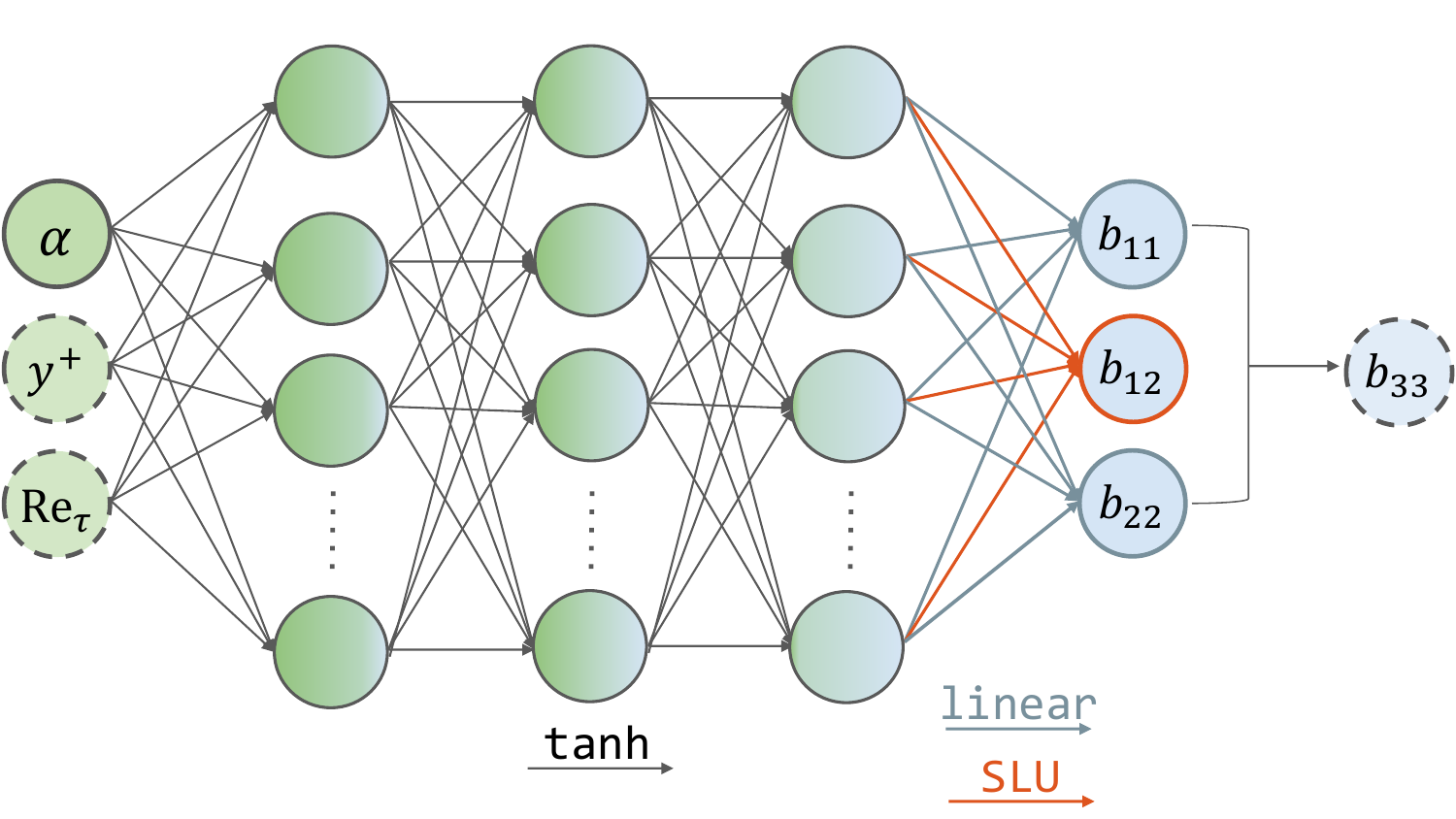}
\caption{MLP}
\label{fig:sketch_MLP}
\end{subfigure}%
\hfill
\begin{subfigure}{.33\textwidth}
\centering
\includegraphics[width=\linewidth]{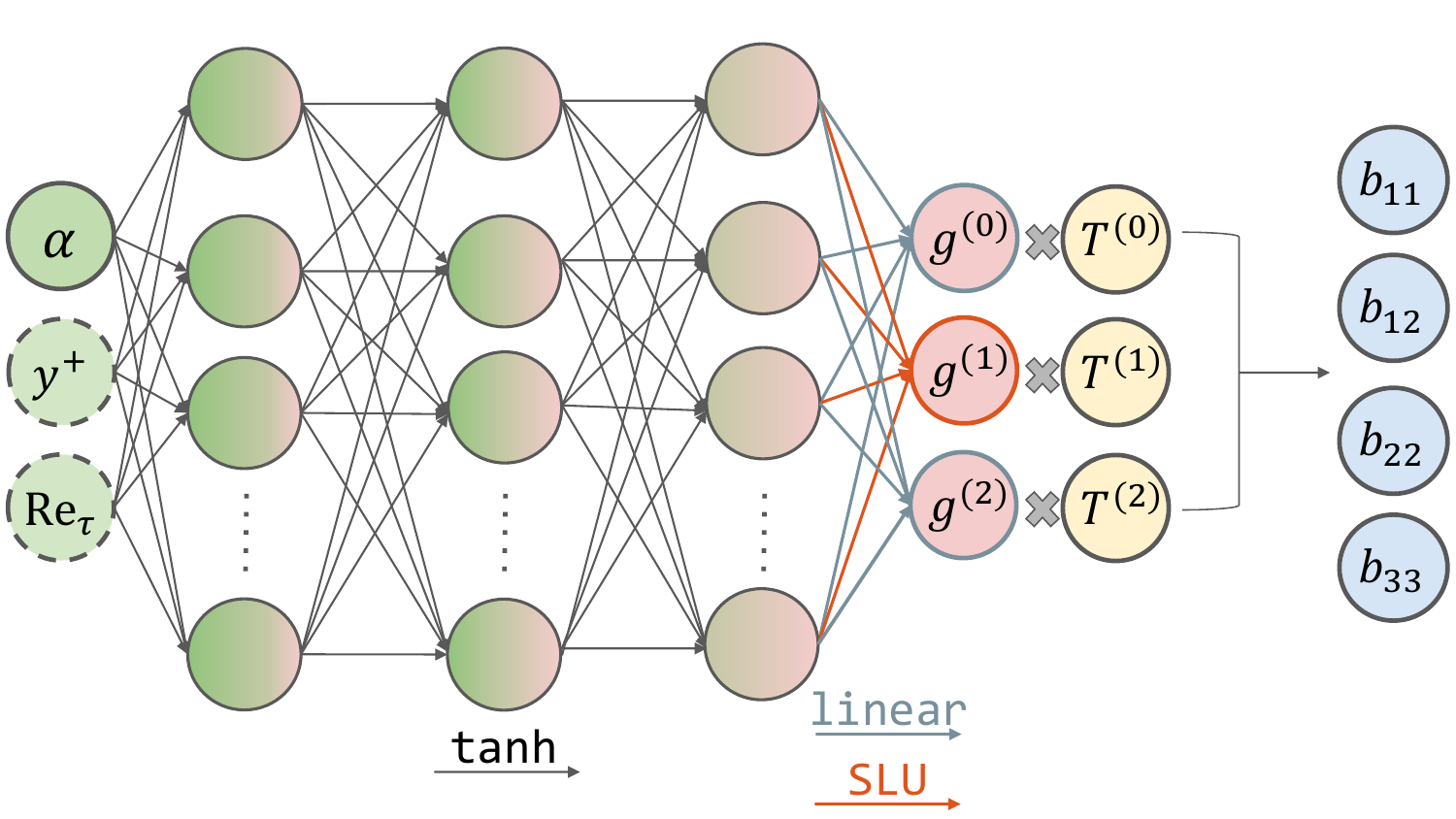}
\caption{aTBNN-1}
\label{fig:sketch_TBNN1}
\end{subfigure}%
\hfill
\begin{subfigure}{.33\textwidth}
\centering
\includegraphics[width=\linewidth]{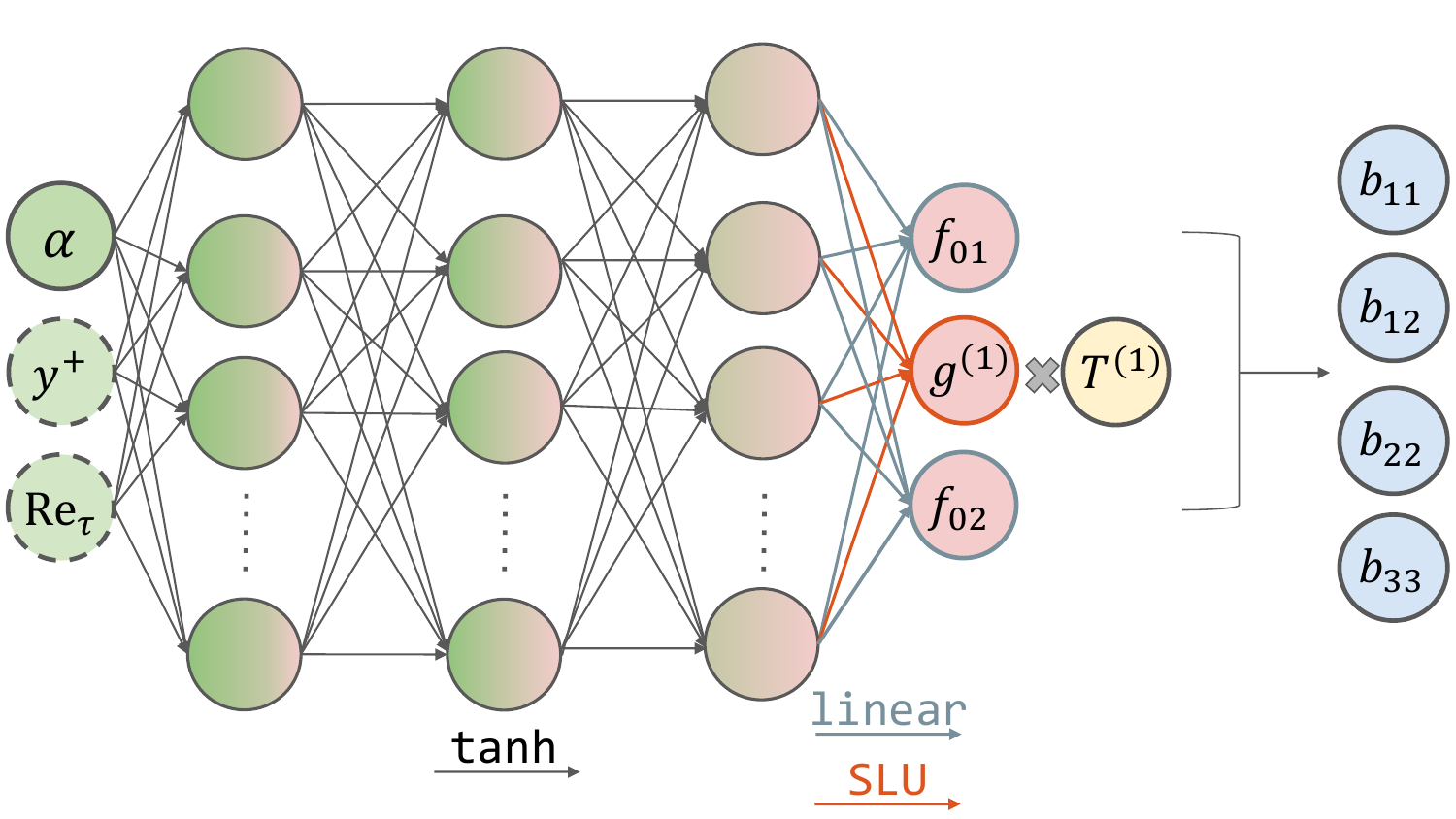}
\caption{aTBNN-2}
\label{fig:sketch_TBNN2}
\end{subfigure}

\caption{Diagrams of NN architectures used for PCF configuration.}
\label{fig:NN}
\end{figure}

Based on these physics models, we build up firstly an MLP for PCF configuration, as illustrated in Figure~\ref{fig:sketch_MLP}, testing different combinations of $\alpha, y^+$ and $\mathrm{Re}_\tau$ as input sets. The outputs of this model are $b_{11}$, $b_{12}$ and $b_{22}$, while $b_{33}$ is evaluated as $-(b_{11}+b_{22})$ to guarantee the zero-trace. There are 3 hidden layers, each with 10 neurons in this MLP model. The hidden layers are activated by the hyperbolic tangent function (tanh). The output node of $b_{12}$ is activated by the Softplus Linear Unit (SLU) to guarantee its negativity, while the others are linearly activated.

The aTBNN models for PCF corresponding to Eq.~\eqref{eq:aTBNN1} and Eq.~\eqref{eq:aTBNN2} are represented in Figure~\ref{fig:sketch_TBNN1} and Figure~\ref{fig:sketch_TBNN2}, and are denoted as aTBNN-1 and aTBNN-2 in the rest of the study. The same number of hidden layers and neurons are used as is in the MLP model. The activation function of the hidden layers remains tanh. The output layer contains three nodes for the three corresponding coefficient functions. Similarly, the output node of $g^{(1)}$ is activated by the SLU to ensure that the predicted $b_{12}$ is negative, the others are linearly activated by default.  

Another augmented model corresponding to Eq.~\eqref{eq:aTBNN3}, denoted as aTBNN-3, is established for SDF configuration. The architecture is similar to the original one, shown in Figure~\ref{fig:tbnn_ling}, only with more input features ($y^+, z^+$ and $\mathrm{Re}_\tau$) in addition to the five invariants and one more constant tensor ($\textbf{T}^{*(0)}$). Considering the increasing complexity, a deeper network is used, with 10 hidden layers and 50 neurons per layer. Various activation functions are tested for the aTBNN-3 model, and their performance will be compared in Section~\ref{sec:random}.

\subsection{Data preprocessing and visualization}\label{sec:prepro}
The data quality has a significant impact on the performance of a deep learning framework. During training, the NN tends to assign more weight to the inputs with larger value ranges, especially if there is a noticeable difference among their scales. This is not an ideal scenario since other smaller inputs may also be important for the predictions.

Hence, it is necessary to preprocess the input data before feeding them into the model in order to enhance the training robustness. Max normalization is applied on $\alpha$ and $\mathrm{Re}_\tau$, which aims to divide these features by their maximum. As for $y^+$, $z^+$, and the invariants $\lambda_i$, log-transformation is performed to address the long tail issue observed among these features. 

On the other side, it is typically advisable to individually normalize each regression target of a neural network to prevent one of them from dominating the loss function.  However, we prefer to adopt another strategy in order to preserve the zero trace of the Reynolds stress tensor. A global reduction based on the Frobenius norm of $b_{ij}$ is performed instead. \red{More preprocessing details are presented in \ref{app:A}.}

\begin{figure}[h!]
    \begin{subfigure}[c]{\textwidth}
    \centering
    \includegraphics[width=0.8\linewidth]{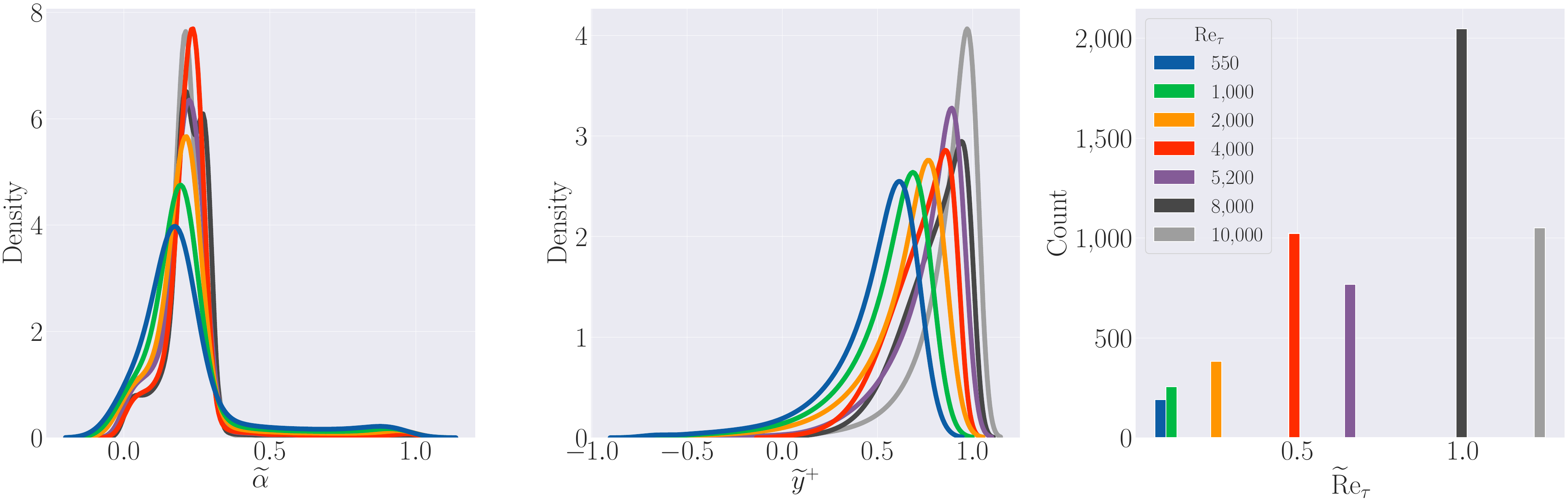}
    \caption{Inputs}
    \label{fig:inputs_pcf_after}
    \end{subfigure}%

    \begin{subfigure}[c]{\textwidth}
    \centering
    \includegraphics[width=0.8\linewidth]{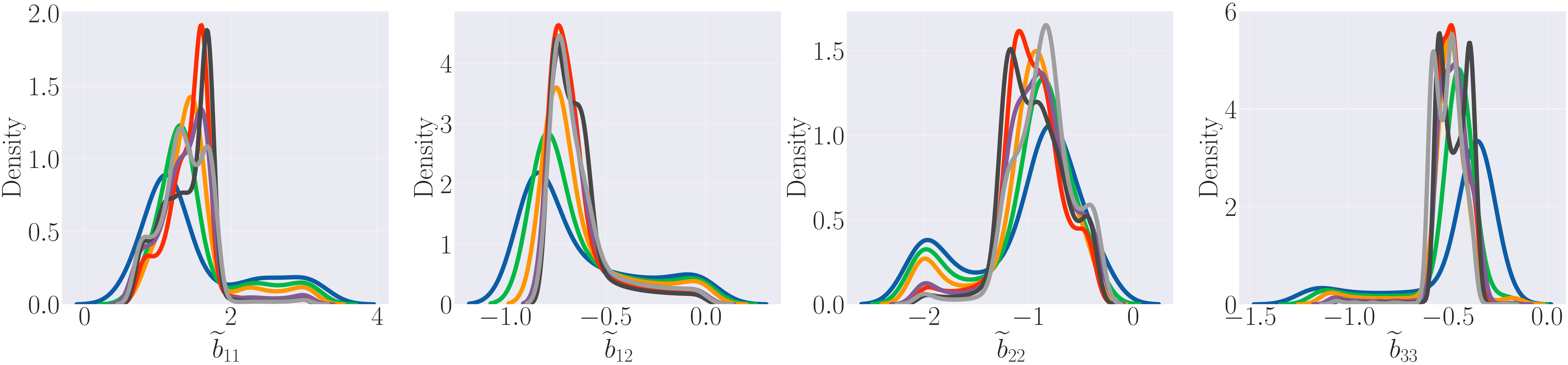}
    \caption{Outputs}
    \label{fig:outputs_pcf_after}
    \end{subfigure}    

\caption{Distributions of the inputs ($\widetilde{\alpha}$, $\widetilde{y^+}$ and $\widetilde{\mathrm{Re}}_\tau$) and outputs ($\widetilde{b}_{ij}$) at different friction Reynolds numbers after preprocessing for PCF study.}
\label{fig:pcf_after}
\end{figure}

\begin{figure}[h!]
 \hspace*{\fill}%
    \begin{subfigure}{.25\textwidth}
    \centering
    \includegraphics[width=\linewidth]{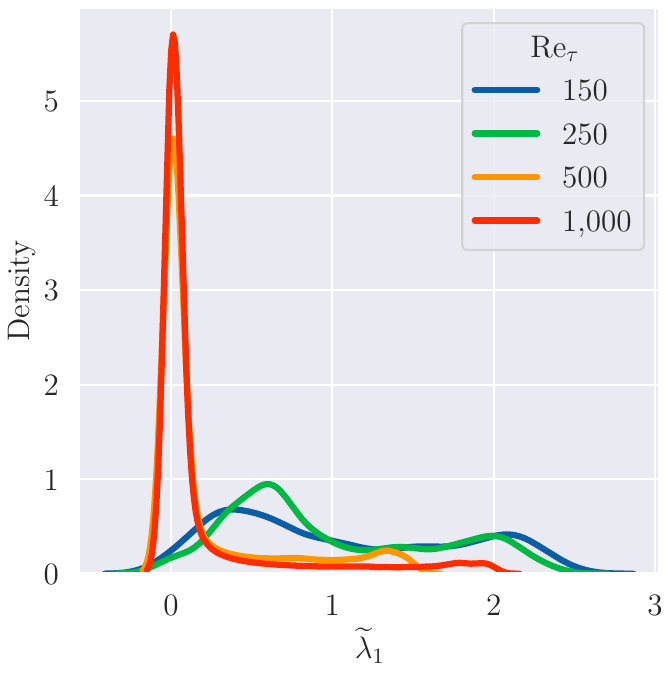}
    \caption{Input}
    \label{fig:inputs_sdf_after}
    \end{subfigure}%
\hspace*{\fill}%
    \begin{subfigure}{.25\textwidth}
    \centering
    \includegraphics[width=\linewidth]{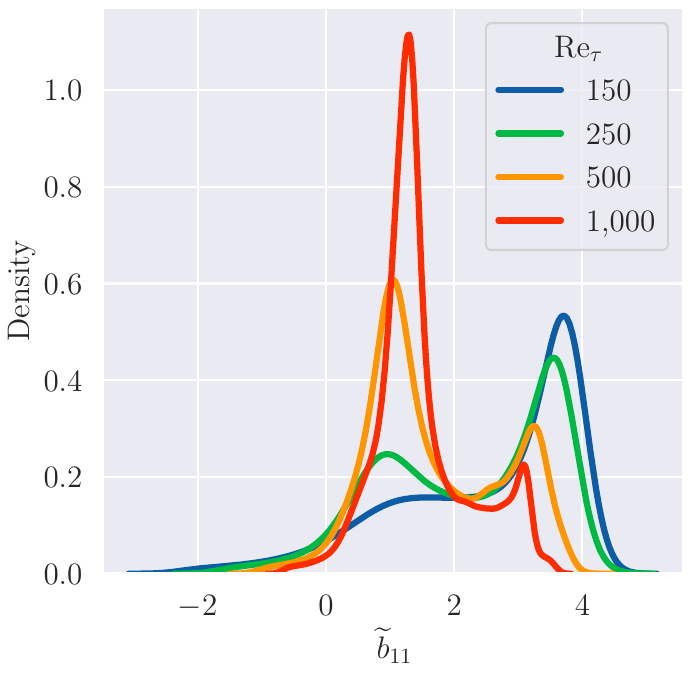}
    \caption{Output}
    \label{fig:outputs_sdf_after}
    \end{subfigure}    
\hspace*{\fill}%
\caption{Distributions of the input ($\widetilde{\lambda_1}$) and the output ($\widetilde{b}_{11}$) at different friction Reynolds numbers after preprocessing for SDF study.}
\label{fig:sdf_after}
\end{figure}

The distributions of the preprocessed inputs and outputs for PCF are shown in Figure~\ref{fig:pcf_after} (we denote preprocessed quantities with a wide-tilde in the following). It can be seen that all inputs and outputs are rescaled into comparable ranges with similar distributions at different friction Reynolds numbers, which is in favor of the anisotropy tensor predictions on a $\mathrm{Re}_\tau$ excluded from the training dataset, i.e., in an interpolation or extrapolation test case. The same scaling effect is observed for SDF data after preprocessing and will not be presented exhaustively here for the sake of compactness. However, some abnormalities are detected while analyzing the distributions of the preprocessed SDF data, represented in Figure~\ref{fig:pcf_after}. Concerning the outputs, a clear trend toward the $\mathrm{Re}_\tau$ can be observed. Taking the $\widetilde{b}_{11}$ as example: there exists two modes, one around $\widetilde{b}_{11} \approx 1$ and the other around $\widetilde{b}_{11} \approx 3$; the density of the former increases with the $\mathrm{Re}_\tau$, while the latter decreases with the $\mathrm{Re}_\tau$. The input distributions, on the other hand, can be a source of concern. Figure~\ref{fig:inputs_sdf_after} illustrates the distribution of $\widetilde{\lambda_1}$ as an example, where no evident trend can be found towards $\mathrm{Re}_\tau$. Consequently, it becomes mathematically challenging to derive a mapping from the invariants to the anisotropy tensor for SDF at different $\mathrm{Re}_\tau$. Hence, NNs may not perform effectively in interpolation or extrapolation scenarios for SDF studies. 

\subsection{Training strategies}\label{sec:strategie}
The goal of training a neural network is to minimize a loss function that compares the predictions with the target outputs. In this work, the loss function $\mathcal{L}$
 is defined as a weighted Mean Squared Error (MSE) based on the Reynolds stress anisotropy tensor components:
\begin{equation}\label{eq:loss_pcf}
\mathcal{L}_{\mathrm{PCF}} = \dfrac{1}{4m} {\sum_{i=1}^{m}} \left[ {\sum_{k=1}^{3}} \mathrm{W}_k (b_{kk}^{(i)} - \widehat{b}_{kk}^{(i)}) ^2 + \mathrm{W}_4 (b_{12}^{(i)} - \widehat{b}_{12}^{(i)}) ^2 \right]
\end{equation}
\noindent for PCF and 

\begin{equation}\label{eq:loss_sdf}
\mathcal{L}_{\mathrm{SDF}} = \dfrac{1}{4m} {\sum_{i=1}^{m}} \left[ {\sum_{k=1}^{3}}  \mathrm{W}_k (b_{kk}^{(i)} - \widehat{b}_{kk}^{(i)}) ^2 +  \mathrm{W}_4 (b_{12}^{(i)} - \widehat{b}_{12}^{(i)}) ^2 +  \mathrm{W}_5 (b_{13}^{(i)} - \widehat{b}_{13}^{(i)}) ^2 +  \mathrm{W}_6 (b_{23}^{(i)} - \widehat{b}_{23}^{(i)}) ^2 \right]
\end{equation}

\noindent for SDF, where the predicted outputs are denoted with a hat, $m$ is the total number of training data, and $\mathrm{W}_k$ ($k = 1, 2, ..., 6$) are weights allocated to each $b_{ij}$ component in the loss function whose sum is equal to unity.

In order to minimize the above loss functions, several state-of-the-art training strategies are applied here and will be presented \red{below}.\\    

\noindent \textit{Learning rate decay} \\

In the back-propagation process involved in the NN training, the weights ($\theta$) and the bias ($b$) of each neuron are updated using a gradient descent algorithm:
\begin{align}
\begin{cases}
\theta = \theta - \eta \frac{\partial \mathcal{L}}{\partial \theta} \\
b = b - \eta \frac{\partial \mathcal{L}}{\partial b}
\end{cases}
\end{align}

\noindent where $\eta$ is the Learning Rate (LR), which adjusts the step size at each training iteration while moving toward the minimum of a loss function with regards to the weights and the bias. A high LR enables the model to learn faster but could overshoot the minimum by taking too big steps; a small LR enables the model to take small updates and learn carefully while taking longer to converge and possibly even getting stuck in an undesirable local minimum of the loss function. Hence, it is usually considered as the most important hyper-parameter \cite{Goodfellow-et-al-2016}. The traditional way is to use a constant LR throughout the learning process and to figure out an optimal value by tuning. However, various works in recent years demonstrate good performance by using a varying LR \cite{smith2017, yu2020a}. Some powerful techniques have been developed and can be divided into two categories: automatically tuning the LR, or decaying the LR globally \cite{chen2021}. The present work adopts the exponential decay schedule whose mathematical form is as follows: 
\begin{equation}
\theta = \theta_0 \cdot k^{\frac {t}{T}}
\end{equation}
\noindent where $\theta_0$ is the initial LR, $k$ the exponential decay rate, $t$ the number of the current epoch and $T$ the decay epoch. Instead of explicitly stating, $\theta_0, k$ and $T$ are respectively set at 0.001, 0.01, and $\np{30000}$ in the present work, meaning that the LR decays every $\np{30000}$ epochs with a base of 0.01, from its initial value at 0.001. 

The wisdom behind LR decay schedules is that a higher LR can be used at the beginning of the learning process to locate quickly the range of good parameters, then a gradually decrease of the LR allows a finer exploration around the local minimum. \\

\noindent \textit{Adaptive loss weighting} \\

It appears from Eqs.~\eqref{eq:loss_pcf} and \eqref{eq:loss_sdf} that the loss functions in our work are multi-part functions that combine different components of the anisotropy tensor. The contribution of each component is measured by the weights $\mathrm{W}_k$ ($k = 1, 2, ..., 6$). Traditionally, the weights in a multi-part loss function are set at the same value, or tuned experimentally to yield near-optimal results. However, recent works have pointed out the need for more sophistical weighting strategies. On the one hand, the learning difficulty of each part could be different; on the other hand, the scaling of each component could also be different despite the preprocessing efforts, which would encourage the optimizer to only look at the component with the largest magnitude. 

The loss weighted SoftAdapt algorithm \cite{heydari2019} is implemented in the present work in order to adaptively update the weights for each loss component throughout the learning process. Let $s_k^i$ be a finite difference approximation of the recent rate of change of the $k^{th}$ component loss $\mathcal{L}_k$ (\textit{e.g.} at iteration $i$, $s_k^i = \mathcal{L}_k^i - \mathcal{L}_k^{i-1}$ by taking the first order approximation). Then the weights of each component are as follows: 
\begin{equation}\label{eq:softadapt}
    W_k^i = \dfrac{ \mathcal{L}_k^i e ^ {\beta s_k^i}}{ \sum_{l=1}^{n} \mathcal{L}_l^i e ^ {\beta s_l^i} }
\end{equation}
\noindent where $\beta$ is a tunable hyper-parameter set at 0.1. 

The mathematical intuition of the algorithm can be clearly seen from Eq.~\eqref{eq:softadapt} and is explained in the original paper. Firstly, the recent performance of each loss component is taken into account. By choosing $\beta > 0$, more weights will be assigned to the worst-performing loss component (the one with the most positive rate of change). On the contrary, a negative $\beta$ favors the best-performing component (with the most negative rate of change). Here we use a positive $\beta$ to improve the worst performing loss component. Secondly, the current values of each loss component also have an impact on the weight allocation. Smaller weights will be assigned to those close to their minima. The performance of the implemented algorithm will be discussed in Section~\ref{sec:random}.

The above techniques represent the current state-of-the-art training strategies, being investigated very recently in the domain of physics-informed deep learning \cite{bischof2021, xiang2022}. Some other modern training techniques are applied in the present work and will not be presented in detail, such as mini-batch training \cite{ruder2017a} and early stopping \cite{prechelt1998}. Regularization, batch-normalization, and drop-out techniques \cite{Goodfellow-et-al-2016} have also been tested but have not shown significant improvements in performance. As a result, these techniques were not retained in the following.. Transfer Learning (TL) \cite{pan2010} will also be tested  and presented subsequently.

\section{Results}\label{sec:results}
In this section, the results of various aTBNN models developed in the present work for PCF case and SDF case are discussed, including the performance of a TL framework, aimed at transferring prior knowledge acquired from a trained aTBNN model. \red{The coefficient of determination $R^2$ is used as another metric other than the MSE to report the model performance, which is defined as follows:}

\begin{equation}
R^2 = 1 - \dfrac{\sum ({y}_{i} - \widehat{y}_{i})^2}{\sum (y_{i} - \overline{Y})^2}
\end{equation}
\red{where $\widehat{y}_{i}$ is the predicted $i^{\text{th}}$ value, $\overline{y}_{i}$ is the actual $i^{\text{th}}$ value and $\overline{Y}$ is the mean of the true values. By definition, the closer the $R^2$ value is to 1, the better the prediction.}

\subsection{Plane channel flow}
Eight case studies are performed for the PCF study (summarized in Table \ref{tab:case}), among which Case 1 - Case 4 are conducted to investigate the impact of input features using the more flexible MLP model (see Figure~\ref{fig:sketch_MLP}), and Case 5 - Case 8 aim to find out the optimal choice of $\textbf{T}^{*(0)}$ among the augmented TBNN models(see Figure~\ref{fig:sketch_TBNN1} and Figure~\ref{fig:sketch_TBNN2}). These models are evaluated on three test sets at different Reynolds numbers: test 1 at $\mathrm{Re}_\tau=550$, test 2 at $\mathrm{Re}_\tau=\np{5200}$ and test 3 at $\mathrm{Re}_\tau=\np{10000}$. Further comparison between these two sets of case studies using respectively MLP and augmented TBNN allows us to select a better neural network model.

\begin{table}
\centering
\caption{\label{tab:case}Summary of case studies on PCF.}
\begin{tabular}{lccccccc}
\toprule
Case & Model & Features & $\textbf{T}^{*(0)}$\\\hline
1 & MLP & $\{\alpha\}$ & /\\
2 & MLP & $\{\alpha, y^+\}$ & /\\
3 & MLP & $\{\alpha, \mathrm{Re}_\tau\}$ & /\\
4 & MLP & $\{\alpha, y^+, \mathrm{Re}_\tau\}$ & /\\
5 & aTBNN-1 & $\{\alpha, y^+, \mathrm{Re}_\tau\}$ & $\textbf{T}^{*(01)}$ \\
6 & aTBNN-1 & $\{\alpha, y^+, \mathrm{Re}_\tau\}$ & $\textbf{T}^{*(02)}$ \\
7 & aTBNN-1 & $\{\alpha, y^+, \mathrm{Re}_\tau\}$ & $\textbf{T}^{*(03)}$ \\
8 & aTBNN-2 & $\{\alpha, y^+, \mathrm{Re}_\tau\}$ & $\textbf{T}^{*(0)}_\mathrm{gen}$ \\
\bottomrule
\end{tabular}
\end{table}

\subsubsection{Input feature selection}
We train the MLP model with different feature combinations in order to figure out the role of each entry and find the ideal set: $\{\alpha\}$ in Case 1, $\{\alpha, y^+\}$ in Case 2, $\{\alpha, \mathrm{Re}_\tau\}$ in Case 3 and  $\{\alpha, y^+, \mathrm{Re}_\tau\}$ in Case 4. The training is respectively stopped at 1,079, 19,940, 2,455, and 5,217 epochs for each case when the loss value evaluated on the validation data set starts to stagnate.

\begin{figure}[h!]
\hspace*{\fill}%
    \begin{subfigure}{0.4\textwidth}
    \includegraphics[width=\linewidth]{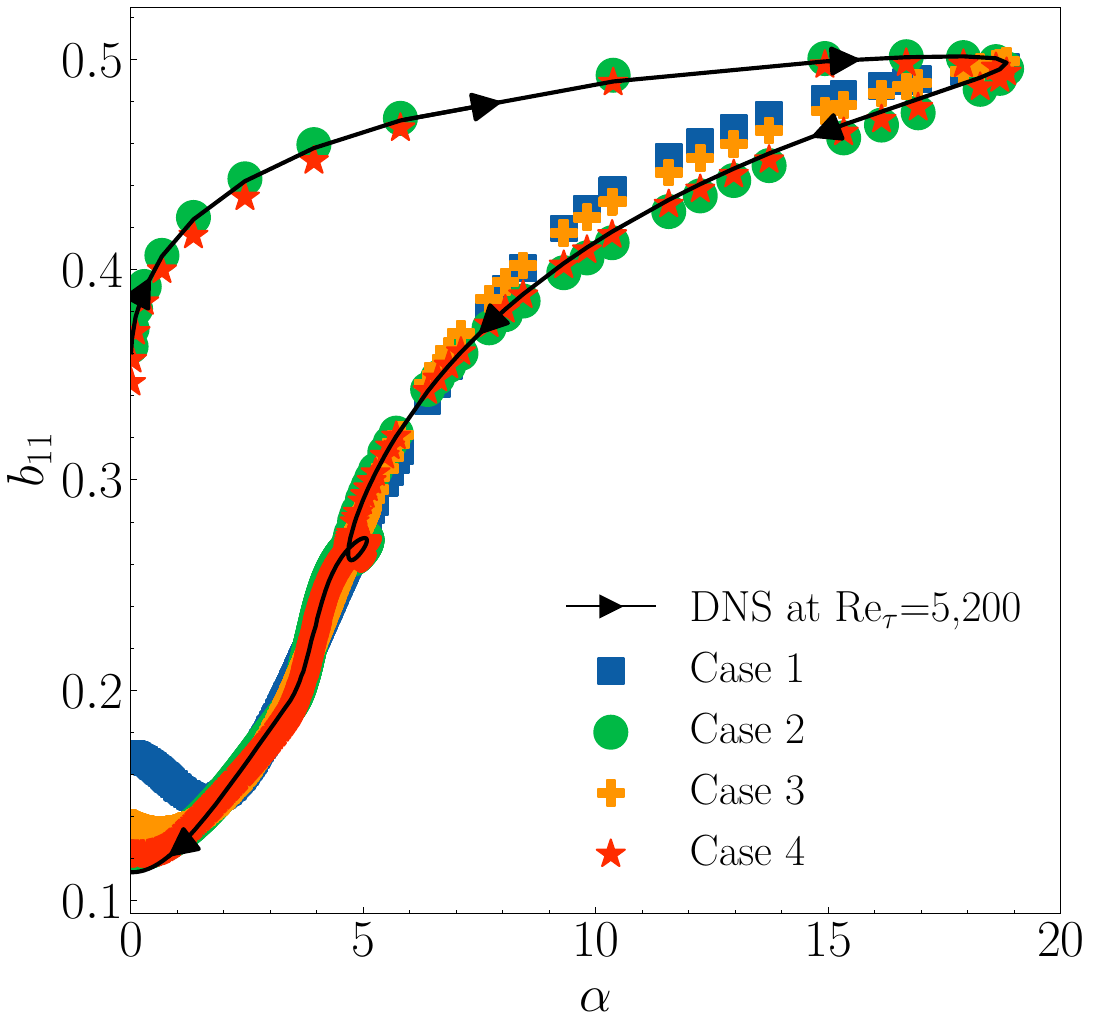}
    \caption{}
    \label{fig:b11_Case1-4_alpha}
    \end{subfigure}% 
\hspace*{\fill}%
    \begin{subfigure}{0.4\textwidth}
    \includegraphics[width=0.99\linewidth]{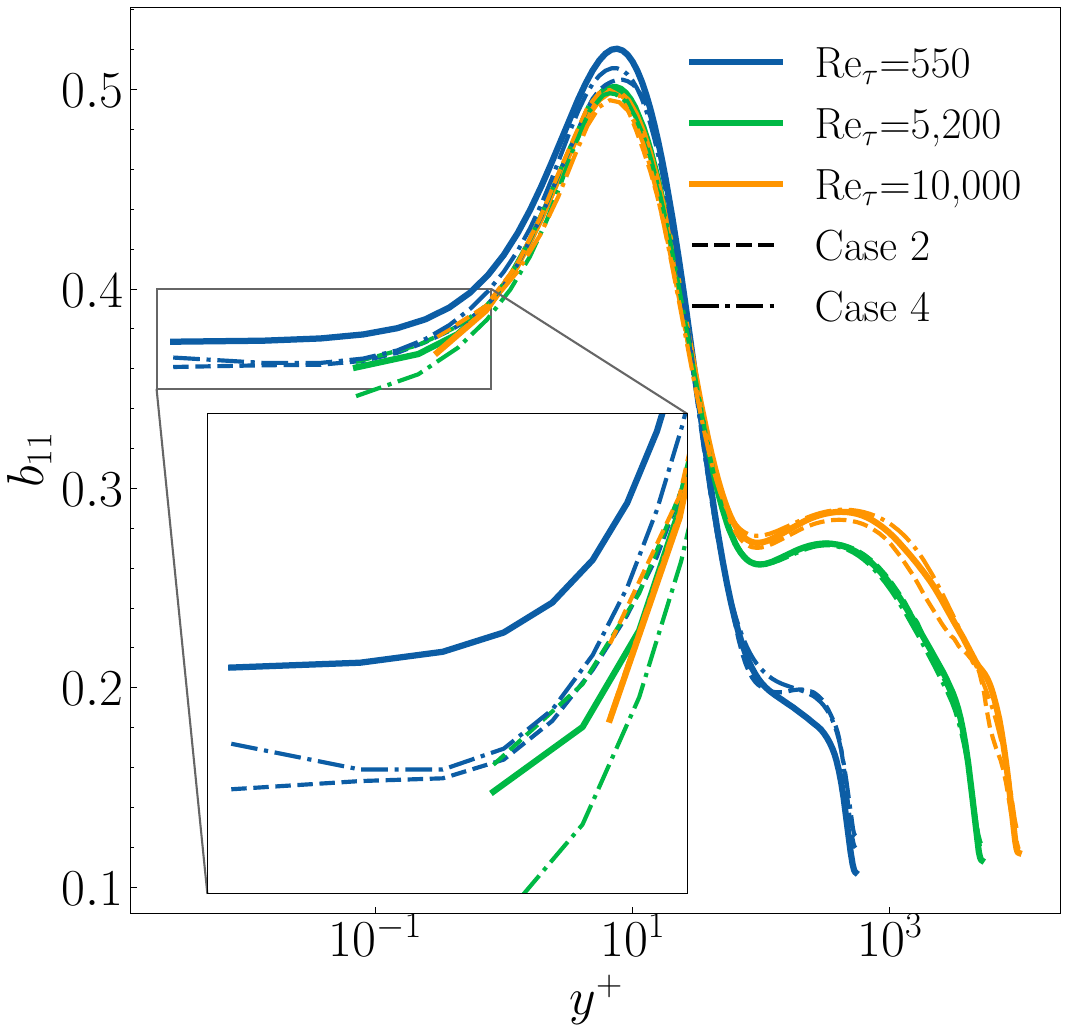}
    \caption{}
    \label{fig:b11_Case1-4_y+}
    \end{subfigure} 
 \hspace*{\fill}%
\caption{$b_{11}$ predictions of PCF compared with DNS data (in solid lines, pointing from the wall to the center of the channel) at (a) $\mathrm{Re}_\tau=\np{5200}$ for Case 1 - Case 4; (b) $\mathrm{Re}_\tau=550; \np{5200}; \np{10000}$ for Case 2 (in dashed lines) and Case 4 (in dotted lines).}
\label{fig:b11_Case1-4}
\end{figure}

Figure \ref{fig:b11_Case1-4} shows the $b_{11}$ predictions in Case 1 - Case 4 for test 2 with $\mathrm{Re}_\tau=\np{5200}$, compared with the DNS data. Noticeably, Case 1 and Case 3 have similar behavior and both completely fail to predict the upper branch of $b_{11}$, corresponding to the near-wall region. The failure of Case 1 confirms the limitation of Pope's model, which relates each $b_{ij}$ component only to $\alpha$ for PCF. Based on such an assumption, the NN built in Case 1 tries to construct a function between $\alpha$ and each learning target $b_{ij}$, which is indeed not feasible according to the Figure~\ref{fig:b_alpha}. 

On the other hand, by including $y^+$ into the feature set, Case 2 and Case 4 are able to overcome the multi-valued issue and to capture the trends and behavior of DNS data with only minor discrepancies, as shown in Figure~\ref{fig:b11_Case1-4_alpha}. Results given in these two cases closely overlap and are in good agreement with the DNS data. To gain a deeper insight into the difference between them, we plot the $b_{11}(y^+)$ curves predicted for all the test sets in Figure~\ref{fig:b11_Case1-4_y+}. Despite the good performance shown in Case 2, we particularly find that it gives the same predictions for all flows at three different friction Reynolds numbers in the near-wall region, which can be observed more clearly in the zoomed-out window. This makes sense because the magnitudes of $y^+$ are universal in the viscous layer (corresponding to the region with $y^+ < 10$) and vary from one $\mathrm{Re}_\tau$ to another while approaching to the center of the channel. More precisely, the lower limit of the $y^+$ is always set at the order of $0.1$ for flows at all the Reynolds numbers in order to guarantee the resolution of a given DNS experiment, while by definition its upper limit is the friction Reynolds number $\mathrm{Re}_\tau$, which differs from one flow to another. Consequently, having only $\{\alpha, y^+\}$ in the feature set, the NN trained in Case 2 can not distinguish flows at different Reynolds numbers in the near-wall region where the $y^+$ value is small. 

In Table \ref{tab:R2results} are reported the $R^2$ values for each case study. An obvious improvement of $R^2$ values is achieved in Case 2 and Case 4 by adding $y^+$ into the feature set. The difference between Case 2 and Case 4 is not that pronounced in terms of $R^2$ values and especially lies in the test sets, in particular for higher Reynolds number extrapolation test at $\mathrm{Re}_\tau=\np{10000}$. However, a specific attention is drawn to the learning difficulty of Case 2 since we had to push the training epochs till 19,940. This is understandable since the NN trained in Case 2 was fed with data mixed with different $\mathrm{Re}_\tau$ without explicitly having this information as is in Case 4, and so it could take a long time for the network to reflect this fact. Hence, we conclude that both $y^+$ and $\mathrm{Re}_\tau$ are critical for our model, and we keep the feature set  $\{\alpha, y^+, \mathrm{Re}_\tau\}$ in the following aTBNN models in PCF to yield better performance in terms of both accuracy and ease of convergence.    

\subsubsection{$\textbf{T}^{*(0)}$ selection}\label{sec:T0select}
As is mentioned in Section \ref{sec:pope_pcf}, we question whether there exists an optimal choice of $\textbf{T}^{*(0)}$ for the aTBNN models. On the basis of former results, we train the aTBNN model using different $\textbf{T}^{*(0)}$ with the feature set $\{\alpha, y^+, \mathrm{Re}_\tau\}$, which yields the best results according to the previous section. The training for Case 5 - Case 8 of Table \ref{tab:case} is respectively stopped at  9,993, 34,215, 30,990, and 10,943 epochs. 

\begin{figure}[h!]
\centering
\includegraphics[width=\textwidth,center]{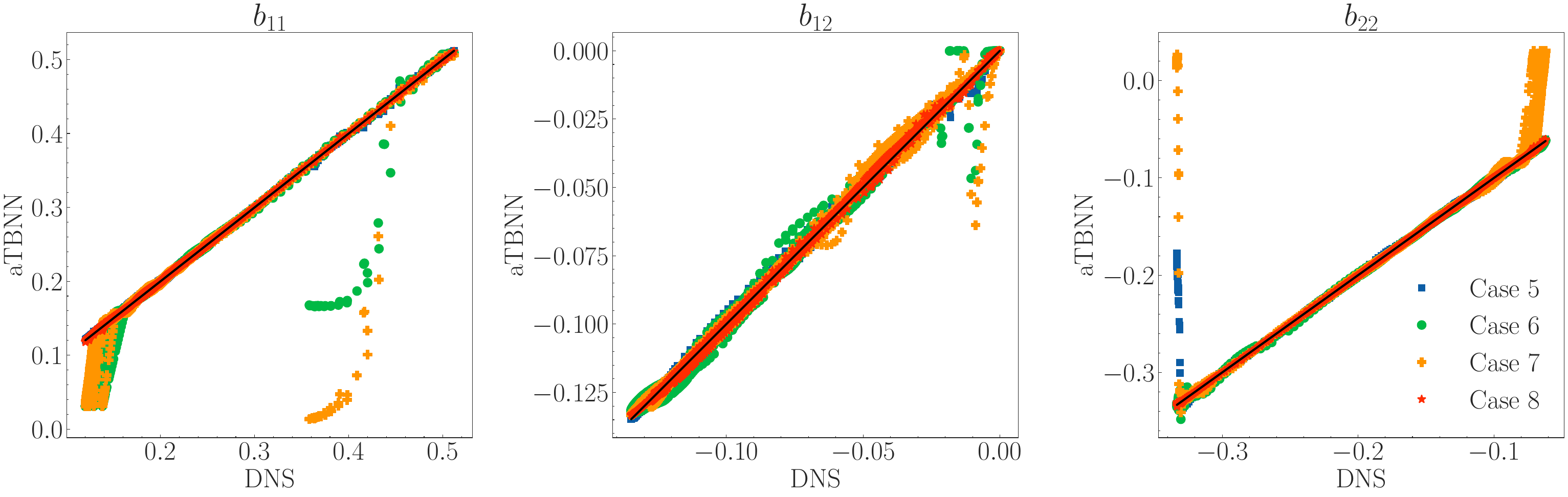}
\caption{$b_{11}$, $b_{12}$ and $b_{22}$ predictions for Case 5 - Case 8 using aTBNN models on PCF, compared to DNS data, in dispersion plots.}
\label{fig:T0}
\end{figure}

Figure \ref{fig:T0} shows the training results for Case 5 - Case 8. Since $b_{33}$ is calculated from $b_{11}$ and $b_{22}$ in the learning process, the corresponding values are not shown. We discover that the performance of the aTBNN models is strongly affected by the choice of $\textbf{T}^{*(0)}$ in each case. This is an important finding as it sheds light on the potential influence of the chosen tensor basis, which has rarely been discussed in the existing literature. In particular, the aTBNN-1 models in Cases 5, 6, and 7 utilizing constant $\textbf{T}^{*(0)}$ proposed by Pope learn well in general but perform notably poorly in some specific intervals, especially when it comes to diagonal components, $b_{11}$ and $b_{22}$. As can be observed, Case 6 and Case 7 totally fail to predict values near the boundaries of $b_{ij}$ intervals, which appears to be a systematic error. By comparing with Figure~\ref{fig:b_diag}, we identify that those zones refer to either the center of the channel ($y=1$) or the near-wall region. Given that the model trained in Case 5 manages to learn values near the channel center, we deduce that the failure at the center for Case 6 and Case 7 is caused by the chosen $\textbf{T}^{*(0)}$ in the model. This is in agreement with our previous analysis based on Figure \ref{fig:b_diag} in Section \ref{sec:pope_pcf}. Nevertheless, the model of Case 5 still fails to predict the $b_{22}$ component in the near-wall region for unclear reasons. It should be emphasized that the aTBNN-2 model used in Case 8 performs perfectly well by using our newly proposed generalized $\textbf{T}^{*(0)}$, as clearly demonstrated in Figure \ref{fig:T0}. The $R^2$ values for Case 5 - Case 8 can also be found in Table \ref{tab:R2results}. Eventually, the model trained with the generalized $\textbf{T}^{*(0)}$ in Case 8 outperforms those with constant $\textbf{T}^{*(0)}$ values in all aspects. 

\begin{singlespace} 
\begin{table}[h!]
\centering
\caption{\label{tab:R2results}$R^2$ \red{coefficient} of $b_{ij}$ predictions for case studies on PCF: test 1, 2 and 3 respectively for test set at $\mathrm{Re}_\tau=550; \np{5200}$ and $\np{10000}$.}
\begin{tabular}{lcccccc}
\toprule
\multicolumn{2}{l}{Case} & $b_{11}$ & $b_{12}$ & $b_{22}$ & $b_{33}$ & Global \\\hline
\multirow{5}{*}{1} & Train & 0.8007 & 0.8755 & 0.7300 & 0.7164 & 0.7807 \\ 
& Test   & 0.8116 & 0.8713 & 0.7340 & 0.5956 & 0.7531\\
& Test 1 & 0.6025 & 0.7808 & 0.3785 & 0.7446 & 0.6266\\
& Test 2 & 0.7990 & 0.8915 & 0.7209 & 0.7854 & 0.7992\\
& Test 3 & 0.8517 & 0.8760 & 0.8062 & 0.4575 & 0.7492\\
\hline
\multirow{5}{*}{2} & Train & 0.9994 & 0.9988 & 0.9996 & 0.9961 & 0.9985 \\ 
& Test & 0.9670 & 0.9837 & 0.9955 & 0.6108 & 0.8892 \\
& Test 1 & 0.9938 & 0.9935 & 0.9985 & 0.9607 & 0.9866 \\ 
& Test 2 & 0.9979 & 0.9990 & 0.9988 & 0.9898 & 0.9964\\
& Test 3 & 0.9440 & 0.9729 & 0.9930 & 0.3255 & 0.8089 \\
\hline
\multirow{5}{*}{3} & Train & 0.8501 & 0.9128 & 0.8027 & 0.7606 & 0.8315 \\ 
& Test & 0.8164 & 0.9095 & 0.7719 & 0.4778 & 0.7439 \\
& Test 1 & 0.6772 & 0.8193 & 0.5494 & 0.7799 & 0.7065 \\
& Test 2 & 0.7945 & 0.9197 & 0.7128 & 0.7963 & 0.8058\\
& Test 3 & 0.8545 & 0.9199 & 0.8770 & 0.2366 & 0.7145\\
\hline
\multirow{5}{*}{4} & \textbf{Train} & \textbf{0.9997} & \textbf{0.9994} & \textbf{0.9999} & \textbf{0.9983} & \textbf{0.9993} \\ 
& \textbf{Test}  & \textbf{0.9937} & \textbf{0.9743} & \textbf{0.9956} & \textbf{0.9646} & \textbf{0.9820} \\
& Test 1 & 0.9904 & 0.9941 & 0.9952 & 0.9753 & 0.9888 \\
& Test 2 & 0.9968 & 0.9983 & 0.9987 & 0.9852 & 0.9947 \\
& Test 3 & 0.9924 & 0.9567 & 0.9939 & 0.9505 & 0.9734 \\
\hline
\multirow{5}{*}{5} & Train & 0.9997 & 0.9989  & 0.9374 & 0.4657 & 0.8504 \\ 
& Test  & 0.9882 & 0.9822 & 0.9405 & 0.4002 & 0.8278 \\
& Test 1 & 0.9919 & 0.9822 & 0.8252 & 0.3217 & 0.7803 \\
& Test 2 & 0.9975 & 0.9967 & 0.9359 & 0.3864 & 0.8291 \\
& Test 3 & 0.9820 & 0.9737 & 0.9641 & 0.4226 & 0.8356 \\
\hline
\multirow{5}{*}{6} & Train & 0.8371 & 0.9939 & 0.9992 & -0.1143 & 0.4217 \\ 
& Test  & 0.8185 & 0.9787 & 0.9171 & -0.2006 & 0.1771 \\
& Test 1 & 0.7483 & 0.9488 & 0.9906 & -0.1101 & 0.3968 \\
& Test 2 & 0.8382 & 0.9963 & 0.9952 & -0.1364 & 0.3665 \\
& Test 3 & 0.8197 & 0.9739 & 0.8580 & -2.5460 & 0.0264 \\
\hline
\multirow{5}{*}{7} & Train & 0.7222 & 0.9910 & 0.5824 & 0.9858 & 0.8204 \\ 
& Test & 0.7428 & 0.9766 & 0.6140 & 0.7561 & 0.7724 \\
& Test 1 & 0.4239 & 0.9595 & -0.0078 & 0.9524 & 0.5820 \\
& Test 2 & 0.7172 & 0.9934 & 0.5618 & 0.9818 & 0.8135 \\
& Test 3 & 0.8158 & 0.9698 & 0.7576 & 0.5885 & 0.7829 \\
\hline
\multirow{5}{*}{8} & \textbf{Train} & \textbf{0.9997} & \textbf{0.9989} & \textbf{0.9999} & \textbf{0.9985} & \textbf{0.9993} \\ 
& \textbf{Test}  & \textbf{0.9911} & \textbf{0.9793} & \textbf{0.9872} & \textbf{0.7843} & \textbf{0.9355} \\
& Test 1 & 0.9930 & 0.9789 & 0.9982 & 0.9720 & 0.9855 \\
& Test 2 & 0.9970 & 0.9976 & 0.9986 & 0.9892 & 0.9956\\
& Test 3 & 0.9873 & 0.9686 & 0.9786 & 0.6304 & 0.8912\\
\bottomrule
\end{tabular}
\end{table}
\end{singlespace} 

\subsubsection{Best networks comparison}
\begin{figure}[h!]
\centering
\begin{subfigure}{0.3\textwidth}
    \includegraphics[width=\linewidth]{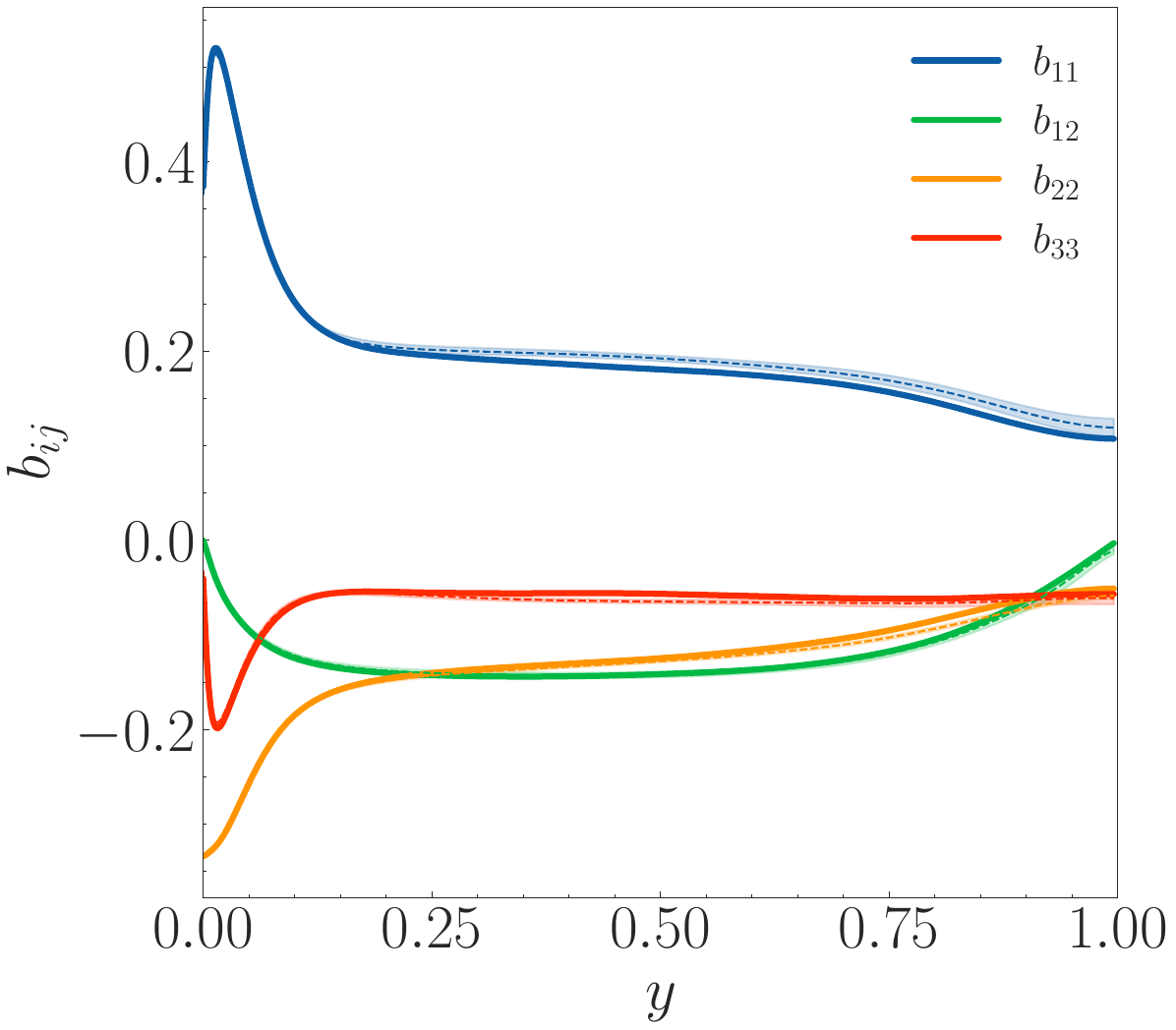}
    \caption{Case 4: Test 1}
    \label{fig:MLP_10runs_T1}
\end{subfigure}
\hfill
\begin{subfigure}{0.3\textwidth}
    \includegraphics[width=\linewidth]{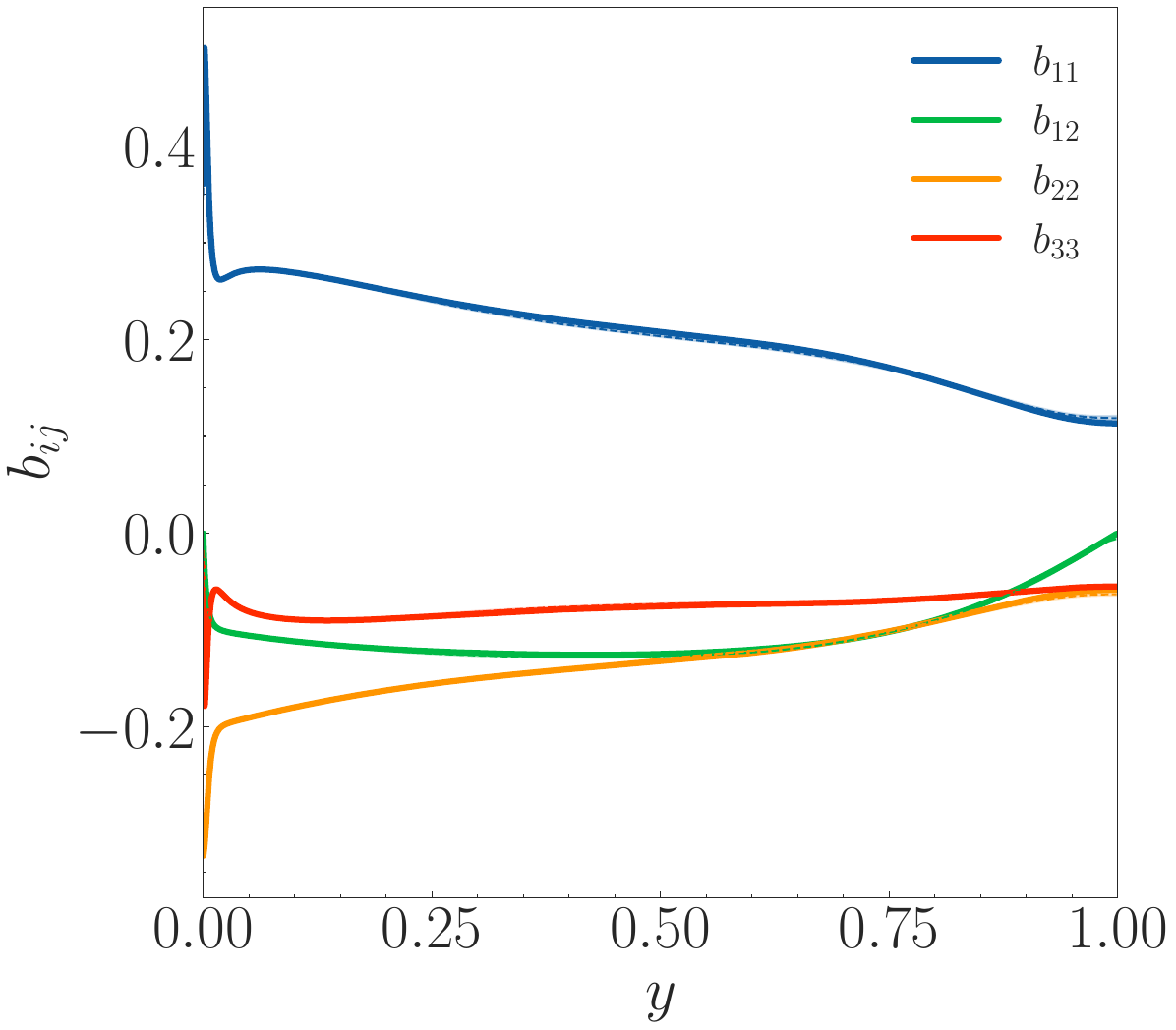}
    \caption{Case 4: Test 2}
    \label{fig:MLP_10runs_T2}
\end{subfigure}
\hfill
\begin{subfigure}{0.3\textwidth}
    \includegraphics[width=\linewidth]{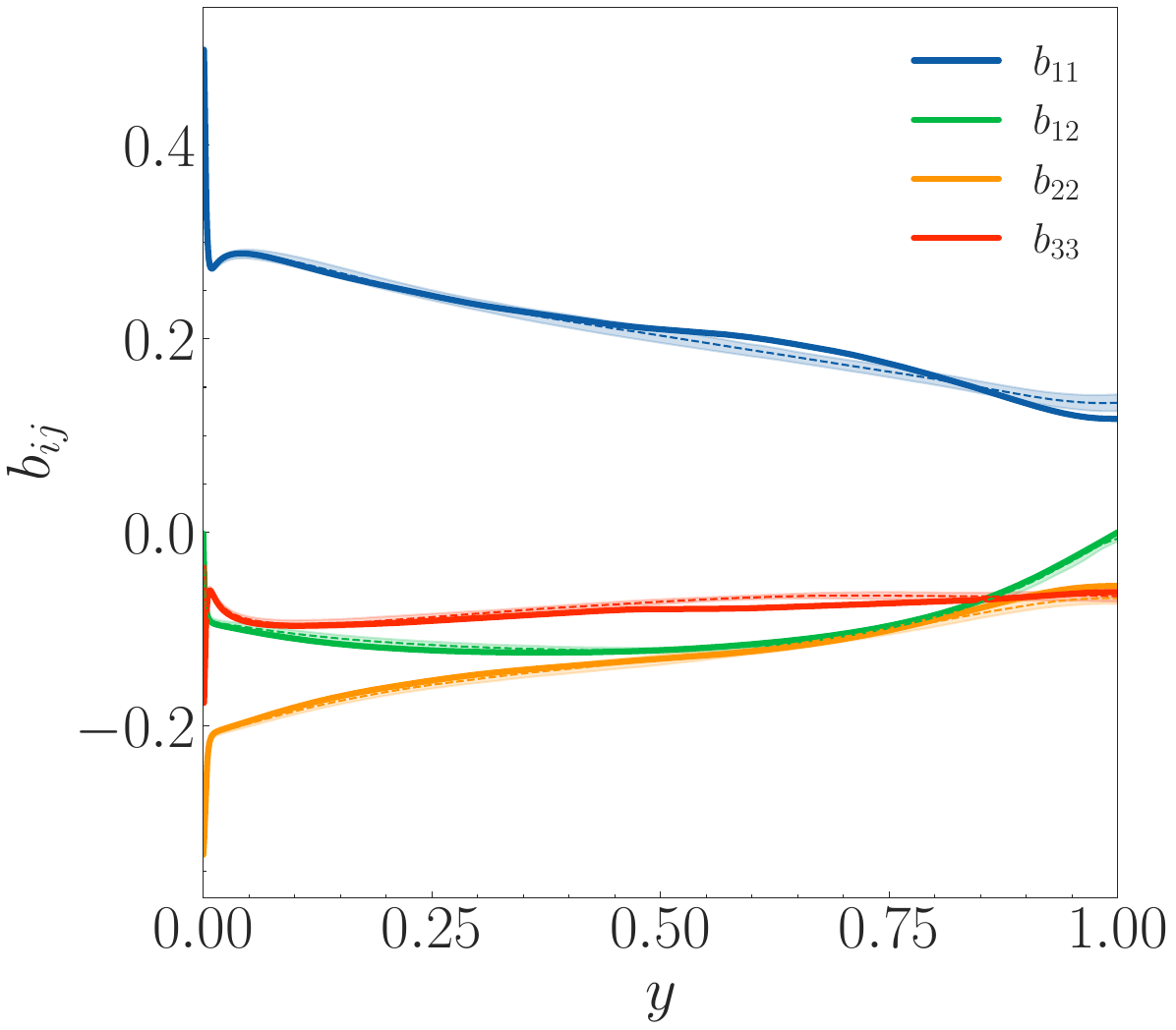}
    \caption{Case 4: Test 3}
    \label{fig:MLP_10runs_T3}
\end{subfigure}
\hfill
\begin{subfigure}{0.3\textwidth}
    \includegraphics[width=\linewidth]{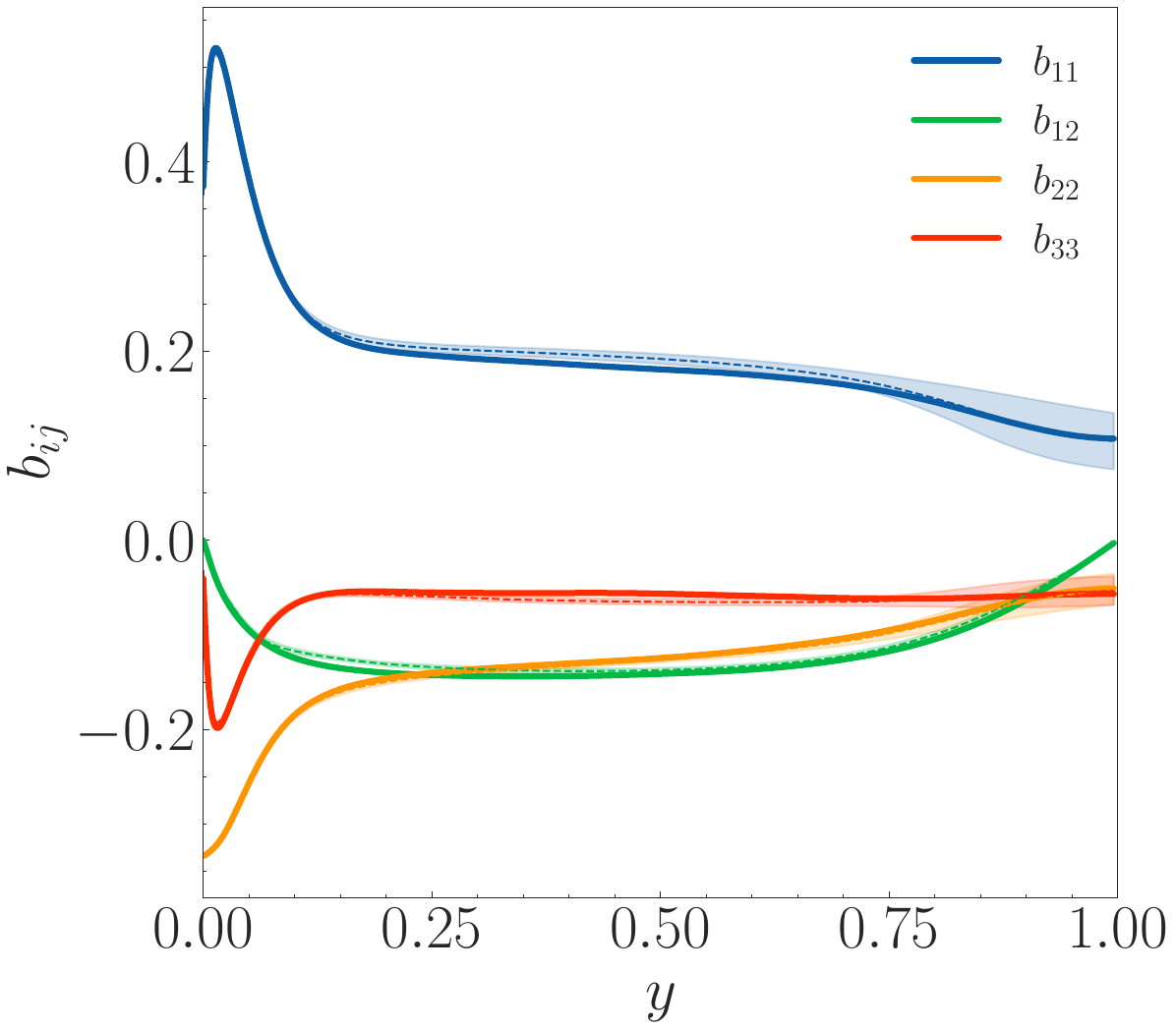}
    \caption{Case 8: Test 1}
    \label{fig:TBNN_10runs_T1}
\end{subfigure}
\hfill
\begin{subfigure}{0.3\textwidth}
    \includegraphics[width=\linewidth]{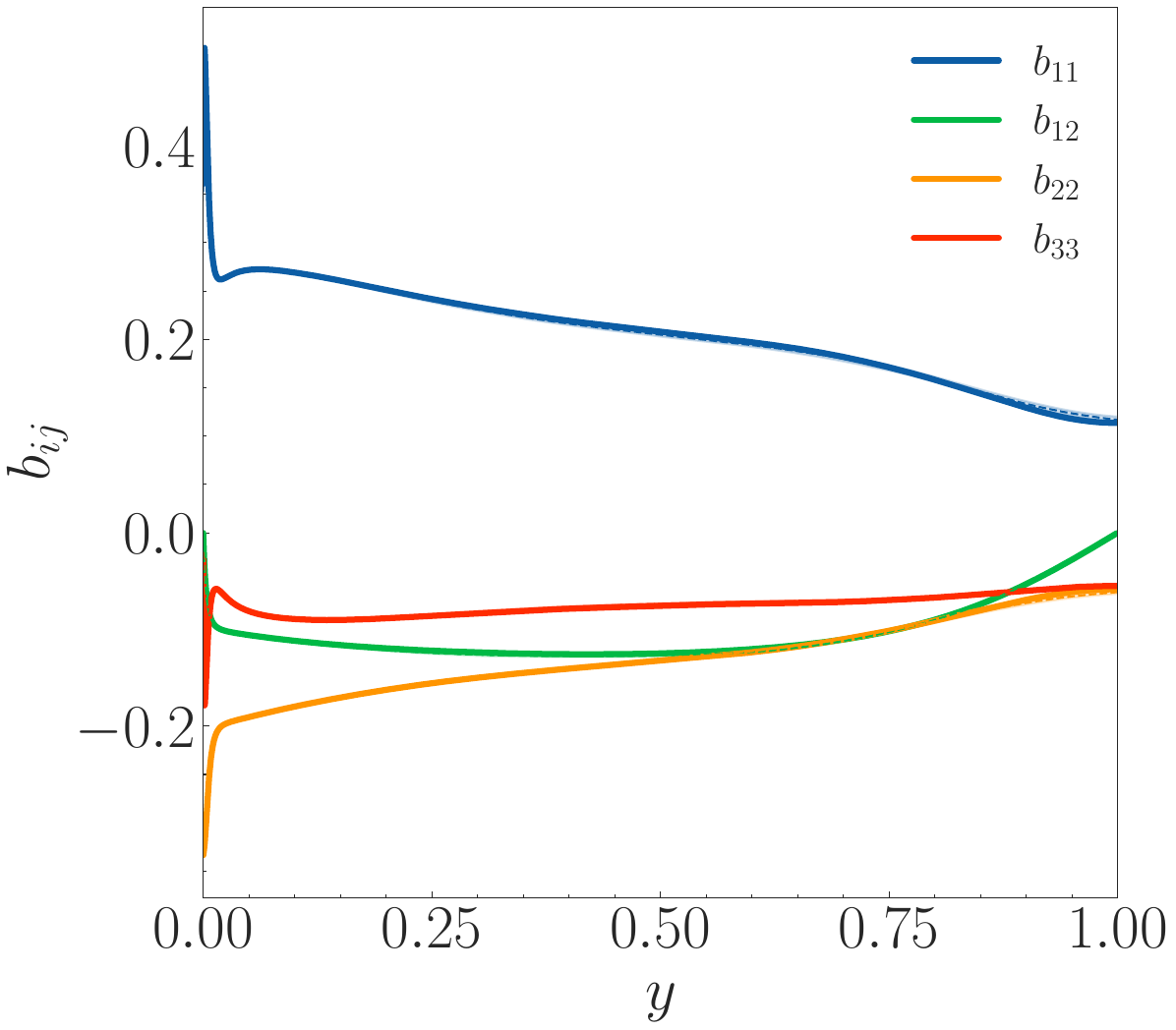}
    \caption{Case 8: Test 2}
    \label{fig:TBNN_10runs_T2}
\end{subfigure}
\hfill
\begin{subfigure}{0.3\textwidth}
    \includegraphics[width=\linewidth]{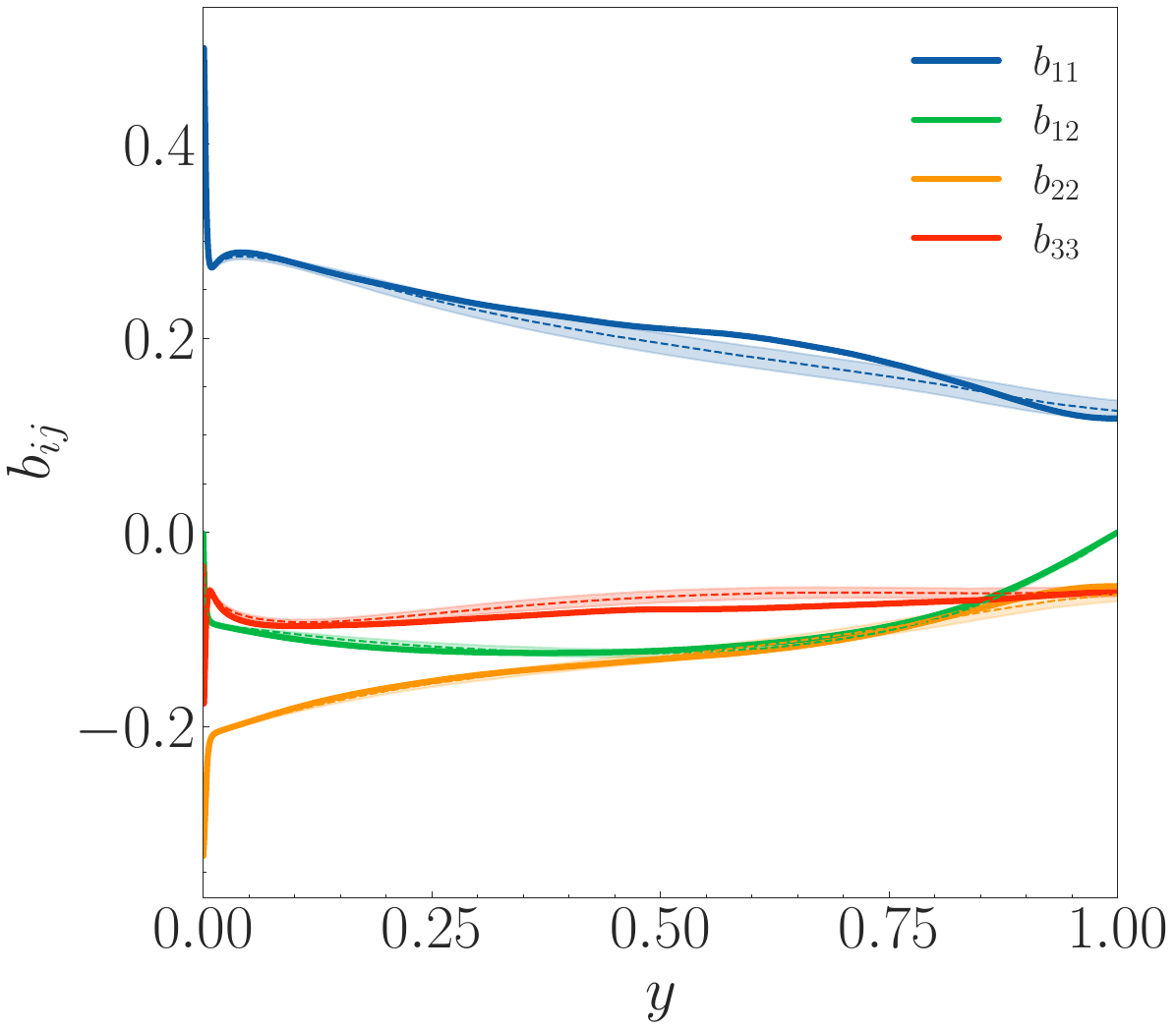}
    \caption{Case 8: Test 3}
    \label{fig:TBNN_10runs_T3}
\end{subfigure}
\hfill        
\caption{Averaged predicted $b_{ij}$ of PCF after ten repeated learnings: the DNS data and the averaged predicted values are shown in solid and dotted lines, respectively. The transparent colored region corresponds to the interval of $\pm1$ standard deviation.}
\label{fig:10runs}
\end{figure}

The ultimate comparison lies between the MLP model trained in Case 4 and the aTBNN-2 model trained in Case 8. As shown in Table \ref{tab:R2results}, both models provide comparable results, except for the prediction of $b_{33}$ component at $\mathrm{Re}_\tau=\np{10000}$. In order to test their robustness with respect to random initialization, the learning process was repeated ten times with different weight initialization for both models. Results are shown in Figure \ref{fig:10runs}, in comparison with the DNS data, where the transparent zones represent the interval of $\pm1$ standard deviation to the averaged predictions of the ten learnings.

According to Figure \ref{fig:MLP_10runs_T2} and Figure \ref{fig:TBNN_10runs_T2}, we can clearly observe that the predicted $b_{ij}$ profiles are nearly identical to the DNS data for the interpolated test, with low standard deviation for both models. Such good predictive performance for the interpolated test at $\mathrm{Re}_\tau=\np{5200}$ has demonstrated the pertinence of our training data, representative enough to yield a good prediction for channel flows with an interpolating Reynolds number, as previously discussed in \ref{sec:prepro}. Test results on two extreme extrapolation tests at $\mathrm{Re}_\tau=550$ and $\mathrm{Re}_\tau=\np{10000}$ are shown in Figure \ref{fig:MLP_10runs_T1} and Figure \ref{fig:MLP_10runs_T3} for Case 4 and in Figure \ref{fig:TBNN_10runs_T1} and Figure \ref{fig:TBNN_10runs_T3} for Case 8. While both models do not perform as strongly as they do in the interpolating case, the prediction results still maintain a high degree of closeness and robustness when compared to the DNS data, with only minor differences. A larger standard deviation of the aTBNN-2 model (shown in Figure \ref{fig:TBNN_10runs_T1}) is observed near the channel center.

A more quantitative comparison between these two models is shown in Table \ref{tab:R2Case48}. The averaged performance of the MLP model is better, especially for the prediction of the $b_{33}$ component at $\mathrm{Re}_\tau=\np{10000}$. Such prediction performance for extrapolating flow configurations is highly satisfactory, given the challenging nature of providing extrapolated predictions for flows with varying turbulence levels and characteristics, which may differ from the training flows. Furthermore, these extrapolation results are promising in terms of boosting RANS accuracy for highly turbulent flows, thereby reducing the need for costly DNS calculations. 

The integration of the ML model into our in-house developed CFD code (TrioCFD \cite{angeli2015, angeli2017}) is currently underway. Although this aspect is beyond the scope of the present paper and will be discussed in future work, it is worth noting that at this stage, we have discovered the superiority of the aTBNN framework over the MLP model. Indeed, Wu \textit{et al.} \cite{wu2019b} showed that a small discrepancy in the Reynolds stress can result in considerable errors on the velocity profiles by injecting explicitly the Reynolds stress as a source term in the RANS equations, due to a conditioning issue. An implicit treatment is therefore necessary by separating the linear and non-linear part of the Reynolds stress tensor. This can be simply achieved by using a TBNN model with the coefficient function $g_1$ in predictions, yet not applicable for the MLP model which provides directly $b_{ij}$ as outputs. Hence, considering the subsequent integration work, only the aTBNN framework will be deployed for the SDF study in the following.

\begin{table}[h!]
\centering
\caption{\label{tab:R2Case48}Averaged $R^2$ \red{coefficient} of $b_{ij}$ predictions for Case 4 and Case 8 after ten repeated learnings.}
\begin{tabular}{lcccccc}
\toprule
\multicolumn{2}{l}{Case} & $b_{11}$ & $b_{12}$ & $b_{22}$ & $b_{33}$ & Global \\\hline
\multirow{5}{*}{4} & Train & 0.9998 & 0.9995 & 0.9999 & 0.9990 & 0.9996 \\ 
& Test  & 0.9857 & 0.9802 & 0.9900 & 0.8612 & 0.9542 \\
& Test 1 & 0.9930 & 0.9931 & 0.9970 & 0.9776 & 0.9902 \\
& Test 2 & 0.9980 & 0.9961 & 0.9990 & 0.9896 & 0.9957 \\
& Test 3 & 0.9771 & 0.9685 & 0.9835 & 0.7649 & 0.9235\\
\hline
\multirow{5}{*}{8} & Train & 0.9998 & 0.9992 & 0.9999 & 0.9989 &  0.9994 \\
& Test   & 0.9690 & 0.9817 & 0.9900 & 0.6770 & 0.9044 \\
& Test 1 & 0.9894 & 0.9832 & 0.9951 & 0.9687 & 0.9841 \\
& Test 2 & 0.9980 & 0.9963 & 0.9990 & 0.9910 & 0.9961 \\
& Test 3 & 0.9483 & 0.9728 & 0.9839 & 0.4403 & 0.8363\\
\bottomrule
\end{tabular}
\end{table}

\subsection{Square duct flow}
The second training case is the SDF as described in Section~\ref{sec:flow}. Two studies are carried out with different splittings previously given in Figure~\ref{fig:data_random_mix} and Figure~\ref{fig:data_interpolation}, whose results are respectively discussed in Section~\ref{sec:random} and \ref{sec:interpolation}. A transfer learning framework is then presented in Section~\ref{sec:tl}.

\subsubsection{Random-mix study}\label{sec:random}
A baseline training is first conducted as an extension of former studies on PCF. This baseline model includes five invariants, ten tensors (without $\textbf{T}^{*(0)}$), $y^+$, $z^+$ and $\mathrm{Re}_\tau$ in the input features and deactivates the adaptive loss weighting technique. 

\begin{figure}[h!]
    \begin{subfigure}{0.16\textwidth}
        \includegraphics[width=\linewidth]{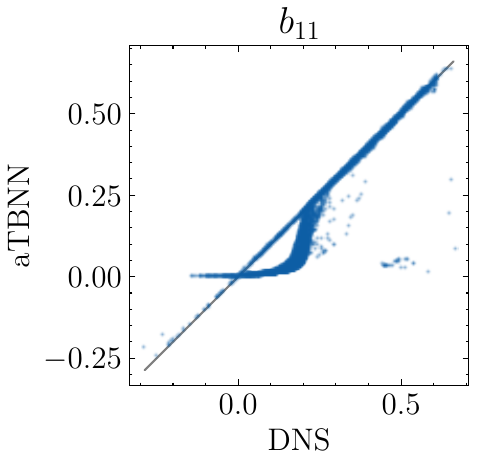}
        \label{fig:noT0_b11}
    \end{subfigure}% 
\hfill
    \begin{subfigure}{0.16\textwidth}
        \includegraphics[width=\linewidth]{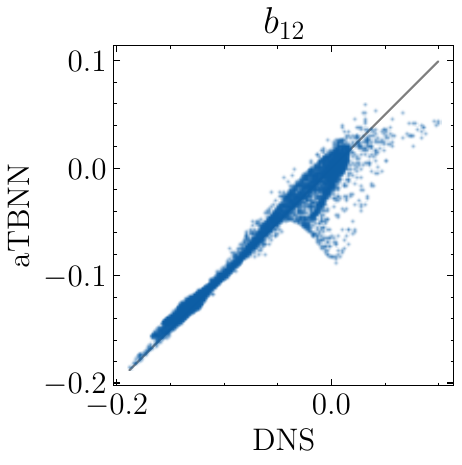}
        \label{fig:noT0_b12}
    \end{subfigure}% 
\hfill
    \begin{subfigure}{0.16\textwidth}
        \includegraphics[width=\linewidth]{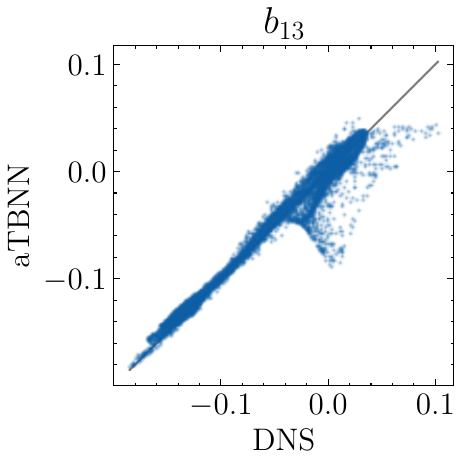}
        \label{fig:noT0_b13}
    \end{subfigure}% 
\hfill 
    \begin{subfigure}{0.16\textwidth}
        \includegraphics[width=\linewidth]{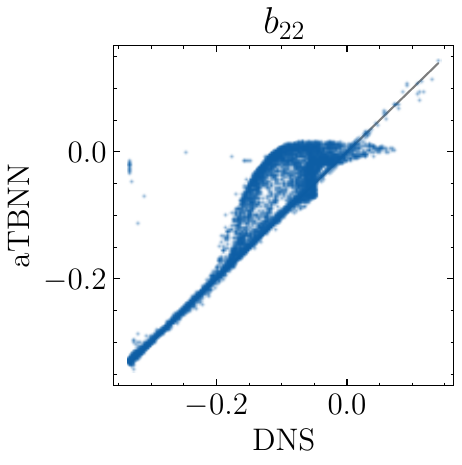}
        \label{fig:noT0_b22}
    \end{subfigure}% 
\hfill
    \begin{subfigure}{0.16\textwidth}
        \includegraphics[width=\linewidth]{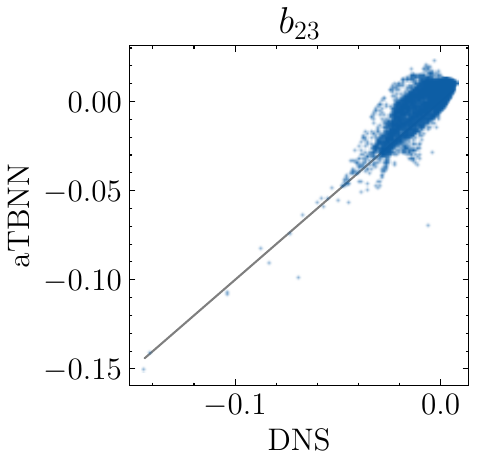}
        \label{fig:noT0_b23}
    \end{subfigure}% 
\hfill
    \begin{subfigure}{0.16\textwidth}
        \includegraphics[width=\linewidth]{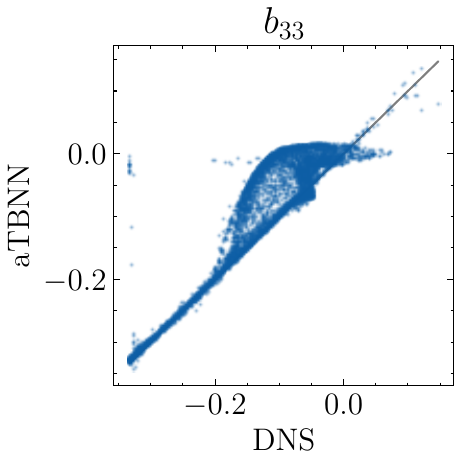}
        \label{fig:noT0_b33}
        \end{subfigure}%
\caption{Baseline aTBNN-3 model performance on SDF resulting from a direct extension of the aTBNN framework used in PCF, without $\textbf{T}^{*(0)}$. The dispersion plots compare the anisotropy tensor $b_{ij}$ predicted by aTBNN with the DNS data on the test set.}
\label{fig:disp_noT0}
\end{figure}

Figure~\ref{fig:disp_noT0} illustrates the performance of this baseline study, which is clearly unsatisfactory and represents the complexity gap between PCF and SDF. Several noteworthy observations can be made, providing vital insights for possible improvements.

Firstly, there are some learning difficulties for a given component: taking the diagonal components as examples, most of the predictions overlap with the DNS data whereas null predictions are generated within specific intervals. Based on the earlier analysis in Section~\ref{sec:popeSDF}, these inaccuracies are likely caused by an eventual incomplete tensor basis. A forward attempt appears to include the constant $\textbf{T}^{*(0)}$ tensor into the basis since its impact has previously been demonstrated in the study on PCF. Table~\ref{tab:T0_square_duct} compares the performance of the model without and with different $\textbf{T}^{*(0)}$. The inclusion of $\textbf{T}^{*(01)}$ into the tensor basis results in a quantitative improvement, aligning with previous DNS data analyses in Section~\ref{sec:popeSDF}. The effect of the other two $\textbf{T}^{*(0)}$ constants is yet negligible. Hence, $\textbf{T}^{*(01)}$ will be included in the following study.
\begin{table}[h!]
\centering
\caption{\label{tab:T0_square_duct}aTBNN-3 model performance on SDF without $\textbf{T}^{*(0)}$ or with the three permutations of $\textbf{T}^{*(0)}$. $R^2$ comparison for training and test sets. The best results are highlighted in bold.}
\begin{tabular}{lcccccccc}
\toprule
&  & $b_{11}$ & $b_{12}$ & $b_{13}$ & $b_{22}$ & $b_{23}$ & $b_{33}$ & Global \\\hline
Train & Baseline & 0.6275 & 0.9633 & 0.9660 & 0.6925 & 0.2702 & 0.6873 & 0.7011\\
 & $\textbf{T}^{*(01)}$ & \textbf{0.99998} & \textbf{0.9908} &\textbf{0.9922} &0.9847 &\textbf{0.5871} &0.9849 & \textbf{0.9233}\\
 & $\textbf{T}^{*(02)}$ & 0.9131  & 0.9767 &0.9694 &\textbf{0.9998} &0.2438 &0.7711 & 0.8123 \\
 & $\textbf{T}^{*(03)}$ & 0.8945  & 0.9659 &0.9783 &0.7148 &0.1179&\textbf{0.9998}  & 0.7785\\
Test& Baseline & 0.6442 	&0.9650 &0.9651 &0.7156 &0.3295 &0.6886  &0.7180\\ 
 & $\textbf{T}^{*(01)}$ & \textbf{0.99996} & \textbf{0.9912} &\textbf{0.9915} &0.9848 &\textbf{0.6224} &0.9842  &\textbf{0.9290}\\
 & $\textbf{T}^{*(02)}$ & 0.9179 &0.9775 	&0.9695 	&\textbf{0.9997} 	&0.3202 	&0.7787 &0.8273 \\
 & $\textbf{T}^{*(03)}$ & 0.9003 &0.9676 	&0.9782 	&0.7379 	&0.2097 	&\textbf{0.9997} &0.7989 \\
\bottomrule
\end{tabular}
\end{table}

\begin{figure}[h!]
 \centering
    \begin{subfigure}{.8\textwidth}
        \centering
        \includegraphics[width=\linewidth]{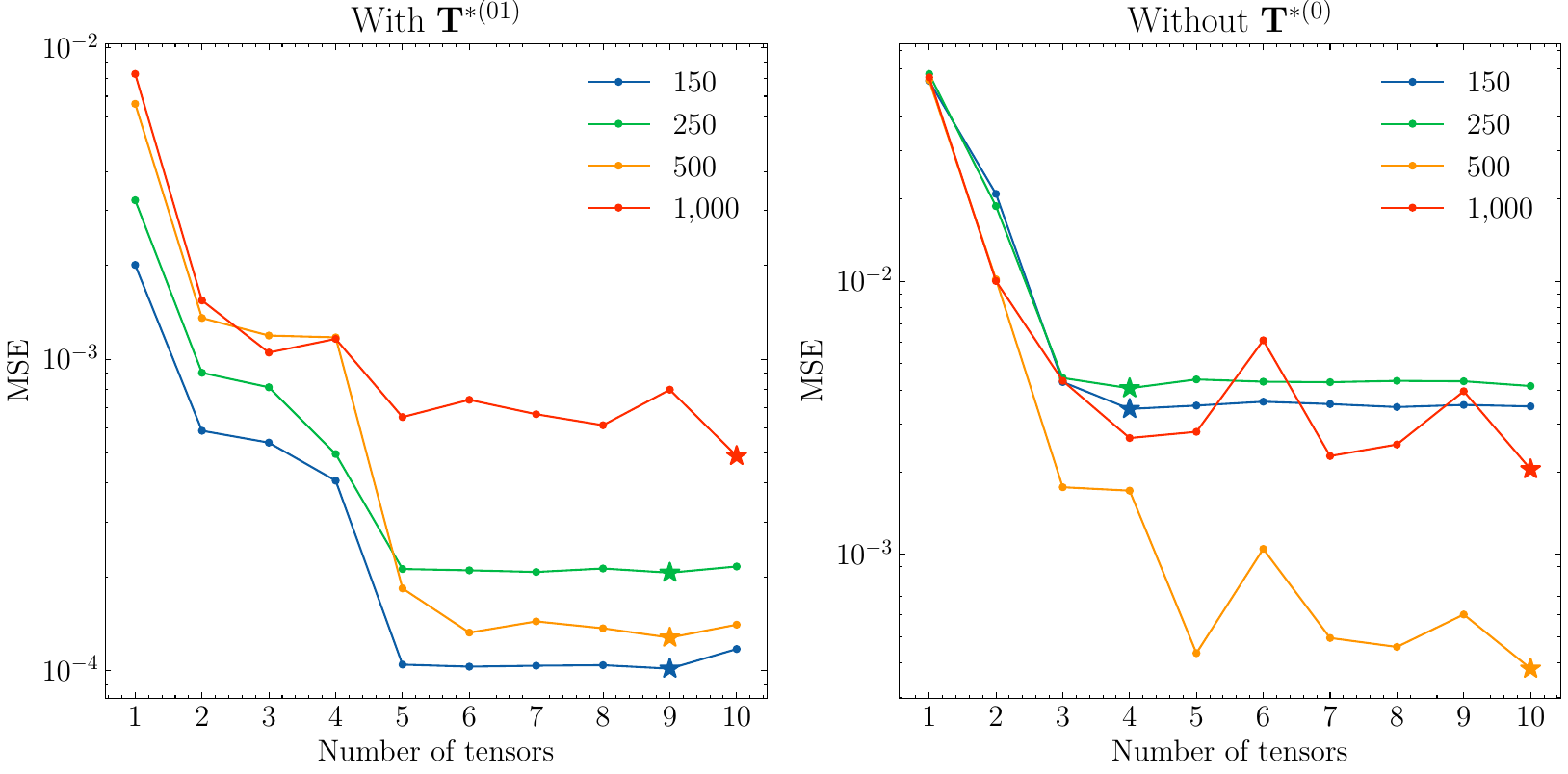}
        \caption{MSE}
        \label{fig:MSE_OptT}
    \end{subfigure}% 
\hfill
    \begin{subfigure}{.8\textwidth}
        \centering
        \includegraphics[width=\linewidth]{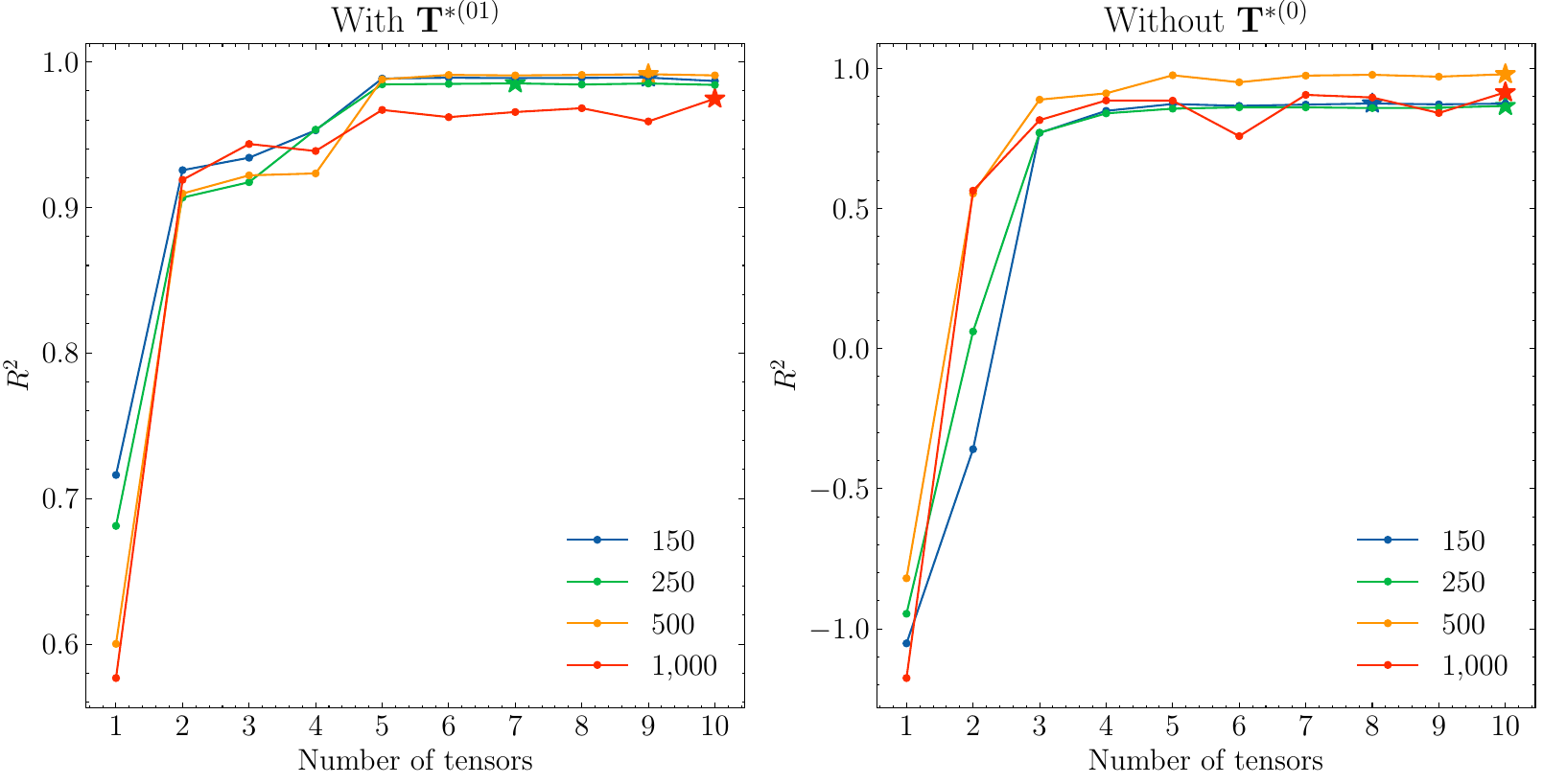}
        \caption{$R^2$}
        \label{fig:Q2_OptT}
    \end{subfigure}  
\caption{aTBNN-3 model performance on SDF using different numbers of basis tensors: with $\textbf{T}^{*(01)}$ and without any $\textbf{T}^{*(0)}$, comparison of (a) MSE and (b)$R^2$ by testing at dataset with different friction Reynolds numbers. Best results (lowest MSE or highest $R^2$) are marked by stars (\FiveStar).}
\label{fig:optT}
\end{figure}

On the other side, an optimization study is carried out on the number of basis tensors. The NN is trained by using different numbers of basis tensors with $\textbf{T}^{*(01)}$ and without any $\textbf{T}^{*(0)}$, then tested on each dataset. The results are shown in Figure~\ref{fig:optT}. Remarkably, the model's performance exhibits a rapid improvement when the number of basis tensors is increased from one to five. However, employing more than five basis tensors does not necessarily result in enhanced accuracy. This finding is supported by numerous works in the literature, deploying a reduced number of basis tensors, the most popular being cubic \cite{craft1996} and quadratic \cite{speziale1987a} models. Different authors \cite{jongen1998, lund1993a, modesti2020} share our view and suggest a basis of five tensors except for some degenerate cases. It should be highlighted that the study in \cite{modesti2020} utilized the same DNS datasets of SDF as the present study and adopted a completely analytical framework like most of the previous studies. To the authors' knowledge, it is the first time that a numerical proof of a five-term representation of the anisotropy tensor is given from a TBNN framework. As a consequence of this optimization study, only five basis tensors as well as $\textbf{T}^{*(01)}$ will be used in the following aTBNN-3 model, instead of the 11 tensors as shown in Eq.~\eqref{eq:aTBNN3}. \red{Nevertheless, we would like to emphasize that the current choice of five tensors is based on numerical experiments on square duct flows and might not be applicable to other more complex cases.}

\begin{figure}[ht!]
    \begin{subfigure}{0.32\textwidth}
        \includegraphics[width=0.98\linewidth]{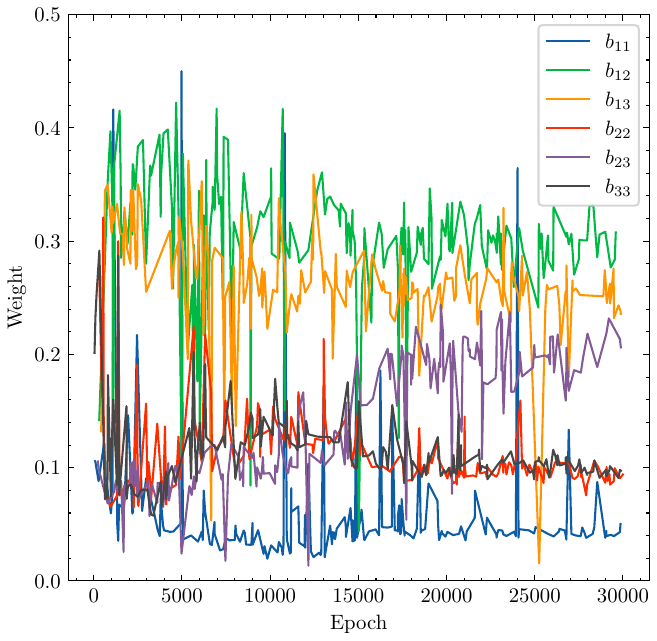}
        \caption{Adaptive weights}
        \label{fig:SoftAdapt}
    \end{subfigure}% 
\hfill
    \begin{subfigure}{0.32\textwidth}
        \includegraphics[width=\linewidth]{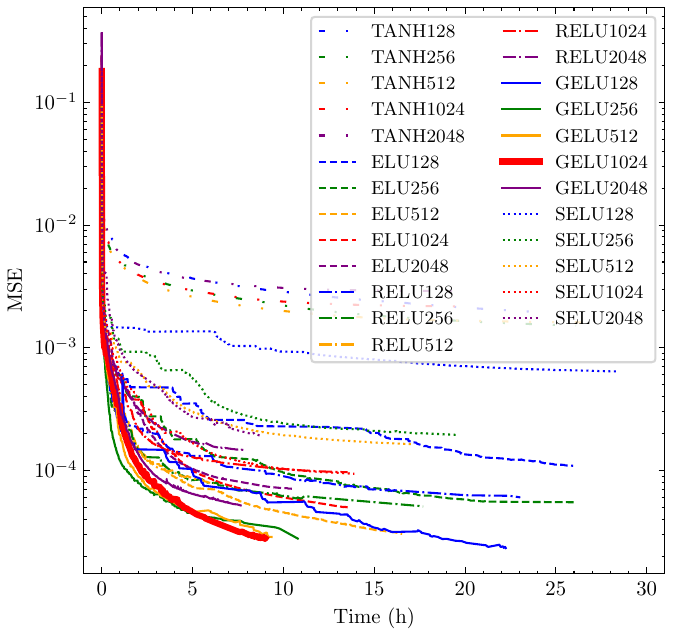}
        \caption{MSE}
        \label{fig:MSE_HyperP}
    \end{subfigure}% 
\hfill
    \begin{subfigure}{0.32\textwidth}
        \includegraphics[width=\linewidth]{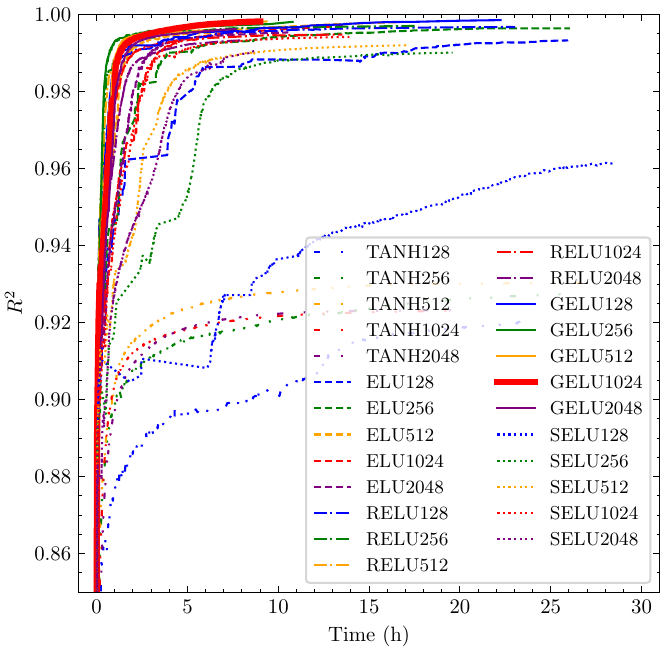}
        \caption{$R^2$}
        \label{fig:Q2_HyperP}
    \end{subfigure}  
\caption{Learning dynamics of aTBNN-3 model trained by SDF dataset: (a) adaptive weights of each $b_{ij}$ component; (b) global MSE using different activation functions and batch sizes; and (c) global $R^2$ using different activation functions and batch sizes.}
\label{fig:hyperP}
\end{figure}

A second issue revealed by Figure~\ref{fig:disp_noT0} is that predictions on different components of the anisotropy tensor are unbalanced due to the multi-part loss functions. To overcome this problem, an adaptive loss weighting algorithm called SoftAdapt \cite{heydari2019}, described in Section~\ref{sec:strategie} is implemented in the model training framework. The weight of each component in the loss function is initially set to be equal while they evolve during the learning process as illustrated in Figure~\ref{fig:SoftAdapt}. Interestingly, the easiest learned component $b_{11}$, the one with the highest $R^2$ in the baseline model, is generally assigned the lowest weight by the algorithm. On the contrary, the most sticky component $b_{23}$, the one with the lowest $R^2$ in the baseline model, gradually accumulates more weight during the learning process. This underscores the effectiveness of the implemented adaptive loss weighting algorithm, which will be utilized in the subsequent study.

The influence of the activation function and batch size on the performance is also investigated, as reported in Figure~\ref{fig:hyperP} and Table~\ref{tab:hyperP}. It can be seen that the tanh function formerly used in the PCF study performs relatively badly compared to other tested activation functions upon the present data on SDF, which could explain the saturation regions observed from Figure~\ref{fig:disp_noT0}. Note that this might be caused by the problem of vanishing gradients commonly observed from NNs using tanh as an activation function, and it has been previously demonstrated that other tested activation functions can help prevent this problem \cite{nguyen2021}. Considering the trade-off between convergence time and model performance, we choose the Gaussian Error Linear Unit (GELU) \cite{hendrycks2023} as the activation function along with a batch size of 1024 in our aTBNN-3 model, whose results are shown in bold red in Figure~\ref{fig:hyperP} and in Table~\ref{tab:hyperP}. 

\begin{table}[ht!]
\centering
\caption{\label{tab:hyperP} aTBNN-3 model performance on SDF: $R^2$ comparison using different activation functions and batch sizes, best results in bold black, chosen framework in bold red.}
\begin{tabular}{ccccc}
\toprule
Activation & Batch  &  Global $R^2$   & Global $R^2$ & Training time per epoch \\
function & size & Train  & Test & (second) \\\hline
TANH & 128 & 0.9146 & 0.9202 & 6.5917\\
& 256 & 0.9217 & 0.9275 & 3.7913\\
& 512 & 0.9247 & 0.9303 &  2.3194\\
& 1024 & 0.9174 & 0.9231 &  1.6287\\
& 2048 & 0.9176 & 0.9232 & 1.2525\\
RELU & 128 & 0.9969 & 0.9968 & 6.5240\\
& 256 & 0.9972 & 0.9971 & 3.7138\\
& 512 &0.9961 & 0.9960 &  2.3019\\
& 1024 & 0.9950 & 0.9948 &  1.6490\\
& 2048 & 0.9935 & 0.9933 & \textbf{1.2431}\\
ELU & 128 & 0.9933 & 0.9933 & 6.5115\\
& 256 & 0.9964  & 0.9965 & 3.6765\\
& 512 & 0.9980 & 0.9980 & 2.3794\\
& 1024 & 0.9970 & 0.9971 & 1.6153\\
& 2048 & 0.9954 & 0.9954 & 1.4957\\
\textcolor{red}{\textbf{GELU}} & \textbf{128} & \textbf{0.9986} & \textbf{0.9986} & 7.3199\\
& 256 & 0.9982 & 0.9981 & 4.1012\\
& 512 & 0.9982 & 0.9983 & 2.6669\\
& \textcolor{red}{\textbf{1024}} & \textcolor{red}{\textbf{0.9982}} & \textcolor{red}{\textbf{0.9982}} & \textcolor{red}{\textbf{1.9102}}\\
& 2048 & 0.9962 & 0.9962 & 1.4957\\
SELU & 128 & 0.9590  & 0.9614 & 6.5147\\
& 256 & 0.9901 & 0.9902 & 3.6372\\
& 512 & 0.9919 & 0.9921 & 2.3795\\
& 1024 & 0.9941 & 0.9942 & 1.6573\\
& 2048 & 0.9903 & 0.9906 & 1.3132\\
\bottomrule
\end{tabular}
\end{table}

\begin{figure}[h!]
    \begin{subfigure}[c]{\textwidth}
        \centering
        \includegraphics[width=0.75\linewidth]{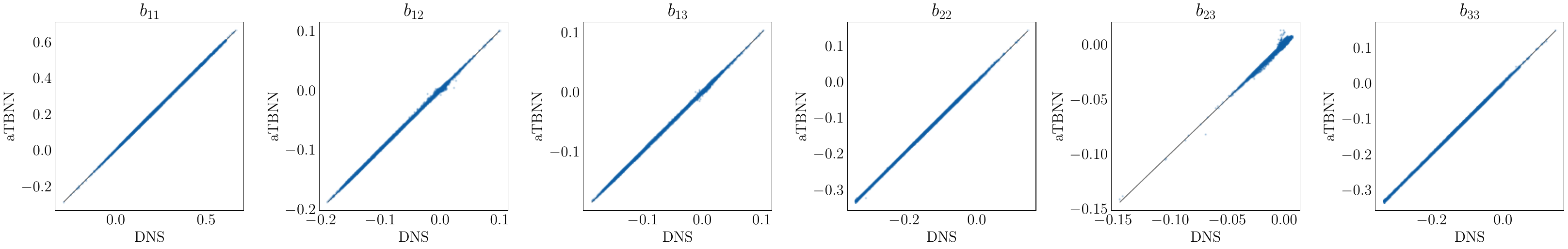}
        \label{fig:RM_disp}
        \end{subfigure}%
\hfill
    \begin{subfigure}[c]{\textwidth}
        \centering
        \includegraphics[width=0.75\linewidth]{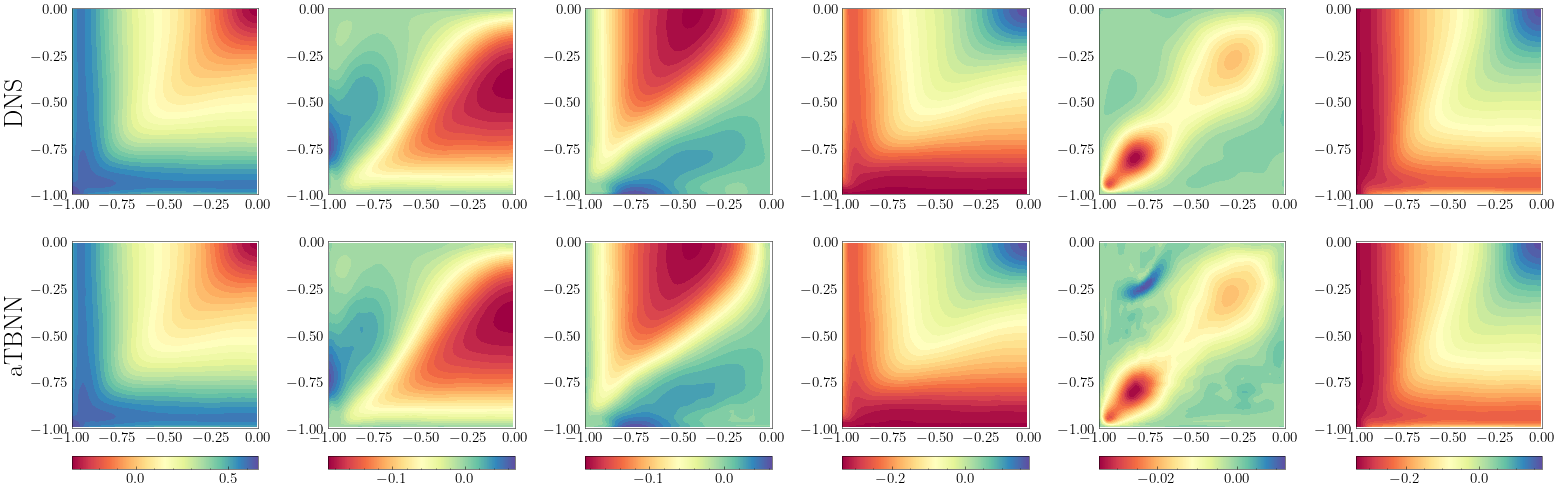}
        \caption{$\mathrm{Re}_\tau = 150$}
        \label{fig:RM_150}
        \end{subfigure}%
\hfill
    \begin{subfigure}[c]{\textwidth}
        \centering
        \includegraphics[width=0.75\linewidth]{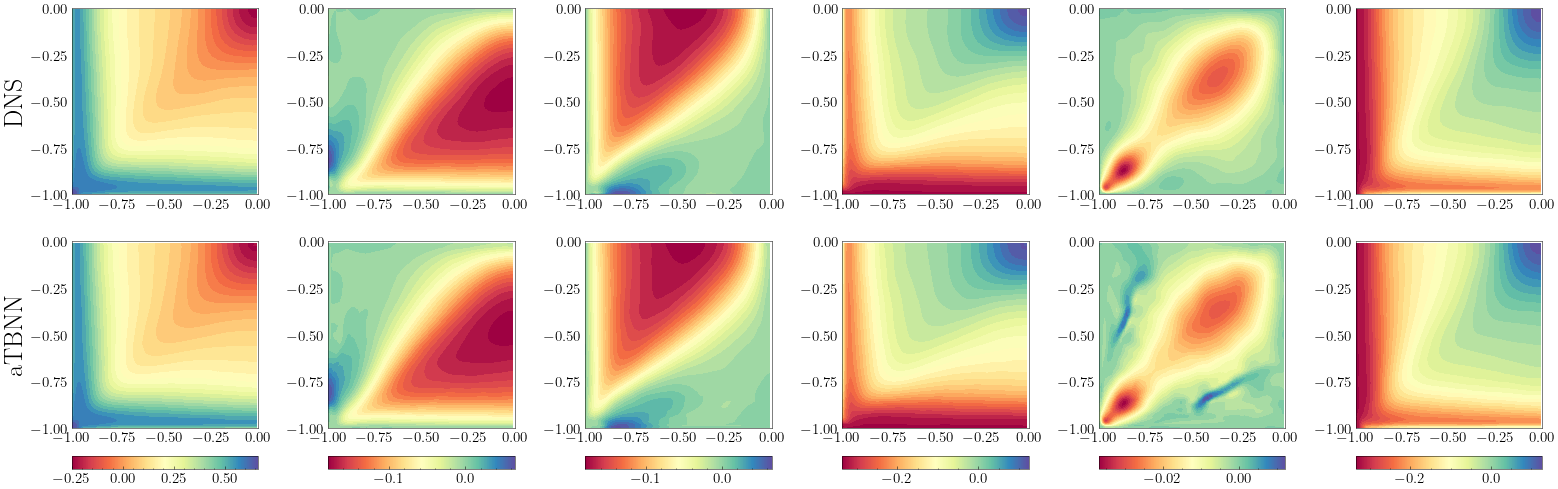}
        \caption{$\mathrm{Re}_\tau = 250$}
        \label{fig:RM_250}
        \end{subfigure}%
\hfill
    \begin{subfigure}[c]{\textwidth}
        \centering
        \includegraphics[width=0.75\linewidth]{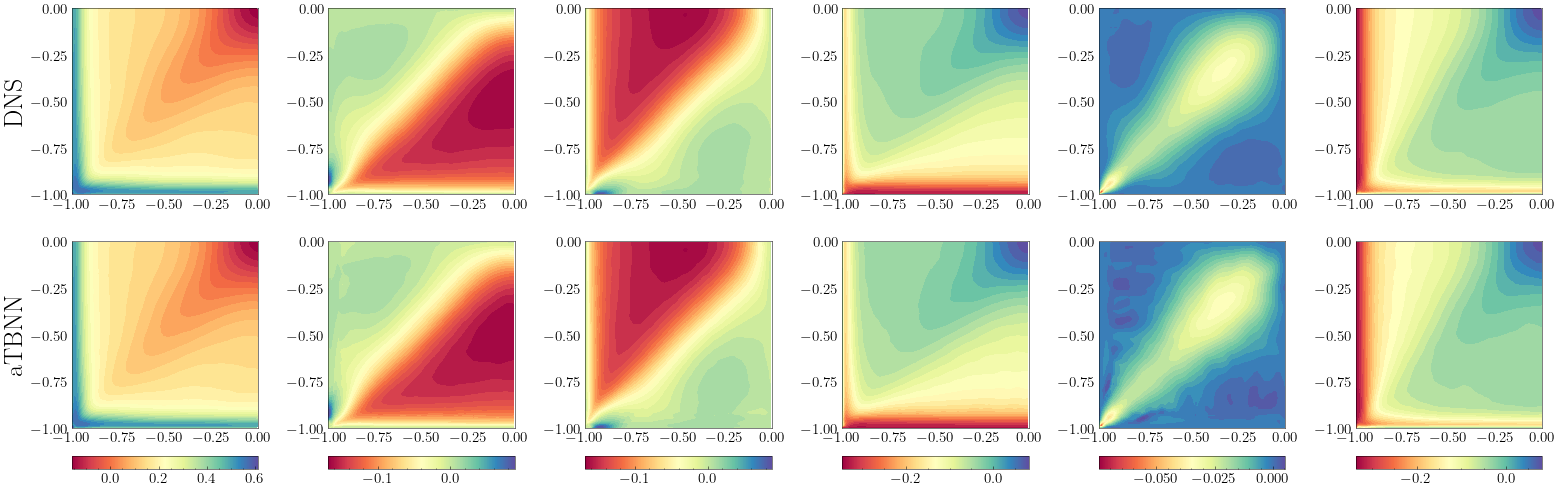}
        \caption{$\mathrm{Re}_\tau = 500$}
        \label{fig:RM_500}
        \end{subfigure}%
\hfill
    \begin{subfigure}[c]{\textwidth}
        \centering
        \includegraphics[width=0.75\linewidth]{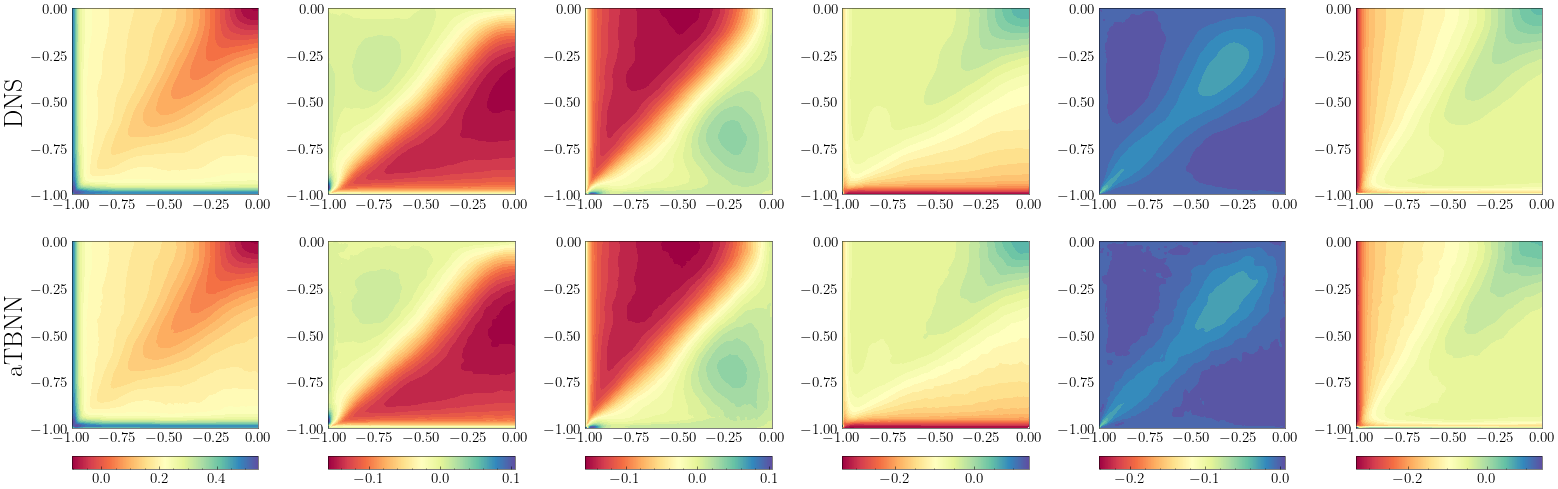}
        \caption{$\mathrm{Re}_\tau = \np{1000}$}
        \label{fig:RM_1000}
        \end{subfigure}%
\caption{Model performance on SDF after optimization. The dispersion plots and the contour plots compare the anisotropy tensor $b_{ij}$ predicted by aTBNN-3 with the DNS data on test set.}
\label{fig:RM_after}
\end{figure}

The improved performance of the present optimized aTBNN-3 model resulting from the above training strategies is reported in Figure~\ref{fig:RM_after}. Considerable improvements are achieved by comparing with the dispersion plots of the baseline model given in Figure~\ref{fig:disp_noT0}. Furthermore, as indicated in Figures~\ref{fig:RM_150}-\ref{fig:RM_1000}, the aTBNN-3 model successfully grasps the profiles of different components of the anisotropy tensor for flows at different $\mathrm{Re}_\tau$. An efficient mapping is therefore established between our input features and the targeted anisotropy tensor, proving the robustness of the following physical model for SDF:

\begin{equation}\label{eq:aTBNN4}
\textbf{b} =g^{(0)}\textbf{T}^{*(01)} + \sum_{n=1}^{5} g^{(n)}\left(\{\lambda_i^*\}_{i=1,2,...,5}, y^+, z^+, \mathrm{Re}_\tau \right)\textbf{T}^{*(n)} 
\end{equation}

The $R^2$ values of this improved framework are listed in Table~\ref{tab:RM_square_duct} in comparison with the baseline model, demonstrating quantitatively the efficiency of our improved framework. Both the global and individual $R^2$ greatly increase. Most notably, the predictive capability of the $b_{23}$ component has been significantly improved and reaches a comparable level with the other components. This is a significant achievement since it is considered to be the most difficult component to learn, due to the complexity of its profiles (see Figure~\ref{fig:RM_after}), which has already been pointed out in \cite{saezdeocarizborde2022}. Nevertheless, we still remark some sorts of asymmetries in the predictions of $b_{23}$ as illustrated in Figure~\ref{fig:RM_after}, which could potentially be improved by introducing additional constraints to the current model. This aspect is currently under investigation.

\begin{table}[ht!]
\centering
\caption{\label{tab:RM_square_duct}aTBNN-3 model performance on SDF after optimization: $R^2$ comparison for training and testing sets. The best results are highlighted in bold.}
\begin{tabular}{lcccccccc}
\toprule
&  & $b_{11}$ & $b_{12}$ & $b_{13}$ & $b_{22}$ & $b_{23}$ & $b_{33}$ & Global \\\hline
Train & Baseline & 0.6275 & 0.9633 & 0.9660 & 0.6925 & 0.2702 & 0.6873 & 0.7011\\
   & \red{\textbf{Improved}}  &   \red{\textbf{0.99998}} &  \red{\textbf{0.9997}} & \red{\textbf{0.9998}} & \red{\textbf{0.9999}}& \red{\textbf{0.9897}}& \red{\textbf{0.9999}} &  \red{\textbf{0.9982}} \\
  Test& Baseline & 0.6442 	&0.9650 &0.9651 &0.7156 &0.3295 &0.6886  &0.7180\\ 
  & \red{\textbf{Improved}}  &   \red{\textbf{0.99998}} &  \red{\textbf{0.9997}}& \red{\textbf{0.9998}}& \red{\textbf{0.9999}}&   \red{\textbf{0.9899}}&  \red{\textbf{0.9999}} &  \red{\textbf{0.9982}}\\
\bottomrule
\end{tabular}
\end{table}

\subsubsection{Interpolation study} \label{sec:interpolation}
\begin{figure}[h!]
    \begin{subfigure}{\textwidth}
        \centering\includegraphics[width=.75\linewidth]{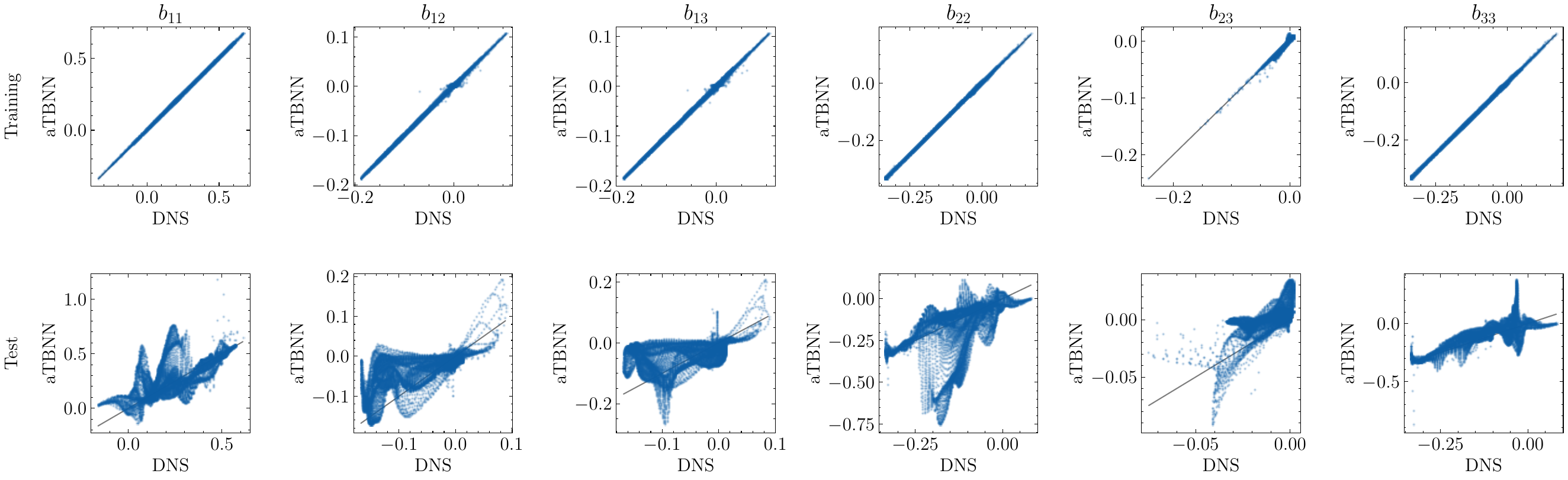}
        \caption{Interpolation study}
        \label{fig:b_inter}
        \end{subfigure}%
\hfill
    \begin{subfigure}{\textwidth}
        \centering\includegraphics[width=.75\linewidth]{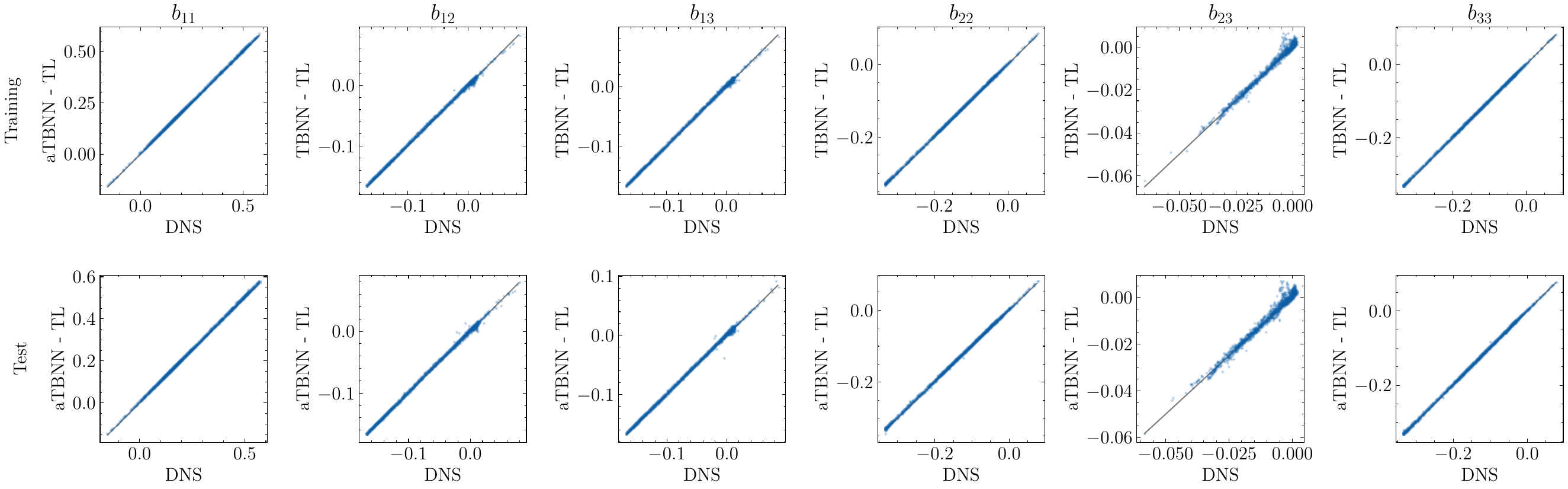}
        \caption{Transfer learning}
        \label{fig:b_TL}
        \end{subfigure}%
\caption{Dispersion plots comparing the anisotropy tensor predictions with the DNS data on training (first line) and test set (second line). (a) Interpolation performance of the aTBNN-3 model trained from SDF data at $\mathrm{Re}_\tau=[150; 250; \np{1000}]$ and tested at $\mathrm{Re}_\tau=500$. (b) Performance of the model making use of the Transfer Learning (see model details in Section~\ref{sec:tl}) fine-tuned with \red{10}\% of the training dataset at $\mathrm{Re}_\tau=500$ after $\np{100000}$ iterations. }
\label{fig:b_inter_TL}
\end{figure}

The second study on SDF aims at testing the interpolation performance of the former optimized aTBNN-3 model based on relationship \eqref{eq:aTBNN4}. The data-splitting of this study is illustrated in Figure~\ref{fig:data_interpolation}: dataset at $\mathrm{Re}_\tau=500$ is completely used in test set, and other data at $\mathrm{Re}_\tau=[150; 250; \np{1000}]$ are used for training. Figure~\ref{fig:b_inter} presents the results of this study, where the dispersion plots comparing the predictions and the training data are shown in the first line and those for the test set in the second line. It can be seen from the figure that the model performs well on training data, whereas it completely fails on the test data. Unlike prior results on PCF, the current model on SDF can not be interpolated, not even to mention extrapolated, at an unobserved $\mathrm{Re}_\tau$. This is not surprising according to the data observation in Figure~\ref{fig:inputs_sdf_after}. \red{From a physics viewpoint, this might be caused by the low-Reynolds-number effect: Zhang \textit{et al.} \cite{zhang2015a} have previously observed from DNS data of SDF that the flow behavior at $\mathrm{Re}_\tau = 300$ is clearly different with higher ones at $\mathrm{Re}_\tau=[600; 900; \np{1200}]$. We posit that our dataset at $\mathrm{Re}_\tau=[150; 250]$ also inherits the low-Reynolds-number effect, resulting in distinct behavior compared to the dataset at $\mathrm{Re}_\tau=\np{1000}$. As a result, it becomes physically impractical for the current neural network model to provide
accurate prediction at $\mathrm{Re}_\tau=500$ due to a learning that has to account for features of flows (at $\mathrm{Re}_\tau=[150; 250]$) bearing the low-Reynolds-number effect and ones  (at $\mathrm{Re}_\tau=\np{1000}$)  without this effect.
} This hurdle could be overcome by including more data at different $\mathrm{Re}_\tau$ in the training set. Unfortunately, DNS data on SDF are still too incomplete due to their high computational effort, most of which omit the physical quantities (especially the dissipation rate) we need. Hence, we posit that there is a detrimental lack of data in the machine-learning-assisted turbulence modeling domain for further development. We propose to alleviate this problem via Transfer Learning (TL), as outlined in the following section.

\subsubsection{Transfer Learning (TL)}\label{sec:tl}
Transfer Learning (TL) \cite{pan2010} is an ML method where a model developed for certain tasks is reused as initial knowledge (or starting point) for the training of a new model in another task. Therefore TL aims to improve understanding of the current task by relating it to other tasks performed at different regimes but through a related source domain. This way, what has been learned in one setting is built upon to improve generalization on a related setting, often at a fraction of the cost compared to isolated training, where each model is independently trained for a specific purpose without any dependency on past knowledge.

As an advanced technique, it has for instance been recently applied in the machine-learning-assisted fluid dynamics research community, to leverage the generalization issue of well-trained NN to a different flow. Guastoni \textit{et al.} \cite{guastoni2020, guastoni2021} applied TL in a series of their work on wall-bounded turbulence predictions from wall quantities via CNN. They showed that the training time of a network that provides predictions at one specific wall-normal location can be significantly reduced, by a factor of 4, if its parameters are initialized with those of a previously trained network at another location. They also demonstrated the efficiency of TL between two CNNs giving predictions at different Reynolds numbers: improved performance was achieved via TL by comparing with classic training with the same amount of training data; they even obtained comparable accuracy using 50\% and \red{10}\% of the training dataset. Similarly, Guan \textit{et al.} \cite{guan2022} acquired accurate and stable \textit{a-posteriori} CNN-LES predictions via TL by using only 1\% of the original data at a new flow with 16$\times$ higher Reynolds number, which is very encouraging.

Despite these prior achievements with CNNs, known for their hierarchical feature extraction particularly appropriate for TL, it is of great interest to investigate the feasibility of TL on the aTBNN models developed in the present study. Moreover, TL might be in our case an alternative solution to efficiently employ the present pre-trained model to another NN at an unobserved Reynolds number. To this end, both the feasibility and efficiency of transferring knowledge among aTBNN models trained by datasets at different Reynolds numbers are investigated in this work.

\begin{figure}[h!]
\centering
\includegraphics[width=0.4\textwidth,center]{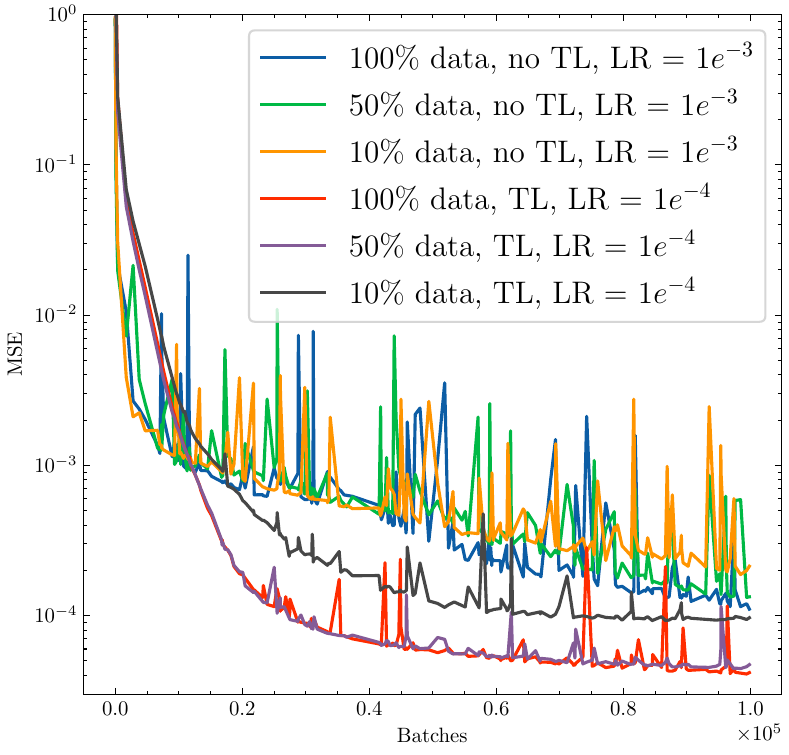}
\caption{Test loss comparison of aTBNN-3 trained with different sizes (100\%, 50\% and \red{10}\%) of SDF dataset at $\mathrm{Re}\tau=500$, using random initialization or initialized with a previously trained model.}
\label{fig:tl}
\end{figure}

The base model used for TL is the one established in the interpolation study mentioned in Sec.~\ref{sec:interpolation}, which is trained on the large dataset collected at $\mathrm{Re}_\tau=[150; 250; \np{1000}]$. The weights of this network are loaded as initial condition to train another network with the same architecture, but on a different and smaller dataset at $\mathrm{Re}_\tau=500$. Subsequently, the same initialization is operated to different networks trained on various downsized datasets, namely comprising 50\% and \red{10}\% of the full dataset at $\mathrm{Re}_\tau=500$. These models are then evaluated on the same test set at $\mathrm{Re}_\tau=500$ (not used in each training dataset). The results, e.g. test losses, are then compared with classically-trained models with random initialization up to $\np{100000}$ iterations, which is shown in Figure~\ref{fig:tl}. Note that even though the learning rate decay scheduler was utilized in every network training, a lower initial LR at $1e^{-4}$ was employed for fine-tuning models with initialization, while the initial LR for classic models with random initialization was set at $1e^{-3}$. Indeed, the final LR of the base model is at the order of $1e^{-5}$, and it was also reported in previous works that a lower LR is needed for the TL models in order to prevent divergence \cite{guastoni2021}. Except for the LR, all other hyperparameters of the TL models are kept identical to those of the classic models, ensuring a fair comparison among these models. It can be seen in Figure~\ref{fig:tl} that all the initialized models using TL outperform classic models with random initialization at the end of $\np{100000}$ iterations. Of particular significance is the performance of the TL model trained on only \red{10}\% of the training dataset. While it shows a slightly higher MSE compared to the other two TL models trained on larger datasets, it still competes effectively with classical models. The dispersion plots of the model with \red{10}\% of the training data after $\np{100000}$ iterations are shown in Figure~\ref{fig:b_TL}, comparing the anisotropy tensor predictions with the referenced DNS data on the training and test sets. A clear improvement can be seen when comparing the current results obtained through the TL process with the earlier interpolation results illustrated in Figure~\ref{fig:b_inter}. Moreover, we also carried out TL experiments using  \red{10}\% of the training data by freezing either the two deepest layers or the two shallowest layers, with the aim of retraining either the shallow or deep layers. The $R^2$ results of these experiments are compared with those of other TL experiments using the same amount of training data  or classic models, as summarized in Table~\ref{tab:TL}. Through the comparison of $R^2$ values, we confirm statistically that all TL models using only \red{10}\% of the training data outperform classic models with random initialization, especially on the predictions of $b_{23}$, which, as mentioned earlier, is the most difficult component to learn. It is also worth mentioning that the TL models exhibit slightly better performance when retraining \red{all the layers than freezing some deep or shallow layers, contrary to the common wisdom guiding TL in the ML community for CNNs \cite{subel2023}}. On the other hand, as illustrated in Figure~\ref{fig:tl}, the losses of TL models generally decrease much more rapidly, and reach convergence levels after $\np{100000}$ iterations, whereas classic models with higher LR still require continued training. Overall, we can conclude that the present TL process is not only feasible to transfer knowledge from a bigger range of $\mathrm{Re}_\tau$ to a specific one, but also advantageous and efficient in terms of training data and training time, especially with additional hyperparameter optimization \red{and fine-tunning techniques, such as learning rate restarts, warmup, and more \cite{gotmare2018a}.} 

\iffalse
\begin{table}[h!]
\centering
\caption{\label{tab:TL}$R^2$ comparison of aTBNN-3 trained with different sizes (100\%, 50\% and \red{10}\%) of dataset at $\mathrm{Re}\tau=500$, using random initialization or initialized with a previously trained model. The statistics are computed after $\np{100000}$ iterations. The best results are highlighted in bold.}
\begin{tabular}{lccc}
\toprule
 & Data size  &  Global $R^2$   & Global $R^2$  \\
 &   & Train  & Test \\\hline
No TL & 100\% & 0.9935 & 0.9928 \\
 & 50\% & 0.9940 & 0.9937 \\
 & 25\% & 0.9944 & 0.9938 \\
TL - retrain all layers & 100\% & 0.9973 & 0.9971  \\
 & 50\% & 0.9975 & 0.9974  \\
 & 25\% & \textbf{0.9977} & 0.9974 \\
TL - retrain shallow layers& 100\% & 0.9971 & 0.9968\\
 & 50\% & 0.9971 & 0.9969 \\
 & 25\% &  0.9973 & 0.9969\\
TL - retrain deep layers& 100\% & \textbf{0.9977} & \textbf{0.9975} \\
 & 50\% & 0.9976 & \textbf{0.9975} \\
 & 25\% &  0.9976 & 0.9971 \\ 
\bottomrule
\end{tabular}
\end{table}
\fi 
\begin{table}[h!]
\centering
\caption{\label{tab:TL}$R^2$ comparison of aTBNN-3 trained with different sizes (100\%, 50\% and \red{10}\%) of SDF dataset at $\mathrm{Re}\tau=500$, using random initialization or initialized with a previously trained model via TL: all layers are retrained in TL-0, the two deepest layers are frozen in TL-1 and the two shallowest layers are frozen in TL-2. The statistics are computed after $\np{100000}$ iterations. The best results are highlighted in bold.}
\begin{tabular}{lcccccccc}
\toprule
 & Data size  &  $b_{11}$ & $b_{12}$ & $b_{13}$ & $b_{22}$ & $b_{23}$ & $b_{33}$ & Global \\\hline
No TL & 100\% &  0.99995 	& 0.99867 	& 0.99888 	& 0.99985 	& 0.95960 	& 0.99984 & 0.99280  \\
 & 50\% &   	0.99963 	&0.99894 	&0.99907 	&0.99940 	&0.96597 	&0.99924 & 0.99371 \\
& \red{10\%} & \red{0.99979}& 	\red{0.99833}& 	\red{0.99871}& 	\red{0.99966}& 	\red{0.95207}& 	\red{0.99964}& \red{0.99137}\\
\\
 
%TL & 100\% &   	0.99985 	&0.99948 	&0.99959 	&0.99979 	&0.98392 	&0.99977& 0.99707  \\
%& 50\% & 0.99997 	&0.99949 	&0.99961 	&0.99992 	&0.98544 	&0.99993 & 0.99739  \\
%TL& 100\% & 	0.99996 	&0.99949 	&0.99958 	&0.99991 	&0.98220 	&0.99991 & 0.99684\\
%  & 50\% &  	0.99996 	&0.99948 	&0.99957 	&0.99992 	&0.98246 	&0.99991& 0.99688 \\
%TL& 100\% & 0.99997 	&0.99952 	&0.99963 	&0.99993 	&0.98607 	&0.99992& \textbf{0.99751} \\
% & 50\% &  0.99998 &0.99952 	&0.99960 	&0.99993 	&0.98577 	&0.99993& 0.9945 \\

\red{TL-0} & \red{10\%} &  \red{\textbf{0.99987}}& 	\red{\textbf{0.99926}}& 	\red{\textbf{0.99923}}& 	\red{\textbf{0.99976}}& 	\red{\textbf{0.98135}}& 	\red{\textbf{0.99976}} & \red{\textbf{0.99753}}\\
\red{TL-1} & \red{10\%} & \red{0.99977}& 	\red{0.99866}& 	\red{0.99854}& 	\red{0.99945}& 	\red{0.97727}& 	\red{0.99973}& \red{0.99557} 	\\
\red{TL-2} & \red{10\%} & \red{0.99980}& 	\red{0.99923}& 	\red{0.99910}& 	\red{0.99958}& 	\red{0.97894}& 	\red{0.99957}& \red{0.99604}\\
\bottomrule
\end{tabular}
\end{table}

\section{Conclusion}\label{sec:conclu}
The foundation of our study originates from Ling \textit{et al.}'s TBNN \cite{ling2016} and its associated papers, which incorporate Galilean and rotational invariances to enhance the modeling of the Reynolds stress anisotropy tensor for RANS simulations using high-fidelity data. While previous studies based on the TBNN architecture showcased its predictive capabilities, some ambiguities persisted at the physical modeling level, particularly concerning the application of Pope's GEVM. Additionally, from a numerical perspective, achieving balanced predictions for the various components of the full anisotropic Reynolds stress tensor had always been a challenge.

Another crucial aspect of the modeling, related to the consistency between data and the model, is the coupling method between the ML turbulence model and the CFD solver. In Ling et al.'s paper, adopting the frozen substitution method, \textit{a posteriori} results demonstrated that TBNN improved predictions of the anisotropy tensor, leading to qualitatively more accurate velocity predictions compared to standard RANS. However, there remained a quantitative discrepancy with DNS results.

To address and surmount these previous limitations, while striving to combine accuracy and realisability, we reevaluated Ling et al.'s work within the context of wall-bounded flows, with a focus on an iterative coupling framework. We introduced aTBNN models, specifically tailored for PCF and SDF, which proved to be more suitable and significantly more accurate. The changes in the present aTBNN models preserve the same Galilean and rotational invariances as the original model. We placed particular emphasis on the selection of input features and basis tensors, as well as the influence of training batch size, activation functions, and other network hyperparameters on model performance.

On one hand, we incorporated additional physics features (normalized wall distances and the friction Reynolds number) into these models based on observations from DNS data and prior domain knowledge. Through cross-comparison with a more flexible MLP architecture in the PCF configuration, we demonstrated the necessity of these additional input features. The former addressed the multi-valued problem in predictions, while the latter enabled discrimination among flows at different turbulent levels in the near-wall region. \blue{However, it should be recalled that flows with very high Reynolds number will no longer subject to this Reynolds number dependency \cite{townsend1976}. As such, one might exercise caution while performing such extrapolations.} On the other hand, we proposed two optimal tensor basis models for our two target flows with different dimensions.

The aTBNN models are trained by state-of-the-art strategies to optimize a multi-part loss function, considering the contribution of each component of the anisotropy tensor. The performance of these models was validated on both the PCF and SDF, and they consistently provided excellent anisotropy tensor predictions in strong agreement with the reference DNS data. Notably, this agreement was achieved for both interpolation and extrapolation scenarios in the PCF using a shallow network of three hidden layers, even when testing at an unobserved friction Reynolds number. However, predicting the duct flow case presented more challenges due to its physical complexity and limited training data at various regimes. More specifically, the current aTBNN model for SDF failed during an interpolation test within the training range, possibly because of the intricate  dependence of the flow to the chosen Reynolds number \cite{pinelli2010}. We proposed a numerical strategy to mitigate this issue based on transfer knowledge from an aTBNN model trained by the SDF dataset at $\mathrm{Re}_\tau=[150; 250; \np{1000}]$ to $\mathrm{Re}_\tau=500$, showing that much improved performance and faster convergence can be achieved through TL with only \red{10}\% of the original dataset.

Several challenges remain for future studies. An extension of the present work would involve \textit{a posteriori} validations of the flow fields predicted by the aTBNN, coupled iteratively with a RANS solver, and a comparison with standard RANS and DNS references. This investigation is currently ongoing, integrating our neural networks into TrioCFD, an in-house developed RANS solver \cite{angeli2015, angeli2017}. Additionally, it would be valuable to expand our study to more complex flow configurations, including three-dimensional flow statistics and intricate phenomena such as recirculations or boundary layer separation and reattachment. These cases might require to turn to promising neural architectures, such as CNN, GNN, and transformers that have been utilized in some recent works \cite{saezdeocarizborde2021, saezdeocarizborde2022, quattromini2023, xu2023}. Finally, the field of machine-learning-assisted turbulence modeling offers numerous avenues for exploration, including interpretability and generalizability analysis \cite{saezdeocarizborde2021, jiang2021}, uncertainty quantification \cite{xiao2019, agrawal2023, gawlikowski2023}, conditioning problems \cite{wang2018, wu2019b}, full Partial Differential Equation (PDE) modeling \cite{jin2021, lucor2022}, and more \cite{cheng2023}.

\section*{CRediT authorship contribution statement}
\textbf{Jiayi Cai}: Conceptualization, Methodology, Software, Formal analysis, Investigation, Writing - original draft. \textbf{Pierre-Emmanuel Angeli}: Conceptualization, Methodology, Formal analysis, Supervision, Funding acquisition, Writing - Review \& Editing. \textbf{Jean-Marc Martinez}: Methodology, Software, Investigation. \textbf{Guillaume Damblin}: Conceptualization, Supervision, Writing - Review \& Editing. \textbf{Didier Lucor}: Conceptualization, Supervision, Writing - Review \& Editing.

\section*{Acknowledgements}
This work was granted access to the HPC resources of TGCC under the allocations 2019-A0062A10806 and 2020-A0092A10806 attributed by GENCI (Grand \'Equipement National de Calcul Intensif).

\appendix

\setcounter{table}{0}

\section{Preprocessing} \label{app:A}
\renewcommand{\thetable}{A.\arabic{table}}

The detailed preprocessing of each input and output of our neural networks are summarized in Table.~\ref{tab:pre-pro}.

\begin{table}[h!]
	\centering
	\caption{\label{tab:pre-pro}Preprocessing of raw inputs and outputs. The notation is as follows: $\widetilde{x}$ is $x$ after preprocessing, $\mathrm{max}(x)$ indicates the maximum value of x in the training set, $\mathrm{sgn}(x)$ indicates the sign of x, $\mathrm{abs}(x)$ indicates the absolute value of x, and $m$ is the number of training samples. }
\begin{tabular}{lccc}
	\toprule
	Case & Raw input/output & Description & Preprocessed input/output   \\ \hline
	\\[-0.5em]
	
	PCF & $\alpha$ & Normalized mean velocity gradient & $\widetilde{\alpha}=\dfrac{\alpha}{\mathrm{max} (\alpha)}$ \\
	\\[-0.5em]
	& ${y}^+$ & Normalized wall distance & $\widetilde{y}^+=\dfrac{\mathrm{log}(y^+)}{\mathrm{max} (\mathrm{log}(y^+)}$ \\
	\\[-0.5em]
	& $\mathrm{Re}_\tau$ & Friction Reynolds number & $\widetilde{\mathrm{R}}\mathrm{e}_{\tau}=\dfrac{\mathrm{Re}_\tau}{\mathrm{max} (\mathrm{Re}_\tau)}$ \\
	\\[-0.5em]
	& $\textbf{b}$ & Reynolds stress tensor & $\widetilde{\textbf{b}} = \dfrac{\textbf{b}}{\sqrt{\dfrac{1}{m} \sum\limits_{k=1}^m \sum\limits_{\substack{p \leq q \\ p, q \in \{1, 2, 3\}}} \left({b}_{k, pq} \right)^2}}$\\
	\\[-0.5em]
	SDF & $\lambda_i^*$ & Invariants & $\widetilde{\lambda_i^*}=\mathrm{sgn}(\lambda_i^*) \times \mathrm{log}(1 + \mathrm{abs}(\lambda_i^*))$ \\
	\\[-0.5em]
	& ${y}^+$ & Normalized wall distance & $\widetilde{y}^+=\mathrm{log}( 1 + y^+)$ \\
	\\[-0.5em]
	& ${z}^+$ & Normalized wall distance & $\widetilde{z}^+=\mathrm{log}( 1 + z^+)$ \\
	\\[-0.5em]	
	& $\mathrm{Re}_\tau$ & Friction Reynolds number & $\widetilde{\mathrm{R}}\mathrm{e}_{\tau}=\dfrac{\mathrm{Re}_\tau}{\mathrm{max} (\mathrm{Re}_\tau)}$ \\
	\\[-0.5em]
	& $\mathbf{T}^{*(i)}$ & Basis tensors & $\widetilde{\textbf{T}^{*(i)}} = \dfrac{\textbf{T}^{*(i)}}{\sqrt{\dfrac{1}{m} \sum\limits_{k=1}^m \sum\limits_{\substack{p \leq q \\ p, q \in \{1, 2, 3\}}} \left({T}_{k, pq}^{*(i)} \right)^2}}$	\\
	\\[-0.5em]
	& $\textbf{b}$ & Reynolds stress tensor & $\widetilde{\textbf{b}} = \dfrac{\textbf{b}}{\sqrt{\dfrac{1}{m} \sum\limits_{k=1}^m \sum\limits_{\substack{p \leq q \\ p, q \in \{1, 2, 3\}}} \left({b}_{k, pq} \right)^2}}$\\	
	
	\bottomrule
\end{tabular}
\end{table}

\color{black}
%\printbibliography

\newpage
\bibliographystyle{elsarticle-num}
\bibliography{article_try}

\end{document}